\documentclass[preprint,aps,showpacs,preprintnumbers,amsmath,amssymb,nofootinbib]{revtex4}
\usepackage{graphicx}
\usepackage{epsfig}

\begin{document}

\begin{flushright}
\end{flushright}

\newcommand{\be}{\begin{equation}}
\newcommand{\ee}{\end{equation}}
\newcommand{\bea}{\begin{eqnarray}}
\newcommand{\eea}{\end{eqnarray}}
\newcommand{\nn}{\nonumber}
\def\CP{{\it CP}~}
\def\cp{{\it CP}}

\title{\large A comprehensive study of the discovery potential of NO$\nu$A, T2K and T2HK experiments }
\author{Soumya C., K. N. Deepthi and R. Mohanta }
\affiliation{School of Physics, University of Hyderabad, Hyderabad - 500 046, India }


\begin{abstract}
With the recent measurement of reactor mixing angle $\theta_{13}$ the knowledge of neutrino oscillation parameters that describe PMNS matrix has improved 
significantly except the CP violating phase $\delta_{CP}$. The other unknown parameters in neutrino oscillation studies are  mass hierarchy and the octant of the atmospheric mixing angle $\theta_{23}$.
Many dedicated experiments are proposed to determine these parameters which may take at least 10 years from now to become operational. 
It is therefore very crucial to use the results from the existing experiments to see whether we can get even partial answers to these questions. 
In this paper we study the discovery potential of the ongoing NO$\nu$A and T2K experiments  as well as the forthcoming T2HK experiment in addressing 
these questions. In particular, we evaluate the sensitivity of NO$\nu$A to determine neutrino mass hierarchy, octant degeneracy and to obtain CP violation 
phase after running for its scheduled period of 3 years in neutrino mode and 3 years in anti-neutrino mode. 
We then extend the analysis  to understand the discovery potential if the experiments will run for (5$\nu$+5$\bar{\nu}$) years and (7$\nu$+3$\bar{\nu}$) years. 
We also show how the sensitivity improves when we combine  the data from (3$\nu$+3$\bar{\nu}$) years of NO$\nu$A run with (3$\nu$+2$\bar{\nu}$) years of T2K  and (3$\nu$+7$\bar{\nu}$) years of T2HK
experiments.  The CP violation sensitivity is marginal for T2K and NO$\nu$A experiments  even for ten years data taking of NO$\nu$A. 
T2HK has a significance above 5$\sigma$ for a fraction of two-fifth values of the $\delta_{CP}$ space. 
We also find that $\delta_{CP}$ can be determined to be
better than $35^\circ $, $21^\circ $ and $9^\circ $ for all values of $\delta_{CP}$ for T2K, NO$\nu$A and T2HK respectively. 
\end{abstract}

\maketitle

\section{Introduction}
The discovery of neutrino oscillations has firmly established that neutrinos are massive. It has marked the beginning of many neutrino oscillation experiments. 
The mixing of three neutrino flavors can be described by Pontecorvo-Maki-Nakagawa-Sakata matrix $U_{PMNS}$ \cite{pmnsa1,pmnsa2}, which is parameterized 
in terms of
three rotation angles $\theta_{12}^{}$, $\theta_{23}^{}$, $\theta_{13}^{}$ and three CP-violating phases, one Dirac type  ($\delta_{CP}$) and two Majorana types 
($\rho$ and $ \sigma$). The neutrino oscillation data accumulated over many years allows us to determine the solar and atmospheric neutrino oscillation 
parameters with very high precision. Recently, the reactor mixing angle $\theta_{13}$  has been  measured precisely 
\cite{daya-bay,daya-bay1,reno,t2k,t2k-result} with a  moderately large value, quite close to its previous upper bound.
The observation of the largish value of  $\theta_{13}$ has attracted a lot of attention
in recent times to understand the mixing pattern in the lepton sector. Furthermore, it also  opens up promising prospects
for the observation of  leptonic CP violation \cite{branco, valle}.
As $\theta_{13}$  is non-zero, there
could be CP-violation in the lepton sector, analogous to the quark sector, provided the CP
violating phase $\delta_{CP}$ is not vanishingly small.

The current results from recent neutrino oscillation experiments \cite{minos, minos1,chooz,t2k colb} and their global analysis 
\cite{gfit1,gfit2,gfit3,gfit4} performed by several groups, have implied that the minimal three neutrino framework is adequate to describe the 
observed oscillation phenomenology. The best-fit values and the $3\sigma$ ranges of the oscillation parameters from Ref. 
\cite{gfit4}, are presented in Table-1.

\begin{table}[htb]
\begin{center}
\vspace*{0.1 true in}
\begin{tabular}{|c|c|c|}
\hline
 Mixing Parameters & Best Fit value & $ 3 \sigma $ Range  \\
\hline
$\sin^2 \theta_{12} $ &~ $0.323$ ~& ~$ 0.278 \to 0.375 $~\\
&&\\
$\sin^2 \theta_{23} $ (NH) &~ $0.567$ ~& ~$ 0.392 \to 0.643 $~\\

$\sin^2 \theta_{23} $ (IH) &~ $0.573$ ~& ~$ 0.403 \to 0.640 $~\\
&&\\
$\sin^2 \theta_{13} $ (NH) &~ $0.0234$ ~& ~$ 0.0177 \to 0.0294 $~\\

$\sin^2 \theta_{13} $ (IH) &~ $0.0240$ ~& ~$ 0.0183 \to 0.0297 $~\\
&&\\
$\Delta m_{21}^2/ 10^{-5}~{\rm eV}^2 $ & $ 7.6 $ & $ 7.11 \to 8.18 $ \\
&&\\
$|\Delta m_{31}|^2/ 10^{-3}~ {\rm eV}^2 ~({\rm NH}) $~ &~ $ 2.48
$ & $ 2.30 \to 2.65 $ \\

$|\Delta m_{31}|^2/ 10^{-3} ~{\rm eV}^2 ~({\rm IH}) $ ~&~ $ 2.38 $ & $ 2.20 \to 2.54 $ \\

\hline
\end{tabular}
\end{center}
\caption{The best-fit values and the  $3\sigma$ ranges of the neutrino oscillation parameters from Ref.  \cite{gfit4}.}
\end{table}

Another important discovery in recent times is the precision measurement of $\sin^2  \theta_{23}$ by MINOS experiment \cite{minos2}, which is 
found to be non-maximal. Using the complete set of accelerator and atmospheric data they disfavored the maximal mixing by $-2 \Delta \log ({\cal L})=1.54$. 
They obtained the best-fit value for the mixing angle $\theta_{23}$ as $\sin^2 \theta_{23}=0.41$, known as lower octant (LO) and 
 $\sin^2 \theta_{23}=0.61$, the so-called higher octant (HO) values.

With these exciting discoveries of non-zero $\theta_{13}$ and non-maximal $\theta_{23}$, the focus of neutrino oscillation studies has now been 
shifted towards the determination of other unknown parameters. These include the determination of  mass hierarchy,  
octant of the atmospheric mixing angle $\theta_{23}$, discovery of CP violation  and the magnitude of the CP violating phase $\delta_{CP}$. 
In this paper we would like to investigate the prospects of  addressing  these issues with the  off-axis long-baseline experiments T2K, NO$\nu$A and T2HK
with updated experimental specifications.
The experimental specifications of these experiments are briefly described below.

$\bullet $ NO$\nu$A \cite{nova} is an off-axis long baseline neutrino oscillation experiment designed to study $\nu_{\mu} \rightarrow \nu_{e}$ appearance measurements using Fermilab NuMI muon neutrino beam ($\nu_{\mu}$). Its secondary aim is to precisely measure $\nu_{\mu}$ disappearance parameters. It uses a high intensity proton beam with a beam power of 0.7 MW with $6 \times 10^{20}$ POT/year. Its detector is a 14 kt totally active liquid scintillator detector (TASD) located at Ash River, 810 km from Fermilab. The detector is located slightly off the centerline (14 mrad) to the neutrino beam where one can find a large flux of neutrinos of 2 GeV energy. The oscillation from $\nu_{\mu} \rightarrow \nu_{e}$ is expected to be maximum at this energy. It is scheduled to run 3 years in $\nu$ mode followed by 3 years in $\bar{\nu}$ mode. The detector properties of NO$\nu$A considered in our simulations  are taken from
Ref. \cite{ska12} with the following characteristics as given in Table-II.

\begin{table}[htb]
\begin{center}
\vspace*{0.1 true in}
\begin{tabular}{|c|c|}
\hline
 Signal efficiency & 45\% for $\nu_{e}$ and $\bar{\nu}_{e}$ signal ~ \\
                           &100\% $\nu_{\mu}$ CC and $\bar{\nu}_{\mu}$ CC ~ \\
\hline
 Background efficiency & 0.83\% $\nu_{\mu}$ CC, 0.22\% $\bar{\nu}_{\mu}$ CC ~ \\
                        & 2\% $\nu_{\mu}$ NC, 3\% $\bar{\nu}_{\mu}$ NC ~\\
                        & 26\% (18\%)  $ \nu_{e}$ ($\bar{\nu}_{e}$) beam contamination ~ \\
\hline
NC background smearing & migration matrices ~ \\
\hline
Systematics & 5\% signal normalization error ~\\
                   & 10\% background normalization error ~\\
\hline
\end{tabular}
\end{center}
\caption{Details of NO$\nu$A detector characteristics considered in this analysis.}  
\end{table}

$\bullet $ T2K (Tokai-to-Kamiokande) is a currently running long-baseline experiment designed to study neutrino oscillations. 
An intense $\nu_{\mu}$ beam of 0.77 MW power is directed from J-PARC to Super-Kamiokande detector,  295 km away. 
It has a 22.5 kt water Cherenkov detector. The details of T2K experiment can be found from \cite{t2k-3}.
We have considered input files for T2K from the General Long Baseline Experiment Simulator (GLoBES) package \cite{t2k-2,t2k-3,t2k-4} 
and the  updated experimental description of T2K 
are taken from \cite{updated-t2k}. We have matched our results with the ones presented in \cite{updated-t2k}. In this analysis,
we have used ($3 \nu$ + $2\bar{\nu}$) running modes for T2K.

$\bullet$ T2HK (Tokai-to-Hyper-Kamiokande) is a future long baseline experiment which is expected to be operational around 2023. 
It can be considered as a natural advancement to the ongoing T2K experiment. It has same baseline and off-axis angle as T2K experiment. 
It uses J-PARC's neutrino experimental facilities with an improved beam power (7.5 MW) and 1 Mt volume water Cherenkov detector, 
Hyper-Kamiokande (Hyper-K). We have considered a fiducial volume of 0.56 Mt, beam power of 7.5 MW and 
other specifications are taken from \cite{updated-t2hk}. 
Hence, T2HK will have high statistics of neutrino events compared to T2K. These  features of T2HK make it as one of the most sensitive 
experiment to probe neutrino CP violation.  
The input files for T2HK obtained from GLoBES package \cite{t2k-2,t2k-3,t2k-4}. 
The primary objective of this experiment is the discovery of CP asymmetry.

Since these experiments use $\nu_\mu$ beam and also will run in antineutrino mode their main focus is to study the appearance 
$(\nu_\mu \to \nu_e)$ and the disappearance channels $(\nu_\mu \to \nu_\mu)$ along with their antineutrino counterparts. 
Since the leading term in the appearance channels $\nu_\mu \to \nu_e ~(P_{\mu e})$ and the corresponding antineutrino mode 
$\bar \nu_\mu \to \bar \nu_e ~(P_{\bar \mu \bar e})$ is proportional to $\sin^2 2 \theta_{13} \sin^2 \theta_{23}$ and 
with the observation of moderately large value of $\theta_{13}$, these experiments are well-suited for the determination of 
mass hierarchy and the octant of $\theta_{23}$. Although, the ongoing NO$\nu$A and T2K experiments are not planned to measure 
$\delta_{CP}$ or to explore CP violation in the neutrino sector, we would like to investigate whether it is possible to constrain 
the $\delta_{CP}$ phase using the data from these two experiments. In other words, how much of the $\delta_{CP}$ space can be ruled out  
by these experiments within the next 10 years. In particular, we would like to investigate

$\bullet $ whether the combination of T2K (3+2) and NO$\nu$A (3+3) provide more quantitative answer on the above posed questions than each one of these experiments.

$\bullet$  how the sensitivity on $\delta_{CP}$,   mass
hierarchy and $\theta_{23}$ octant will improve if NO$\nu$A runs for 10 years in the (5+5) and (7+3)
combination of modes.

$\bullet$ the sensitivities of T2HK experiment for its scheduled run of 3 years in neutrino  and 7 years in anti-neutrino mode.

The paper is  organized as follows. In section II we briefly describe the physics reach of these experiments. 
The prospect of octant resolution and mass hierarchy determination are discussed in section III and IV. Section V contains the CP violation discovery potential
 and the correlations between the CP violating phase $\delta_{CP}$ and the mixing angles $\theta_{13}$ and $\theta_{23}$. We summarize our results in Section VI.

\section{Physics reach}

As discussed before, the determination of the mass hierarchy,  octant of the atmospheric mixing angle $\theta_{23}$ and the search for 
CP violation in the neutrino sector are the important physics goals of the current and future oscillation experiments. A simple way to  
achieve the above three goals is to measure the oscillation probabilities  $P(\nu_\mu \to \nu_e)$  and $P(\bar \nu_\mu \to \bar \nu_e)$. 
This can be seen from the expression for probability of oscillation from $\nu_{\mu}(\bar\nu_{\mu}) \rightarrow \nu_{e} (\bar\nu_{e})$ 
\cite{akhmedov,cervera,freund}, where we have kept terms only first order in $\sin \theta_{13}$ and $\alpha = \Delta m_{21}^2/\Delta m_{31}^2$

\begin{eqnarray}\label{prob}
P(\nu_\mu &\to & \nu_e)  \approx  \sin^22\theta_{13}\sin^2\theta_{23}\frac{\sin^2(\hat A-1)\Delta}{(\hat A-1)^2} \nn\\
                & +&\alpha\cos\theta_{13} \sin 2\theta_{12}\sin2\theta_{13} \sin 2 \theta_{23}\frac{\sin \hat A\Delta}{\hat A}
\frac{\sin(\hat A-1)\Delta}{(\hat A-1)}\cos(\Delta+\delta_{CP})\;,
\end{eqnarray}
 where $\Delta m_{ij}^2=m_i^2-m_j^2$,  $\Delta =\Delta m_{31}^2L/4E$ and   $\hat A = 2 \sqrt 2 G_F n_e E/\Delta m_{31}^2$. $G_F$ is the Fermi coupling constant and $n_e$ is 
the electron number density. The transition probability can be enhanced or suppressed depending on the oscillation parameters 
$\theta_{13}$, $\theta_{23}$, mass hierarchy, i.e., the sign of $\Delta m_{31}^2$ and CP violation phase $\delta_{CP}$. Parameters $\alpha$, 
$\Delta$ and $\hat A$ are sensitive to neutrino mass ordering. For neutrinos, $\hat A$ is positive for normal hierarchy (NH) and negative for 
inverted hierarchy (IH), while its  sign changes  when we go from neutrino to anti-neutrino mode. Moreover, sign of $\delta_{CP}$  
is reversed for anti-neutrinos.

It should be noted from Eq. (\ref{prob}) that the leading term in the transition probability $P(\nu_\mu \to \nu_e)$  is proportional to $\sin^2 2 \theta_{13} \sin^2 \theta_{23}$. Therefore, the observed moderately
large value of $\theta_{13}$ makes it possible for the current generation long-baseline experiments to address the problems of hierarchy and 
the octant of $\theta_{23}$ determination. The second term in Eq. (\ref{prob}) shows the prominence of matter effect on the oscillation probability. The  dependency of all the terms  on a moderately large reactor neutrino mixing angle $\theta_{13}$ suggests that NO$\nu$A detector will be able to collect a good number of $\nu_{\mu}(\bar\nu_{\mu}) \rightarrow \nu_{e} (\bar\nu_{e})$ events.

First, we will try to see whether the energy spectrum information will help us in resolving the octant degeneracy and mass hierarchy. 
We  use GLoBES package \cite{Huber:2004gg,Huber:2009xx}  for the simulation to obtain the energy spectra. In our analysis, we consider the following 
true values for the oscillation parameters as provided in Table-III, unless mentioned otherwise.

\begin{table}[h]
\begin{center}
\vspace*{0.1 true in}
\begin{tabular}{|c|c|}
\hline
 $\sin^2\theta_{12}$ & 0.32 ~ \\
\hline
 $\sin^2 2\theta_{13}$ & 0.1 ~ \\
\hline
 $\sin^2 \theta_{23}$ & 0.41 (LO), 0.59 (HO) ~ \\
\hline
$\Delta m_{atm}^2$ & $2.4 \times 10^{-3} ~{\rm eV}^2$ for NH ~ \\
                                & $-2.4 \times 10^{-3} ~{\rm eV}^2$ for IH ~ \\
\hline
$\Delta m_{21}^2$ & $7.6 \times 10^{-5}~ {\rm eV}^2$ ~ \\
\hline
$\delta_{CP} $ & $0^\circ$  ~ \\
\hline
\end{tabular}
\end{center}
\caption{The true values of oscillation parameters considered in the simulations.}
\end{table}

\begin{figure}[htb]
\begin{center}
\includegraphics[width=7cm,height=5cm, clip]{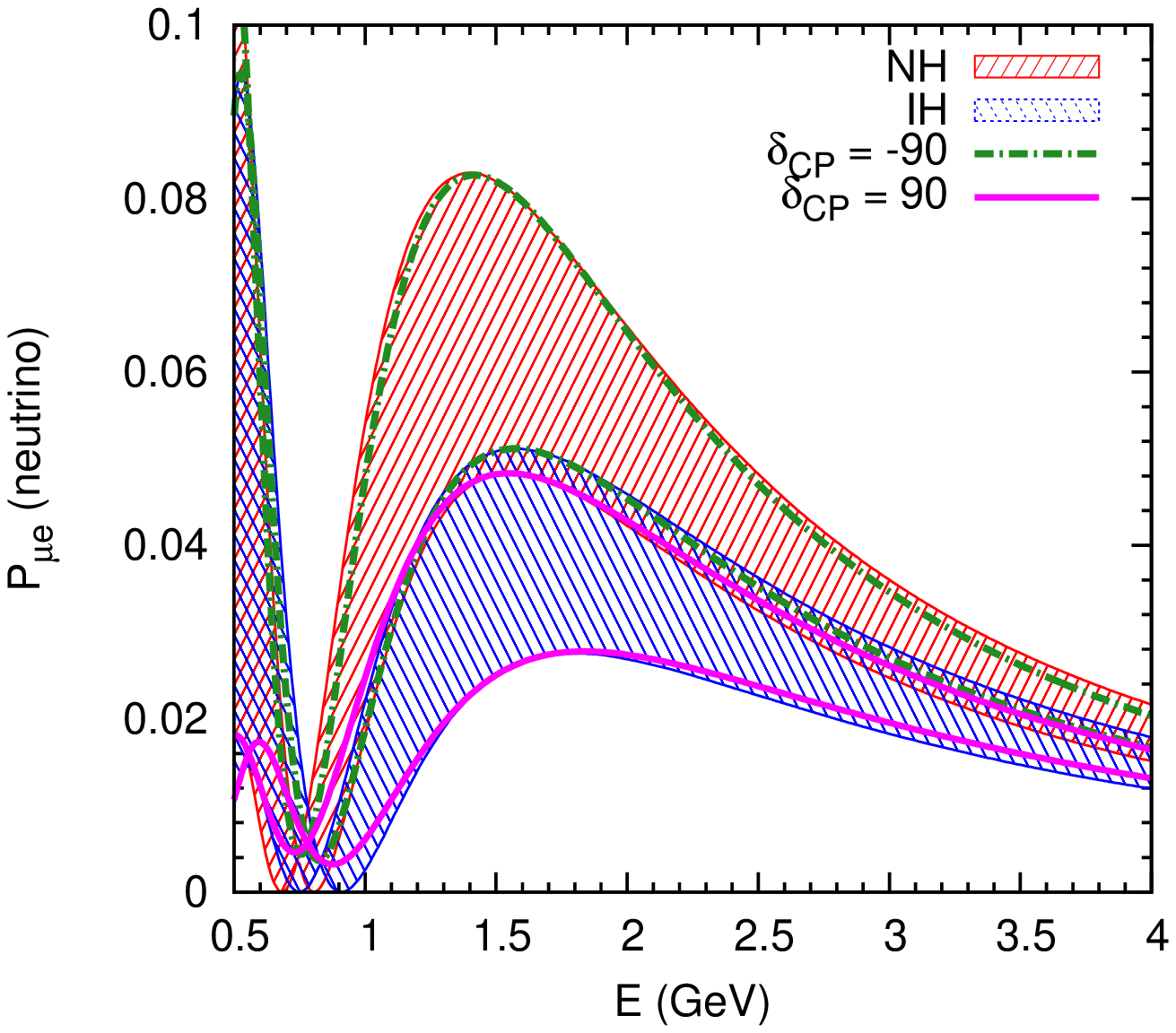}
\hspace{0.2 cm}
\includegraphics[width=7cm,height=5cm, clip]{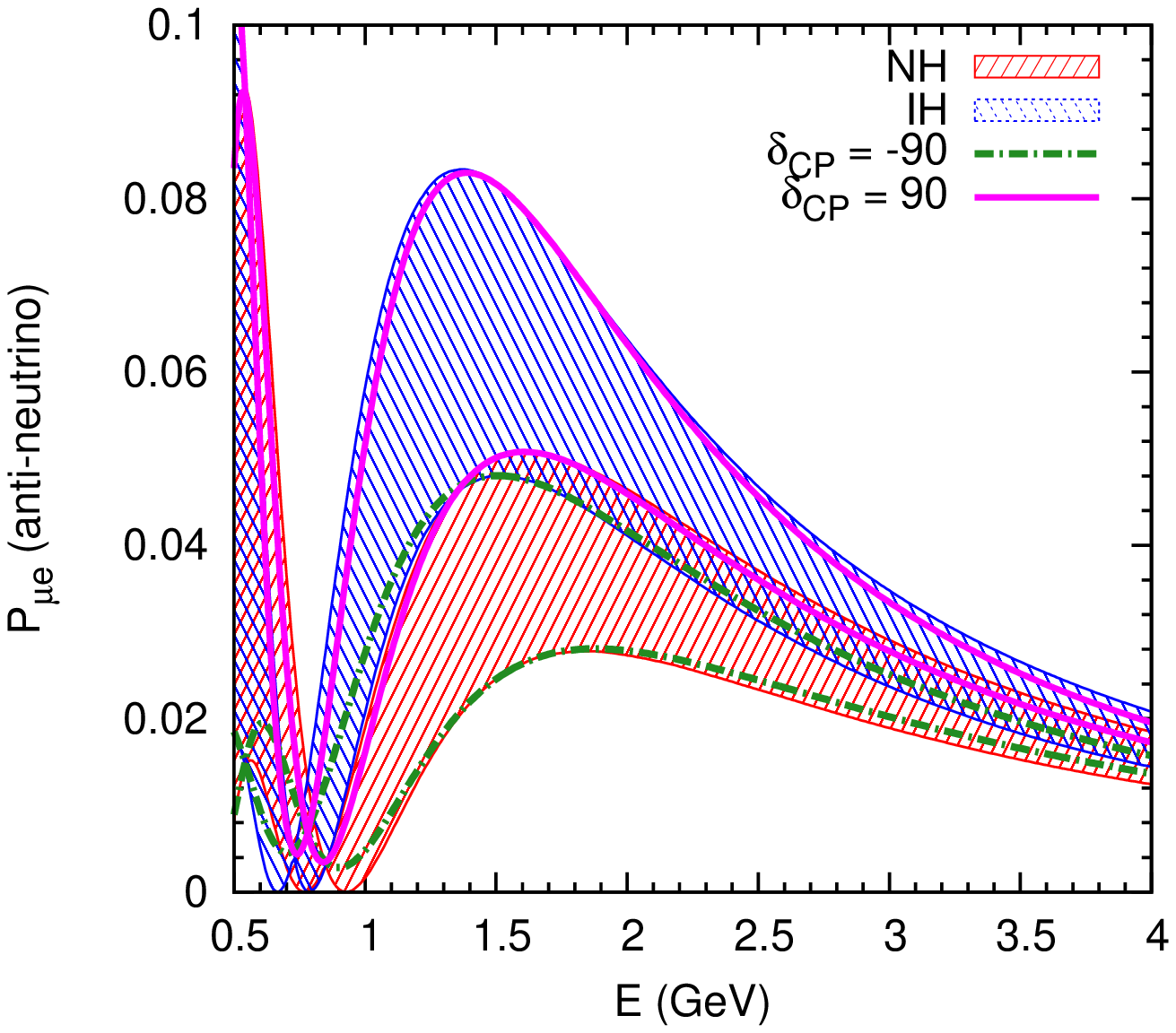}
\includegraphics[width=7cm,height=5cm, clip]{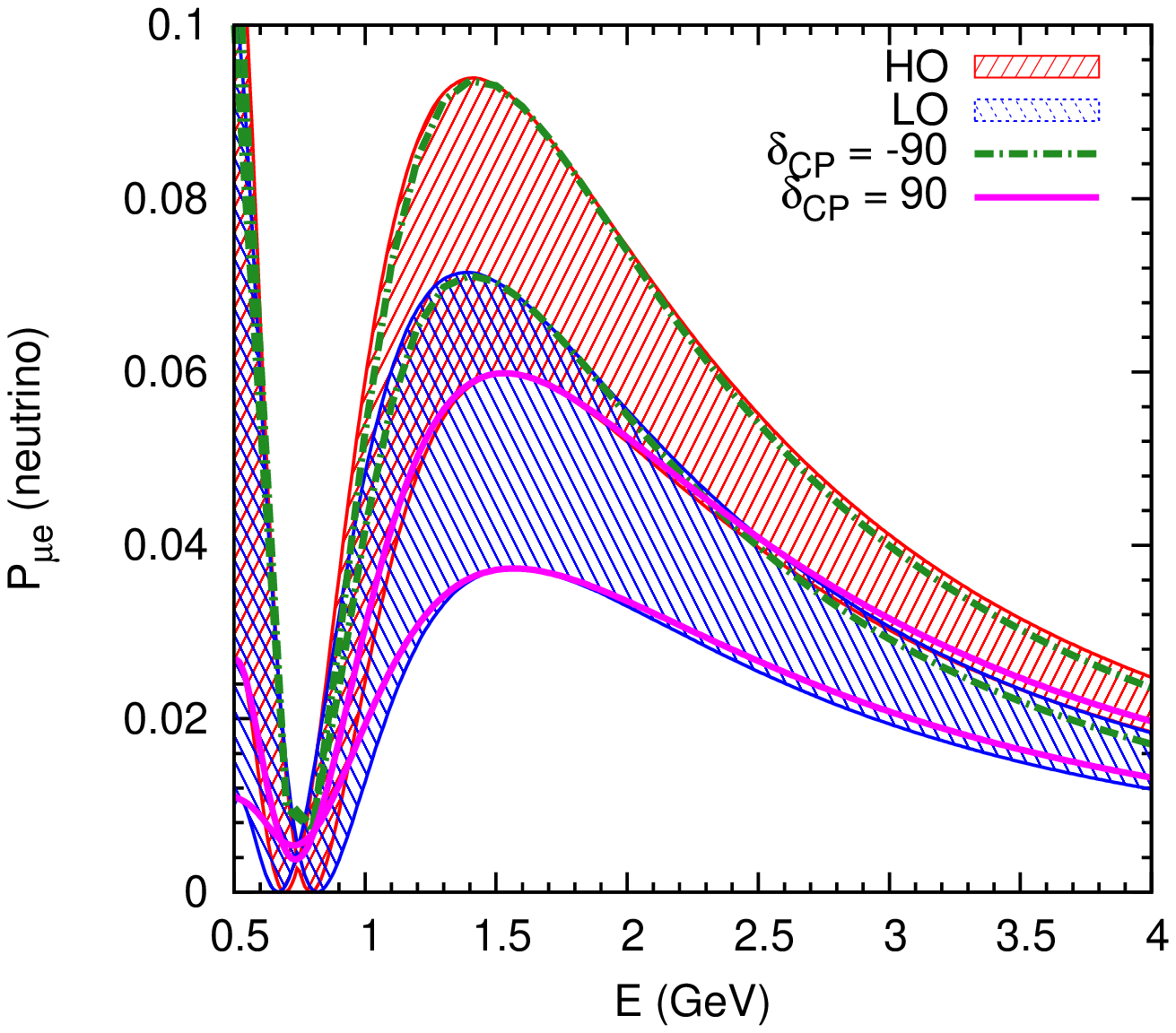}
\hspace{0.2 cm}
\includegraphics[width=7cm,height=5cm, clip]{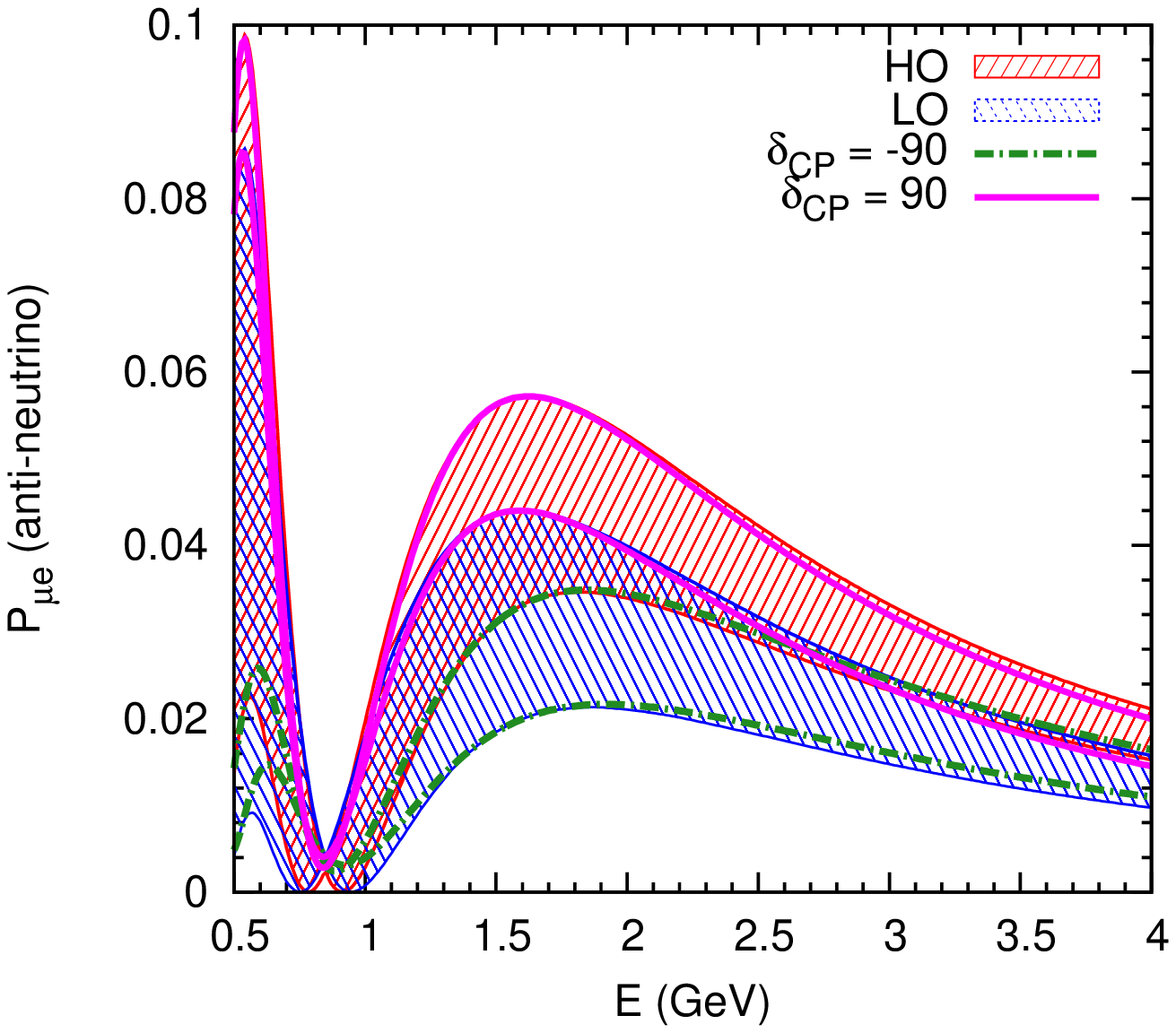}
\end{center}
\caption{$P_{\mu e}$ energy spectrum for NO$\nu$A experiment. The left (right) panel is for neutrino
(antineutrino). The red (blue) band in the top panel corresponds to  NH (IH), where we have used
 $\sin^2 \theta_{23}=0.5$, $\sin^2 2 \theta_{13}=0.1$, baseline $L=810 $ km and vary $\delta_{CP}$
 between ($-\pi$ to $\pi$). The red (blue)  band in the bottom panel is for HO (LO), where we have used
 $\sin^2 \theta_{23}=0.41 ~(0.59)$ for LO (HO) and keep the hierarchy as normal.  Inside each band the probability for $\delta_{CP}= 90^\circ (-90^\circ)$
 case is shown by magenta (green) line. }
\end{figure}

Fig.1, shows the energy spectrum of the appearance probabilities $P(\nu_\mu \to \nu_e)$ for neutrino (left panel) and $P(\bar \nu_\mu \to \bar \nu_e)$ for 
antineutrino (right panel) for NO$\nu$A experiment, where we have varied $\delta_{CP}$ within the range $-\pi$ to $\pi$. In each panel the red (blue) band 
is for NH (IH). Furthermore, in each band the probability for $\delta_{CP}=90^\circ$ and $\delta_{CP}=-90^\circ$ cases are shown explicitly by the magenta and 
green lines. Due to matter effect the probability $P_{\mu e}$ increases for NH and decreases for IH and vice versa for $P_{\bar \mu \bar e}$. Thus, for $\delta_{CP}$ lying in the lower half plane (LHP) i.e., $-180^\circ \leq \delta_{CP} \leq 0$, $P_{\mu e}$ is larger and for $\delta_{CP}$ in the upper half plane (UHP)  ( $0 \leq \delta_{CP}\leq 180^\circ$), $P_{\mu e}$ is much lower. The situations reverse for the antineutrino probability $P_{\bar \mu \bar e}$.  Thus, LHP is the favorable half-plane for NH and UHP is for IH for neutrino mode. However, the most unfavorable condition is (NH, $\delta_{CP}=90^\circ$) and (IH, $\delta_{CP}=-90^\circ$) as the bands almost overlap with each other for the entire energy range.
In the lower panels of Fig. 1, we show  the energy spectrum of $P_{\mu e}$ and $P_{\bar \mu \bar e}$
 for two different values of $\theta_{23}$ assuming NH to be true hierarchy. The blue band in both the panels is for  $\theta_{23}$ in the LO and 
red band is for $\theta_{23}$ in the HO.
As can be seen from the figures that the two bands overlap with each other for some values of $\delta_{CP}$ and distinct for others. The overlap regions are the unfavorable ones for the determination of the $\theta_{23}$ octant.

\section{Octant Resolution as a function of $\theta_{23}$}

In this section we present the results of our analysis on octant sensitivity of $\theta_{23}$ for  T2K, NO$\nu$A and T2HK experiments. 
We also show the results when the data from all the experiments are combined. Although the octant sensitivity of various long baseline experiments has 
been discussed extensively by many authors \cite{Oct, ska, ska1}, here we would like to revisit the octant resolution potential of 
these experiments with the updated specification details. For NO$\nu$A, we consider  its scheduled $(3\nu+3 \bar{\nu})$ years of run, for T2K 
$(3\nu+2 \bar{\nu})$ and for T2HK $(3\nu+7 \bar{\nu})$ years of run. Furthermore, we also investigate the situation if NO$\nu$A continues to run for
next 10 years what would be the potential for resolving octant degeneracy for $(5\nu+5{\bar{\nu}})$ as well as  $(7\nu+3{\bar{\nu}})$ years of running. 
We also see the synergy between T2K, NO$\nu$A and T2HK experiments for their scheduled runs. \\

The indistinguishability of $\theta_{23}$ and $(\pi/2-\theta_{23})$ is known as octant degeneracy. The relevant oscillation probability expressions for long baseline experiments  NO$\nu$A, T2K and T2HK with negligible matter effects are given as
\begin{equation}
{\mathrm{P^{v}_{\mu \mu}}}   =
1 -  \sin^2 2 \theta_{23} \sin^2\left[1.27 ~\frac{\Delta m_{31}^2L}{E} \right]
 +  4 \sin^2 \theta_{13}
\sin^2 \theta_{23}  \cos 2 \theta_{23}  \sin^2 \left[1.27 ~\frac{\Delta m_{31}^2 L}{E} \right],
\label{eq:pmmu}
\end{equation}
\begin{equation}
{\mathrm{P^{v}_{\mu e}}} =
{\mathrm{
\sin^2  \theta_{23} \sin^2  2\theta_{13} \sin^2\left[1.27 ~\frac{\Delta m_{31}^2L}{E} \right]
}}.\hspace{7.8 truecm}
\label{eq:pmmu}
\end{equation}
The leading order term in the $\nu_{\mu}$ survival probability (${\mathrm{P^{v}_{\mu \mu}}}$) depends on $\sin^2 2 \theta_{23}$ and one can't 
distinguish between ${\mathrm{P^{v}_{\mu \mu}}}(\theta_{23})$ and ${\mathrm{P^{v}_{\mu \mu}}}(\pi/2-\theta_{23})$. This kind of degeneracy that 
comes from the inherent structure of  neutrino oscillation probability is called intrinsic octant degeneracy. Whereas in the case of 
${\mathrm{P^{v}_{\mu e}}}$ the degeneracy of the octant with the parameter $\theta_{13}$ comes into play, since it  depends on the parameter 
combination $\sin^2  \theta_{23} \sin^2  2\theta_{13}$ . The  values of $\theta_{23}$ in opposite octant for different values of 
$\theta_{13}$ and $\delta_{CP}$ can have the same probabilities, $i.e,$ ${\mathrm{P^{v}_{\mu e}}}(\theta_{23},\theta_{13},\delta_{CP})=$ 
${\mathrm{P^{v}_{\mu e}}}(\pi/2-\theta_{23},\theta^{\prime}_{13},\delta^{\prime}_{CP})$. This also gives rise to octant degeneracy.


\begin{figure}[!htb]
\begin{center}
\includegraphics[width=5cm,height=5cm]{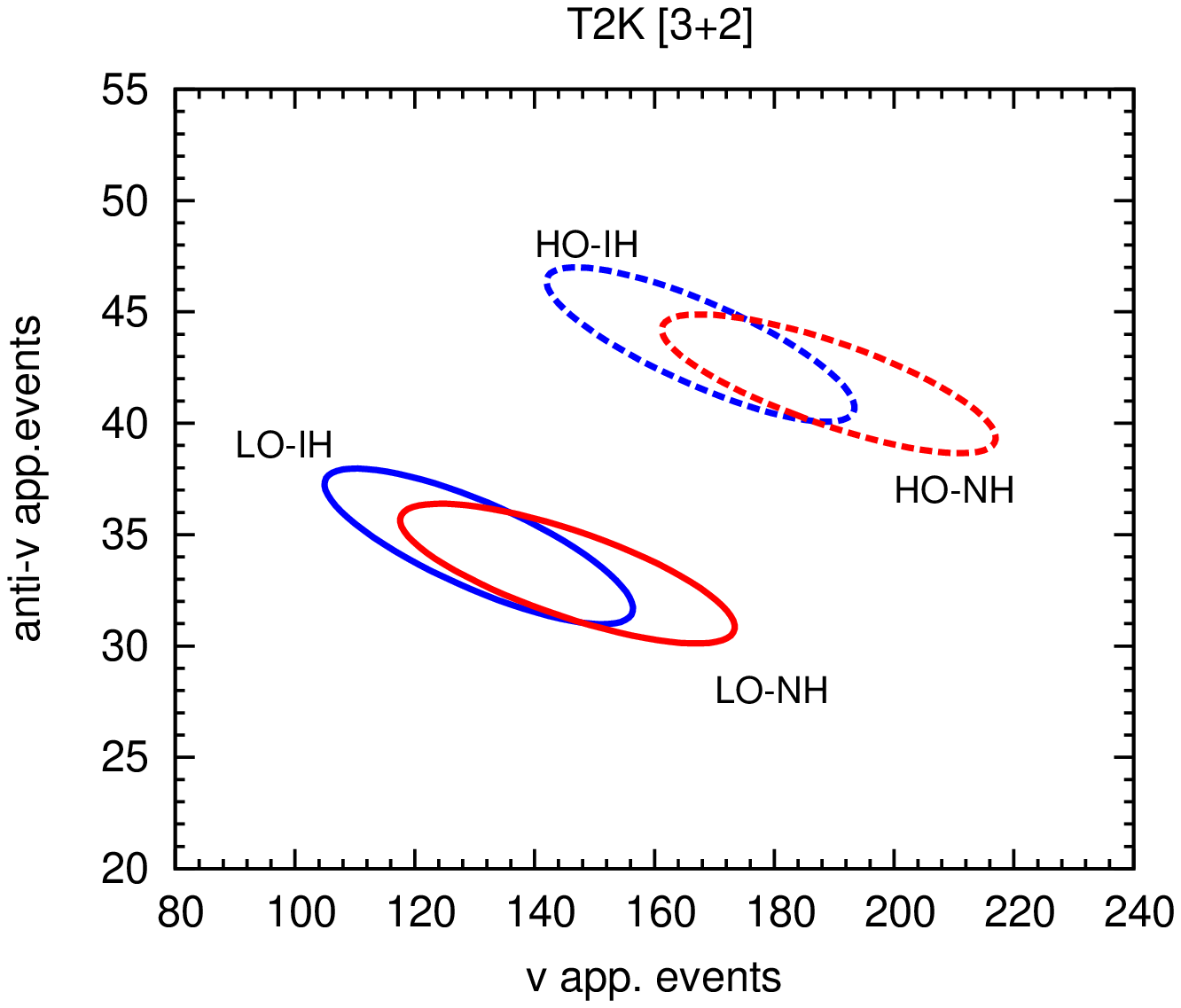}
\includegraphics[width=5cm,height=5cm]{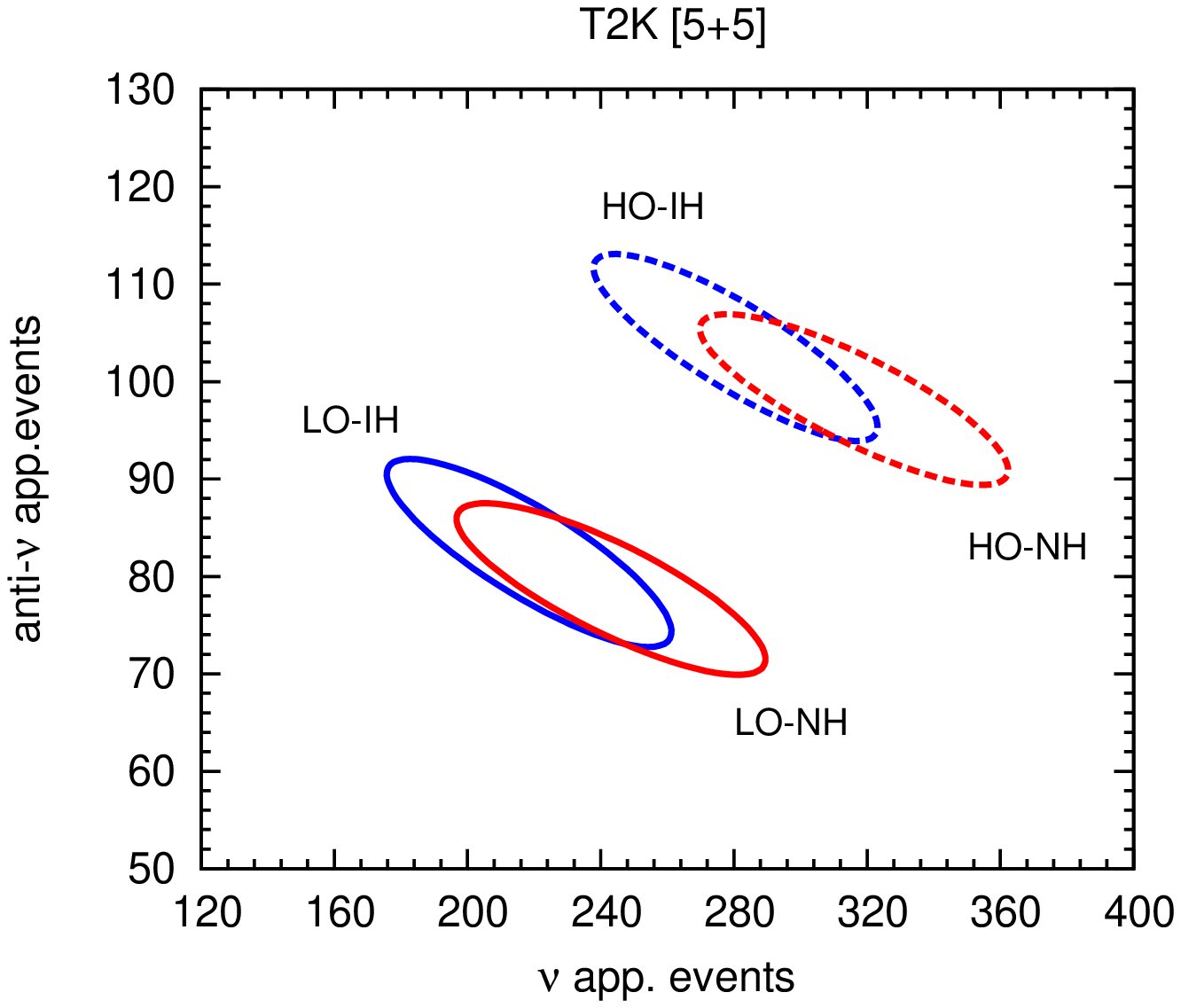}
\includegraphics[width=5cm,height=5cm]{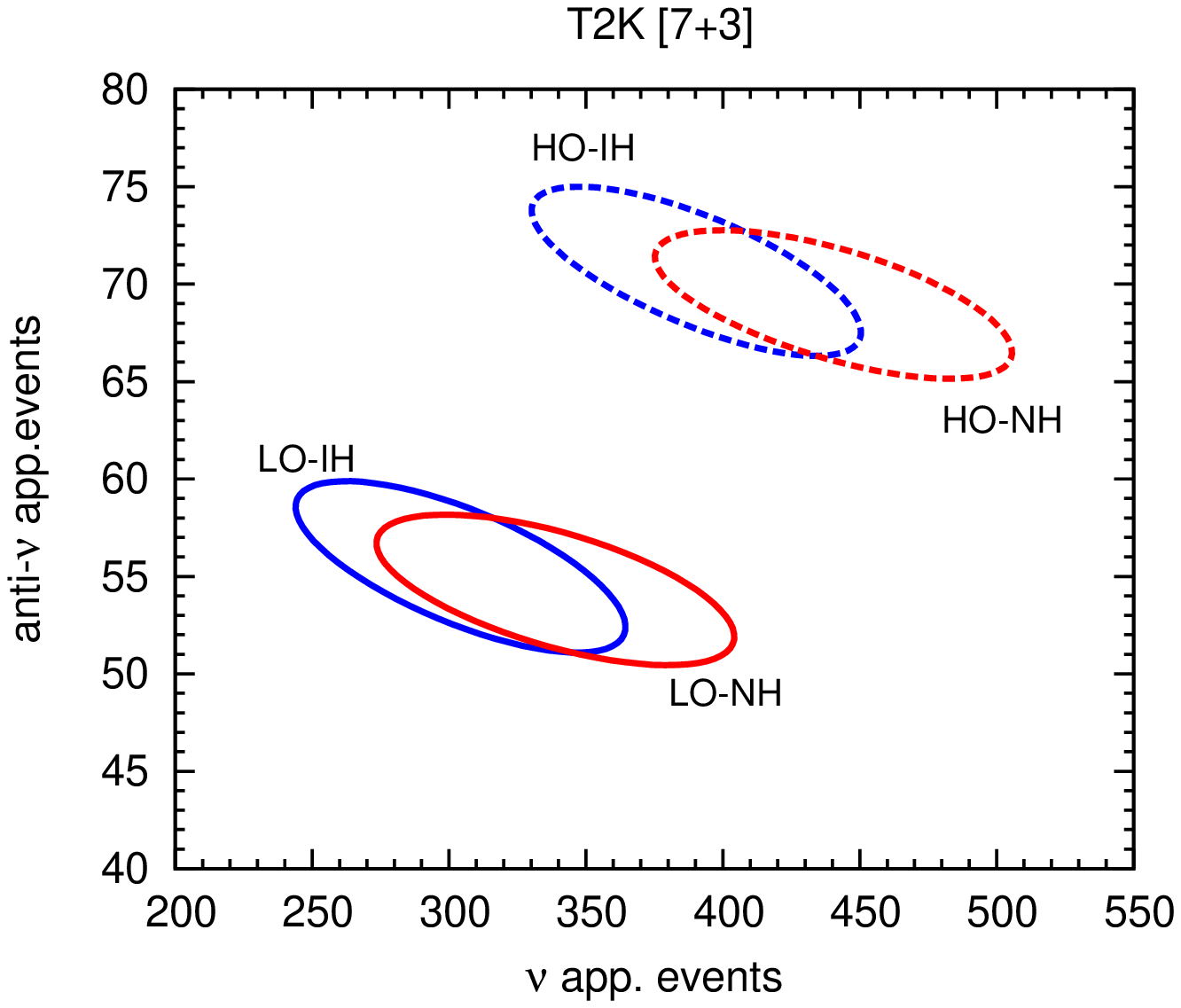}
\includegraphics[width=5cm,height=5cm]{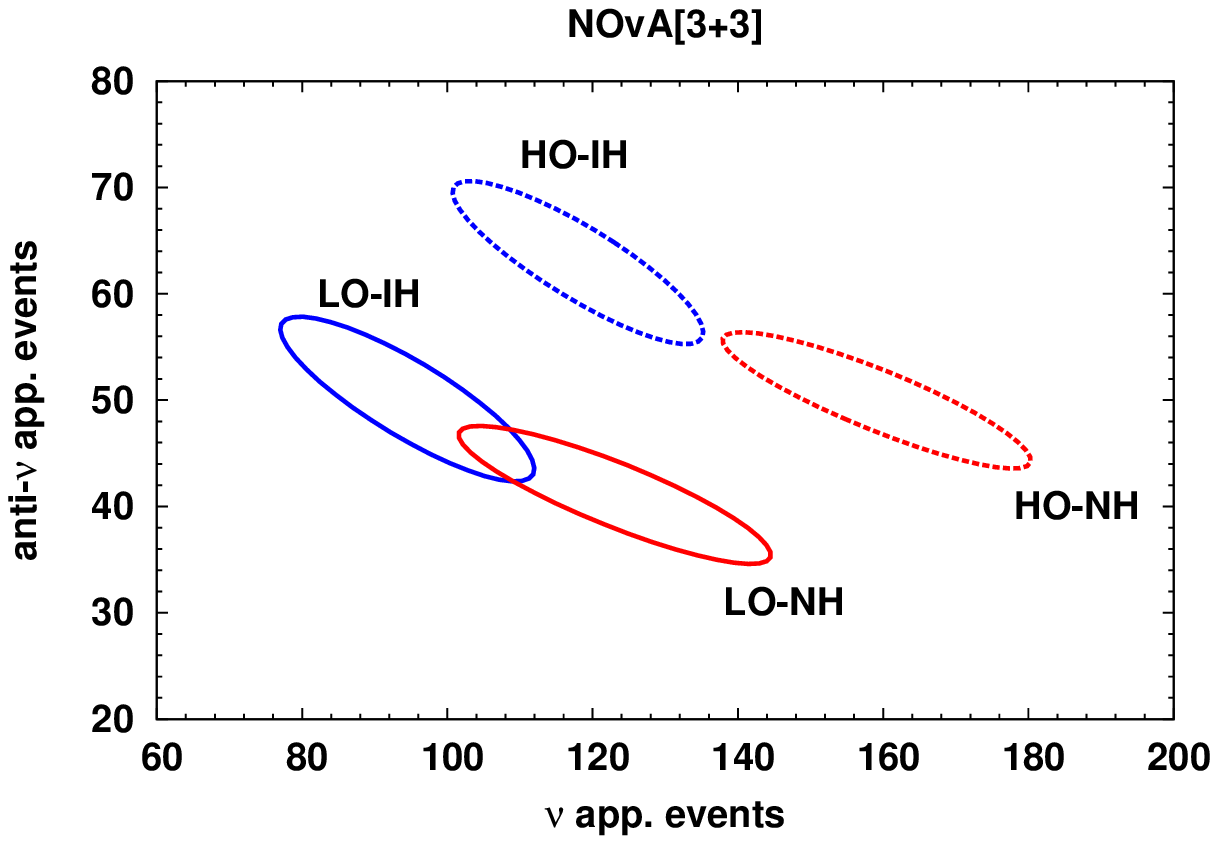}
\includegraphics[width=5cm,height=5cm]{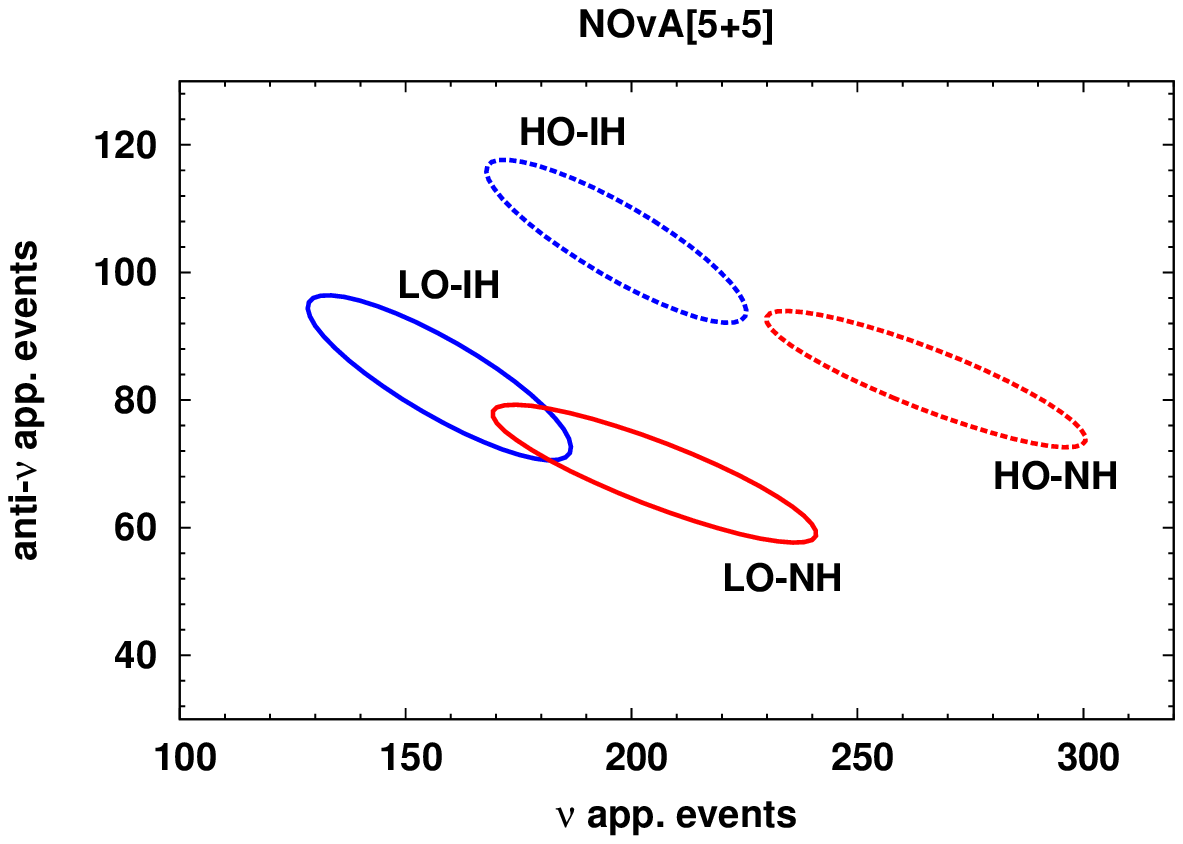}
\includegraphics[width=5cm,height=5cm]{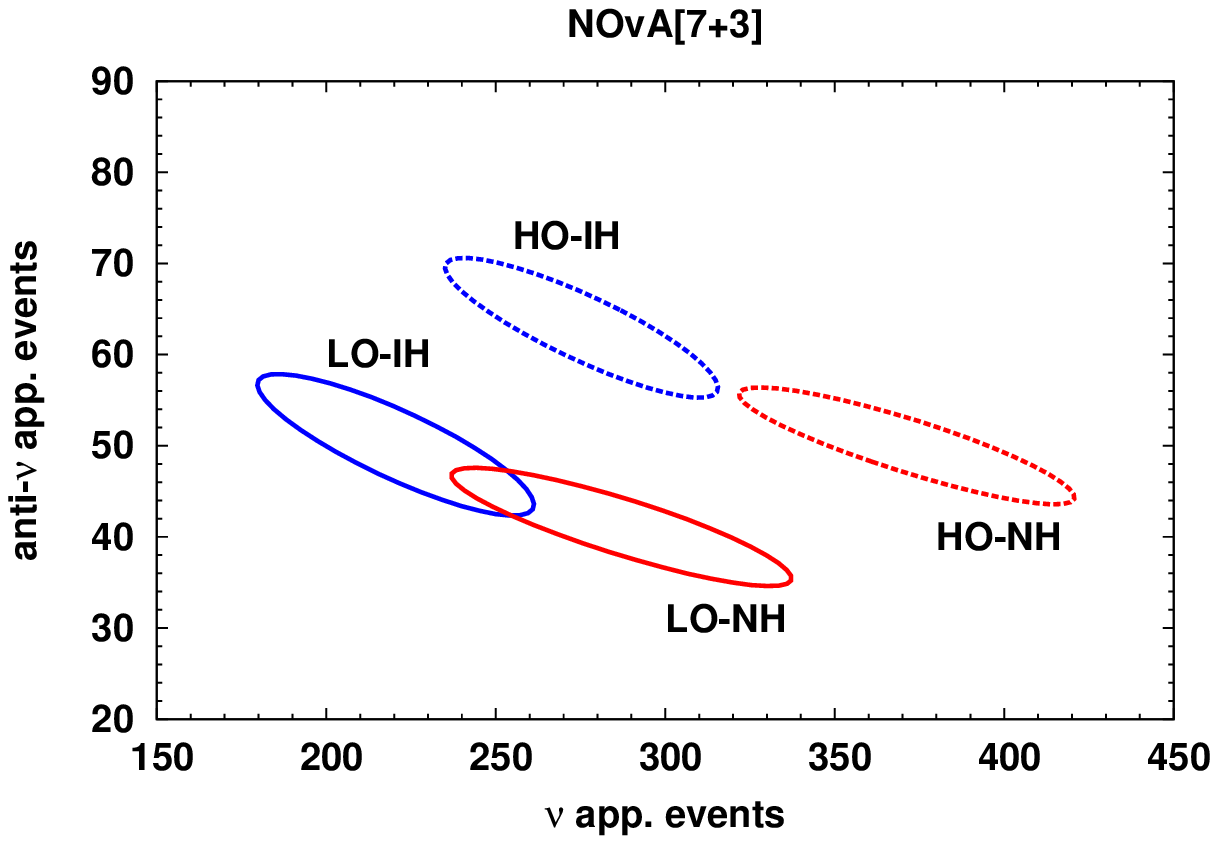}
\includegraphics[width=5cm,height=5cm]{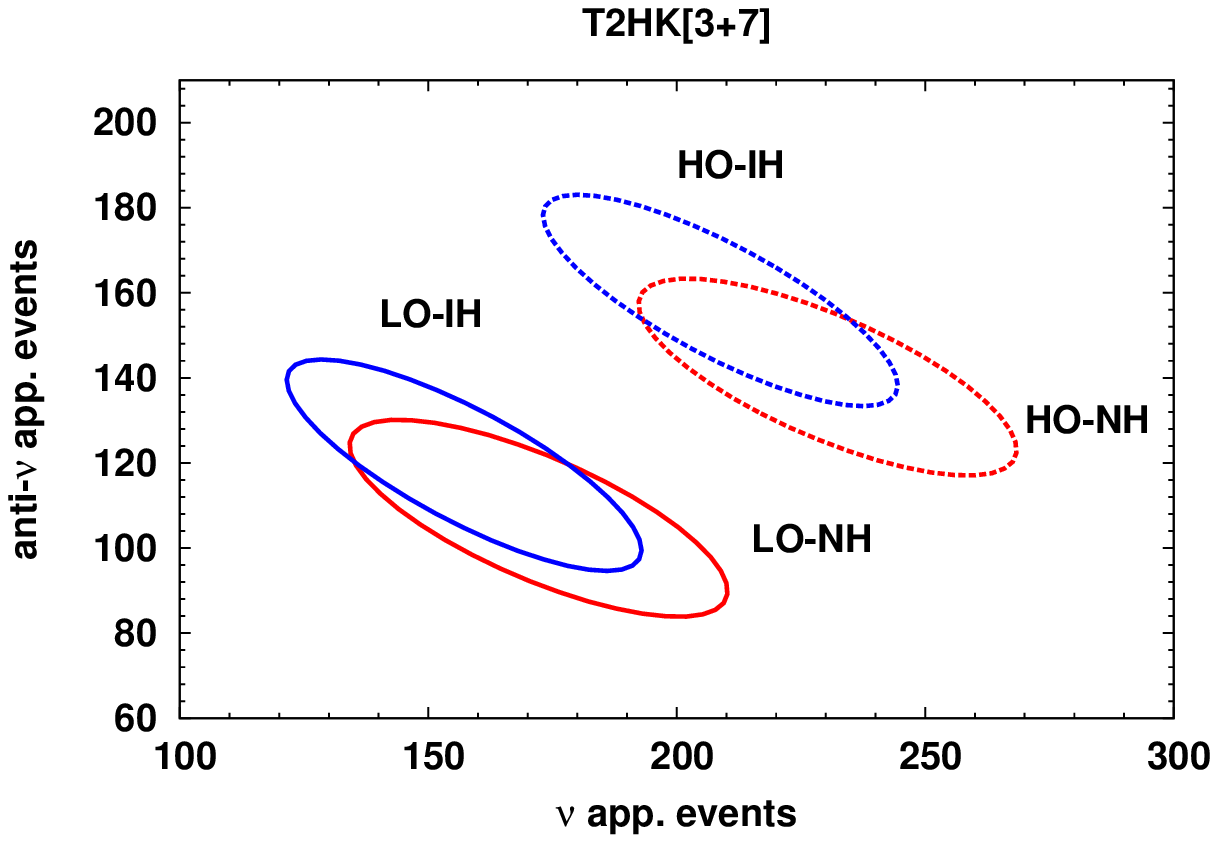}
\end{center}
\caption{Neutrino and antineutrino appearance events for the $\nu_{\mu} \rightarrow \nu_{e}$ versus $\bar \nu_{\mu} \rightarrow \bar\nu_{e}$
channels  by assuming both IH and NH and for lower and higher octants of $\theta_{23}$.}
\end{figure}


Before presenting the main results, we would like to discuss, what one can expect about the determination of
mass hierarchy and octant of $\theta_{23}$ from the bi-probability plots i.e., neutrino-antineutrino appearance event rates. Fig. 2 shows $\nu$ versus 
$\bar{\nu}$ events for all octant-hierarchy combinations. The blue curves are obtained by considering inverted hierarchy mass ordering, 
and both the values of $\theta_{23}$ i.e., $\sin^{2}\theta_{23} = 0.41$ (LO) and ~0.59 (HO). The red curves are obtained by considering normal 
hierarchy mass ordering and LO/HO values for $\sin^{2}\theta_{23}$. These ellipses are plotted by obtaining event spectra for (3+3) yrs, (5+5) yrs, 
(7+3) yrs in $\nu$ and $\bar{\nu}$ mode for all values of $\delta_{CP}$ for NO$\nu$A, T2K and (3+7) yrs for T2HK. Each point on $x$-axis ($y$-axis) 
represents the number of events measured by the respective experiments in neutrino (anti-neutrino) mode. The top panel represents the ellipses for (3+2), (5+5) and (7+3) years of running in neutrino and antineutrino modes for T2K, the second panel represents the NO$\nu$A event rates for (3+3), (5+5) and (7+3) years of run period and the bottom panel represents the T2HK event rates for (3+7) years of run period. For T2K and T2HK experiments the ellipses of both normal as well as inverted mass
orderings overlap with each other for both the octants whereas for NO$\nu$A the overlap region is less (marginal) for LO (HO). Thus, it is very likely that the mass hierarchy and octant degeneracy could be probed better with the NO$\nu$A experiment.

Before doing the simulation, here we would like to emphasize that the relation between  atmospheric parameters $(\Delta m_{atm}^2)$ and $\theta_{\mu\mu}$, 
measured in MINOS, and the  standard oscillation parameters
in nature are given as \cite{rel1,rel2}
\begin{equation}
  \sin\theta_{23} =\frac{\sin\theta_{\mu\mu}}{\cos\theta_{13}}\;,\hspace*{8.6 true cm}
\end{equation}
\begin{equation}
  \Delta m_{31}^2 =\Delta m_{atm}^2 +(\cos^2\theta_{12} - \cos\delta_{CP} \sin\theta_{13} \sin 2\theta_{12} \tan \theta_{23}) \Delta m_{21}^2\;.
\end{equation}
It is clear from the above relations that the observed value of moderately large $\theta_{13}$ significantly affects the oscillation parameters. So here we use  corrected  definitions of these parameters to analyze octant sensitivity. We allocate measured values $\Delta m_{atm}^2 $ and $\theta_{\mu\mu}$ and calculate oscillation probabilities in terms of $\Delta m_{31}^2 $  and  $\theta_{23}$.

We consider the  true values of oscillation parameters given in Table-III and  vary the test  values of $\sin^2\theta_{23}$ in LO (HO) for 
true higher octant (lower octant). We  also marginalize over  $\sin^2 2\theta_{13}$ in the range $[0.07:0.13]$, $\delta_{CP}$ in its full range,
 $\Delta m_{atm}^2$ in the range $[2.05:2.75] \times 10^{-3}~{\rm eV}^2$ for NH. The  parameters $\theta_{12}$ and $\Delta m_{21}^2$  have been kept fixed 
in the analysis and priors for $\sin^2 2\theta_{13}$ and $\sin^2\theta_{23}$ with $\sigma(\sin^2 2\theta_{13})=0.01$ and  $\sigma(\sin^2\theta_{23}) =0.05$
are also added.

We simulate the long baseline experiments T2K and  NO$\nu$A using the GLoBES package. For T2K, we assume 3 years of running in neutrino mode and 
2 years in antineutrino mode and for NO$\nu$A, we consider 3 years of neutrino running followed by 3 years of antineutrino running. 
 Furthermore, we also consider the case if NO$\nu$A continues 
the data taking for ten years beyond its scheduled (3+3) years and perform the analysis for two possible combinations (5+5) and (7+3) years of running.

\begin{figure}[htb]
\begin{center}
\includegraphics[width=7cm,height=5cm, clip]{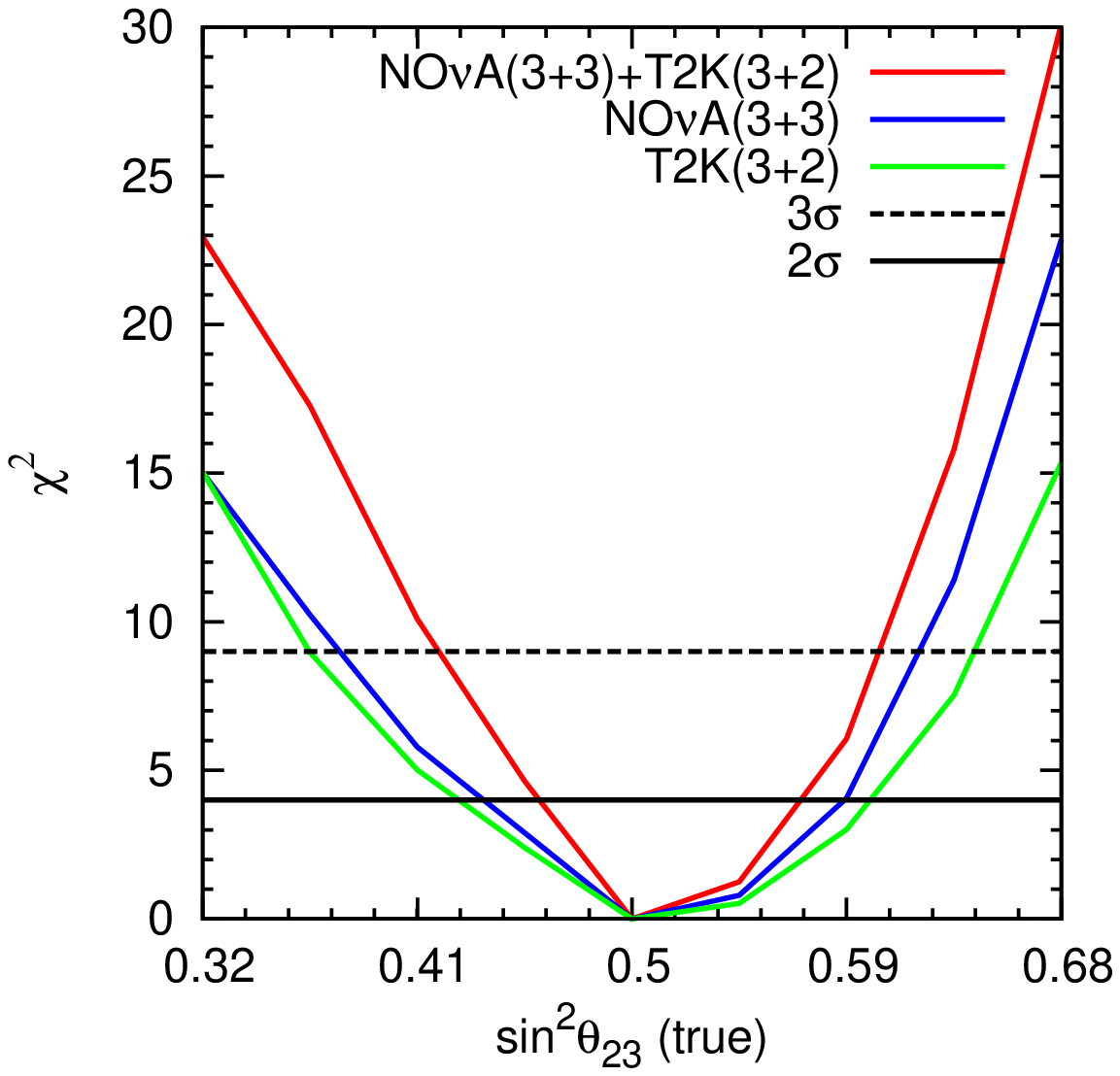}
\hspace{0.2 cm}
\includegraphics[width=7cm,height=5cm, clip]{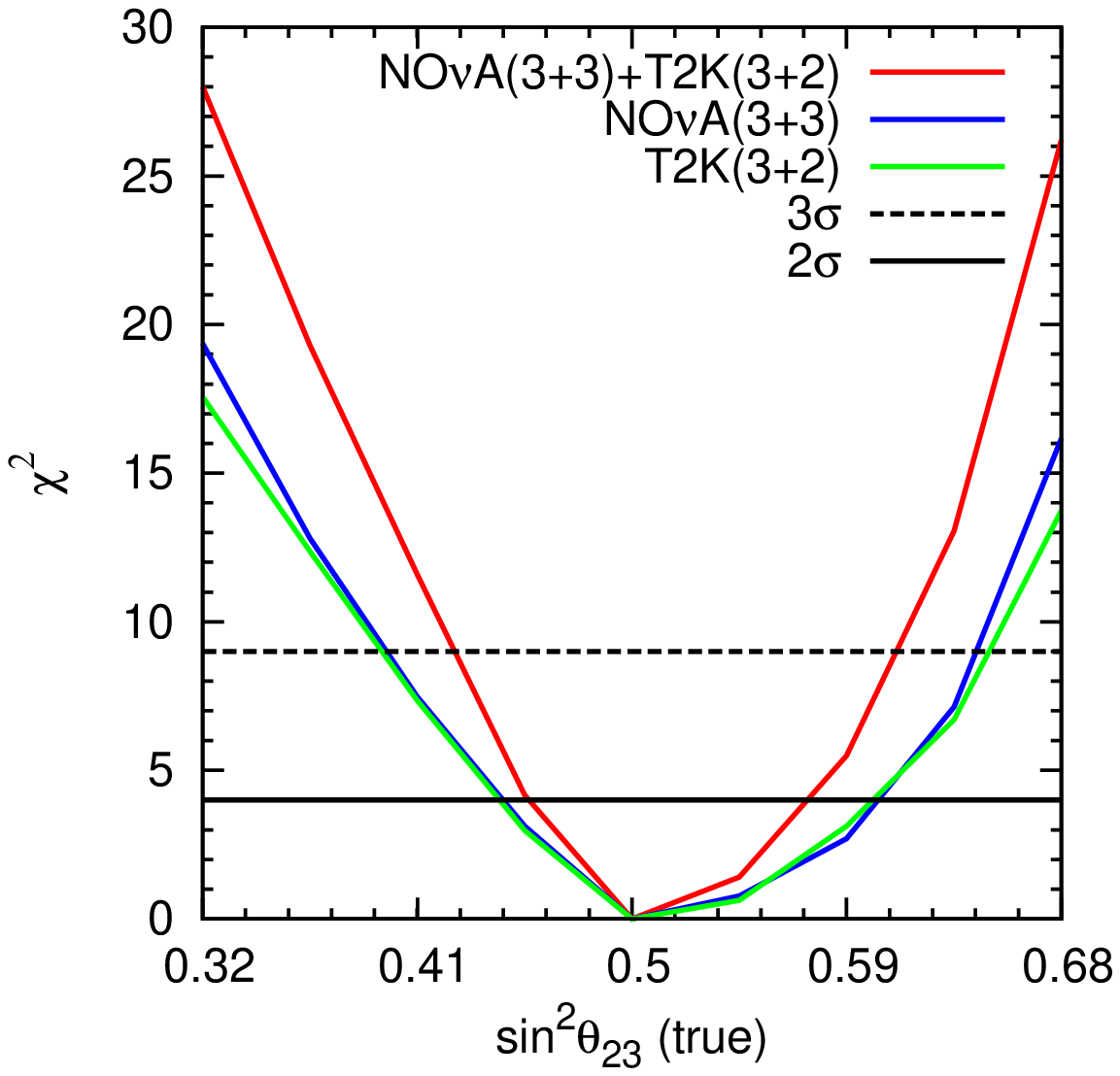}
\includegraphics[width=7cm,height=5cm, clip]{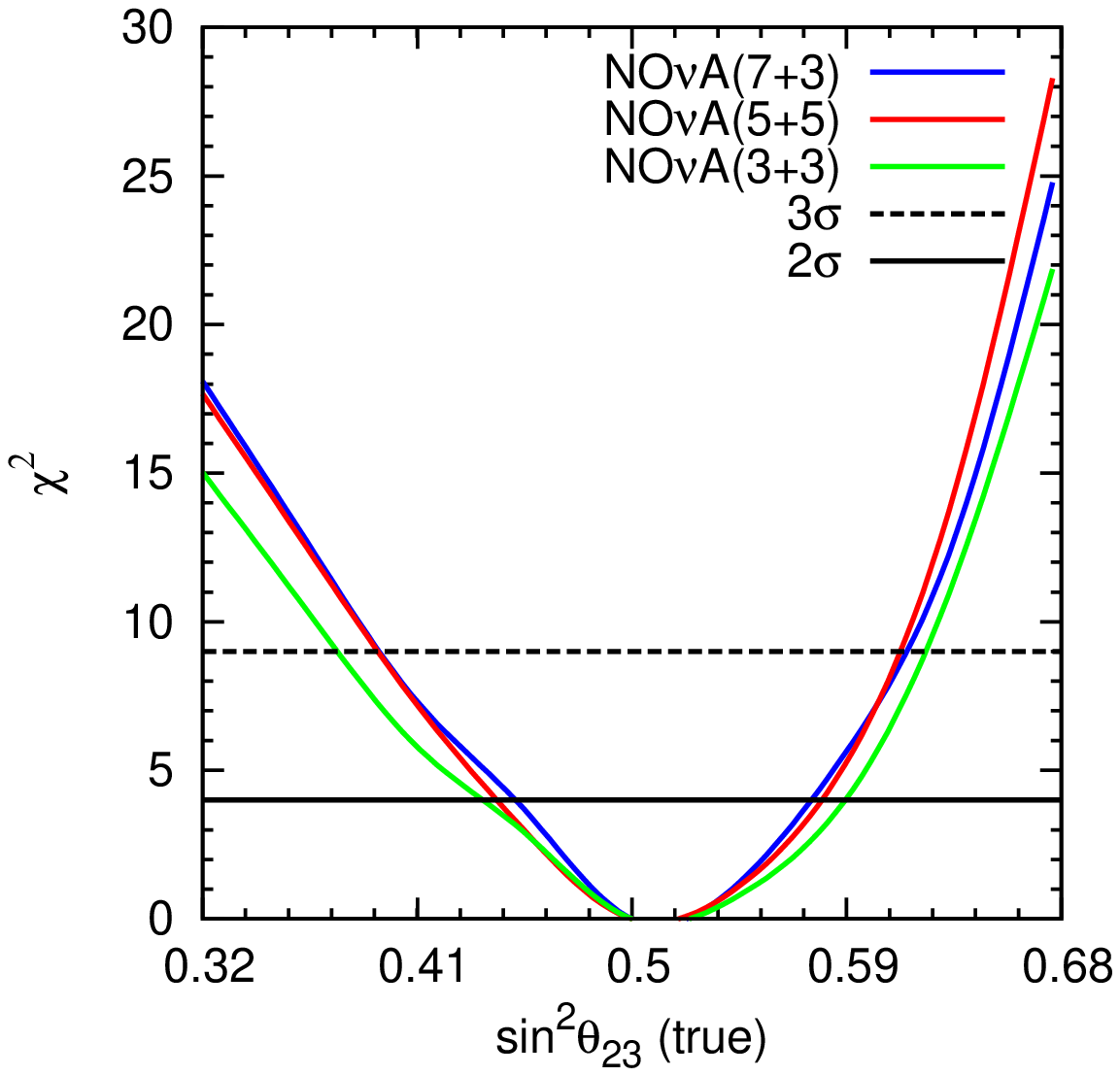}
\hspace{0.2 cm}
\includegraphics[width=7cm,height=5cm, clip]{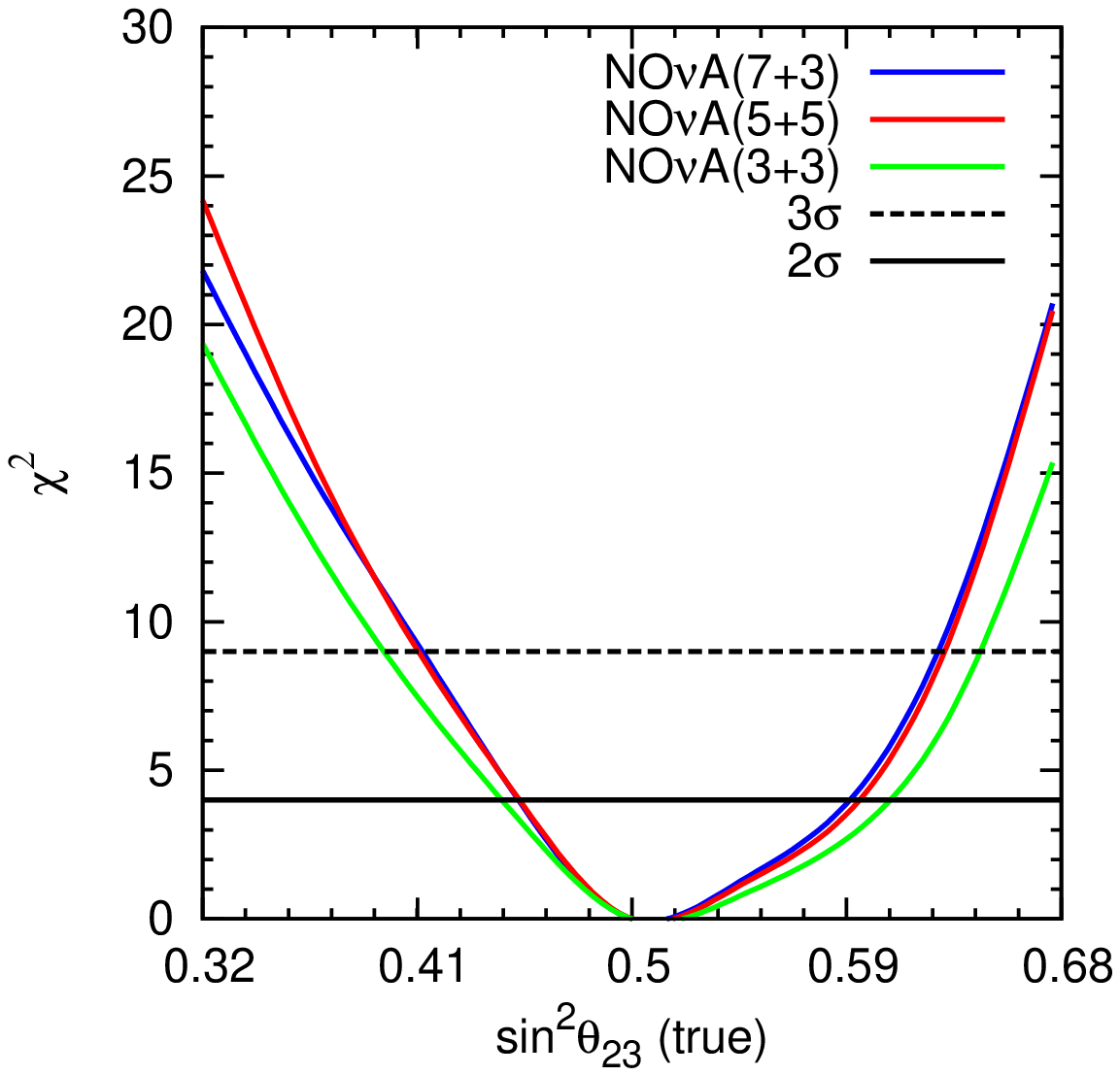}
\end{center}
\caption{Octant sensitivity for a combination of T2K and NO$\nu$A for the case of Normal (left panel) and Inverted (right panel)
hierarchy. }
\end{figure}

In Fig. 3, we illustrate the ability of NO$\nu$A experiment to determine the octant as a function of the true value of $\theta_{23}$. 
The values of  $\chi^{2}$  are evaluated using the standard rules as described in  GLoBES. The green, red and blue curves (in the bottom panel) 
represent the octant resolution of NO$\nu$A with  $(3+3)$,
$(5+5)$ and $(7+3)$ yrs of runs in $\nu$ and $\bar{\nu}$ modes respectively. From  Fig. 3, it can be seen that with only T2K data of (3+2) years of run, 
it is possible to resolve the octant degeneracy with $2 \sigma$ significance if the true $\sin^2 \theta_{23}$ will lie around 0.41 (LO) or 0.59 (HO) 
and one can have a better sensitivity for NO$\nu$A experiment with (3+3) yrs of run period. The significance increases significantly if we 
combine the data from both T2K and NO$\nu$A as seen from the top panels. For ten years of NO$\nu$A run, although we get a better sensitivity than that 
of (3+3) yrs of run,  there is no significant difference between (5+5) yrs and (7+3) years of running.\\

\section{Mass Hierarchy Determination}
Determination of neutrino mass hierarchy is one of the outstanding issues in neutrino oscillation physics.
The conventional method to achieve this is by using matter effects in very long baseline neutrino oscillation experiments, 
as the matter effects enhance the separation between oscillation spectra, and therefore, the event spectra between the normal and inverted hierarchy. In this section we describe the capabilities of T2K, NO$\nu$A and T2HK experiments for the determination of mass hierarchy.

The value of  $\chi^2$ has  been obtained by using the true parameters as listed in Table-2, except that of $\sin^2 \theta_{23}$, which is taken to be 0.5.
The true value of $\delta_{CP}$ is varied within its full range, i.e., between $[-\pi,\pi]$ and test value of 
 $\Delta m_{atm}^2$ is varied in IH (NH) range for true NH (IH). We  also marginalize over 
 $\sin^2 2 \theta_{13}$ and $\sin^2 \theta_{23}$ in their $3\sigma$ ranges and  add prior to $\sin^2 2\theta_{13}$ with 
$\sigma(\sin^2 2\theta_{13})=0.01$.
 
\begin{figure}[htb]
\begin{center}
\includegraphics[width=7cm,height=5cm, clip]{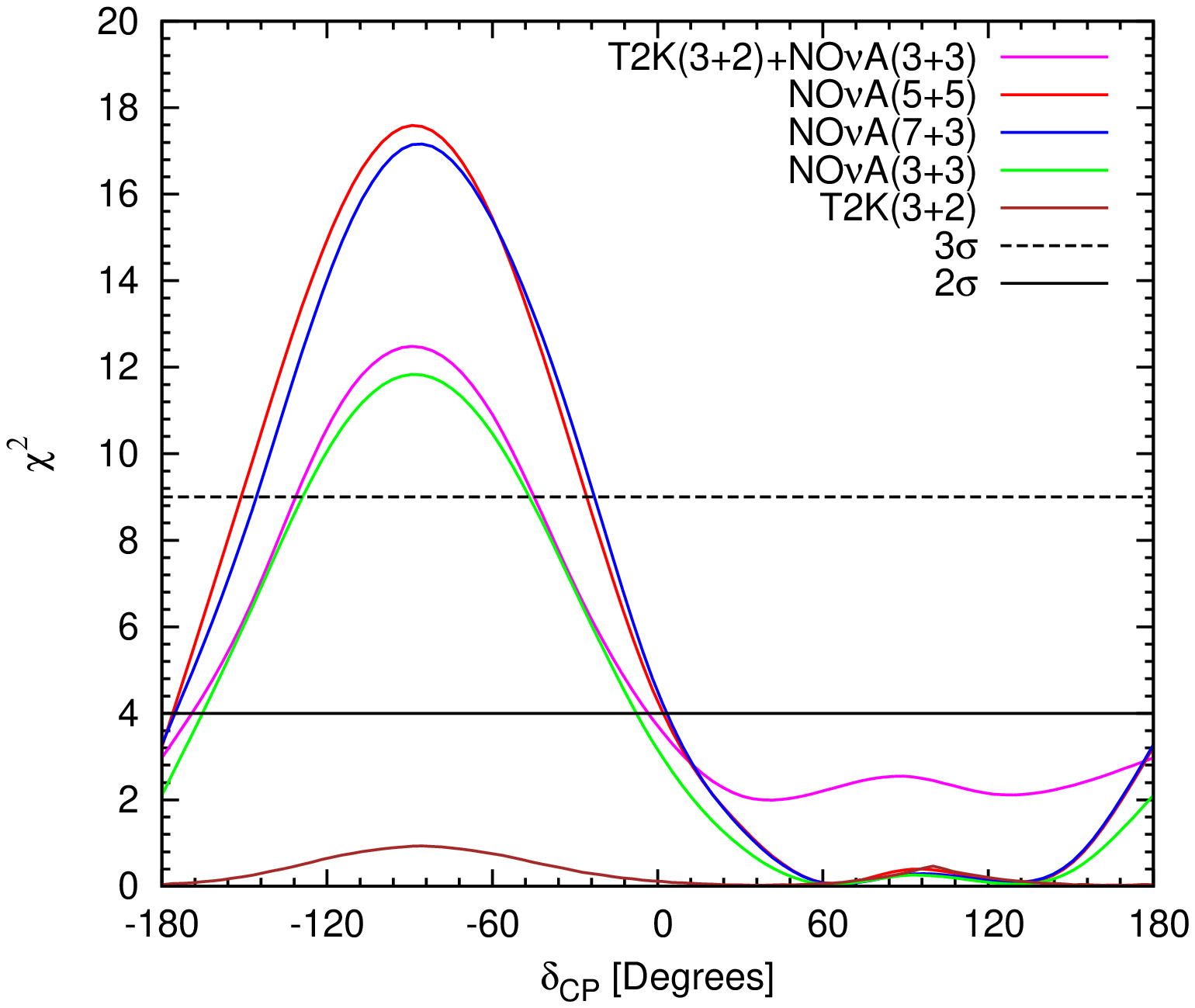}
\hspace{0.2 cm}
\includegraphics[width=7cm,height=5cm, clip]{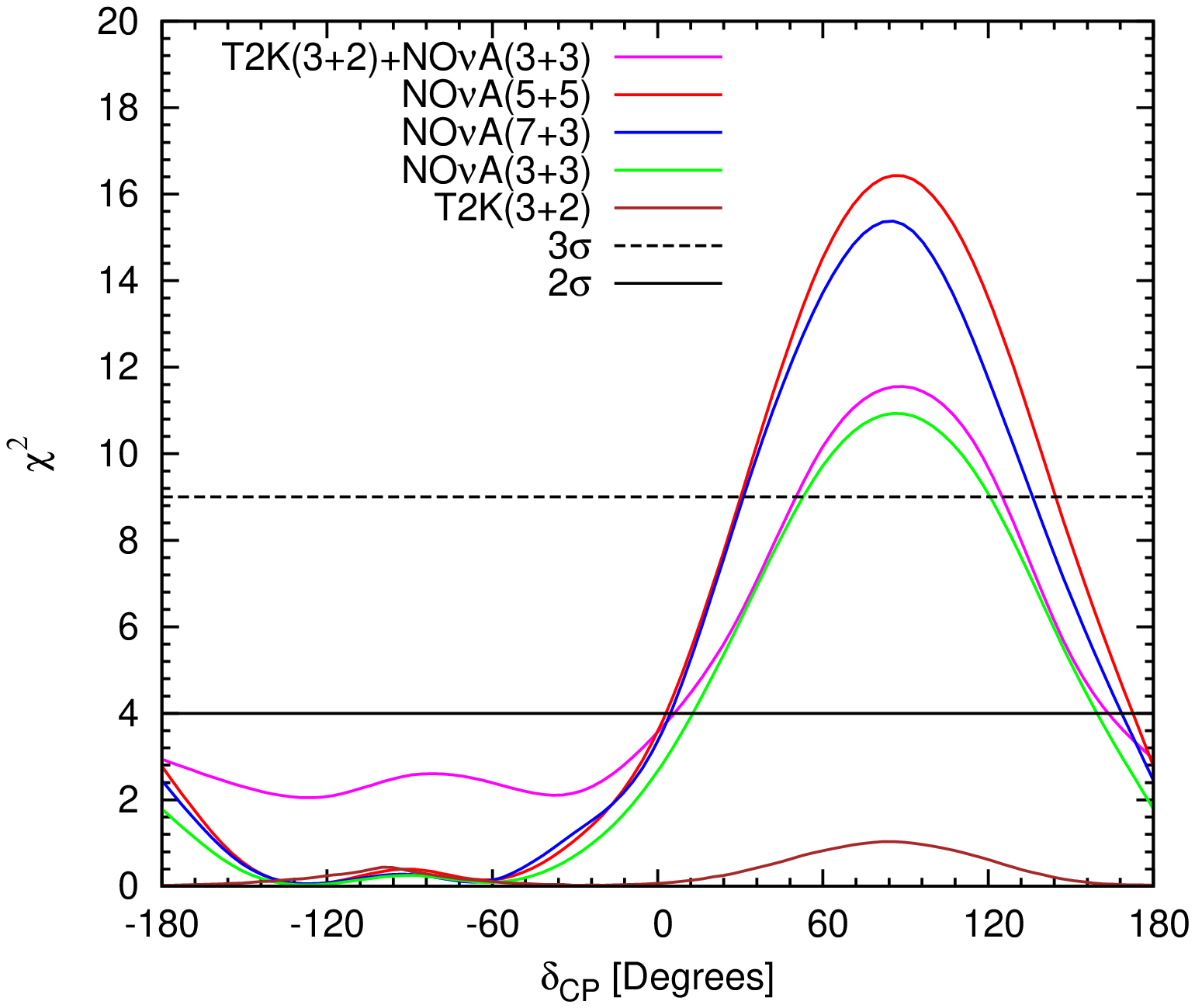}
\includegraphics[width=7cm,height=5cm, clip]{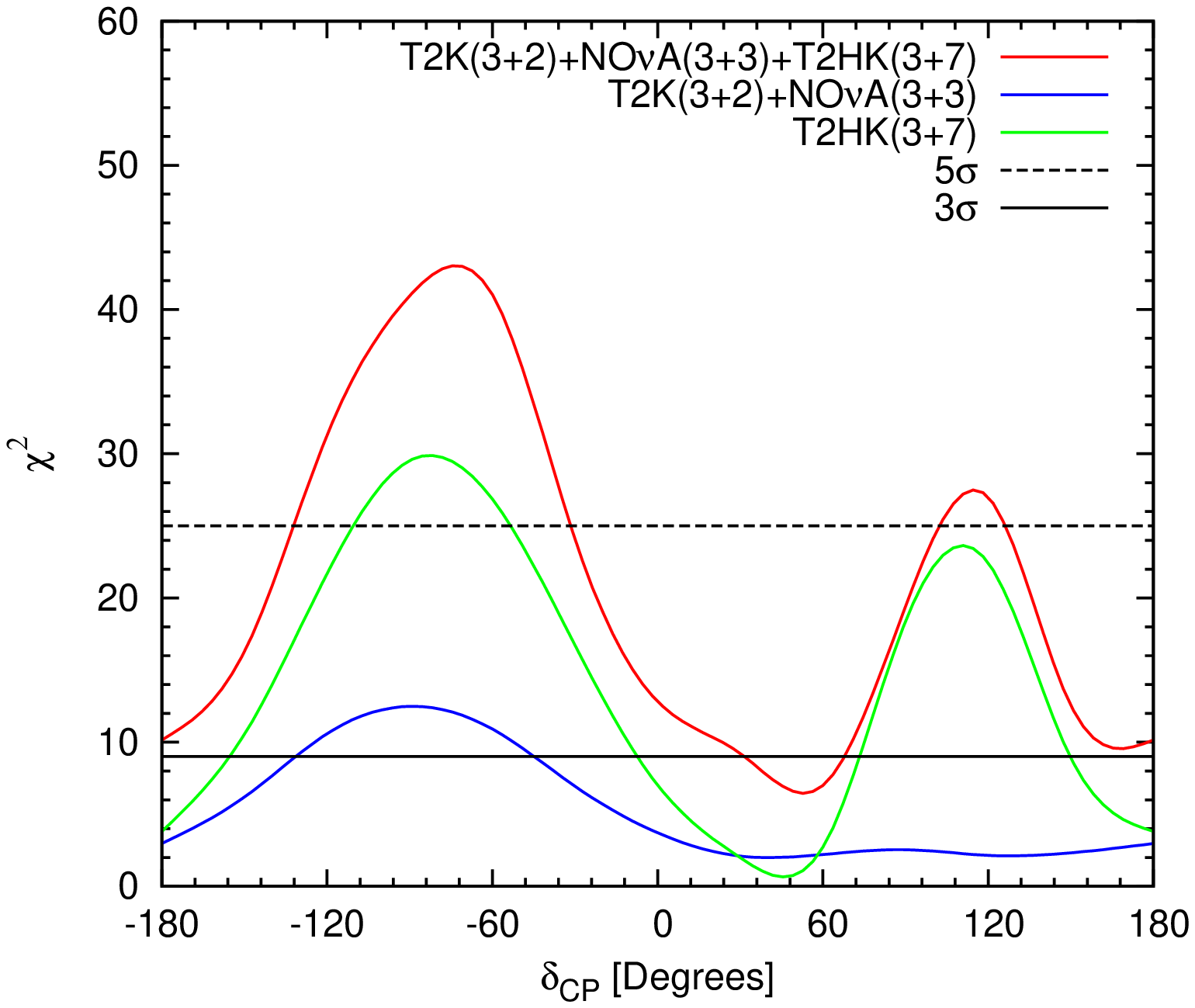}
\hspace{0.2 cm}
\includegraphics[width=7cm,height=5cm, clip]{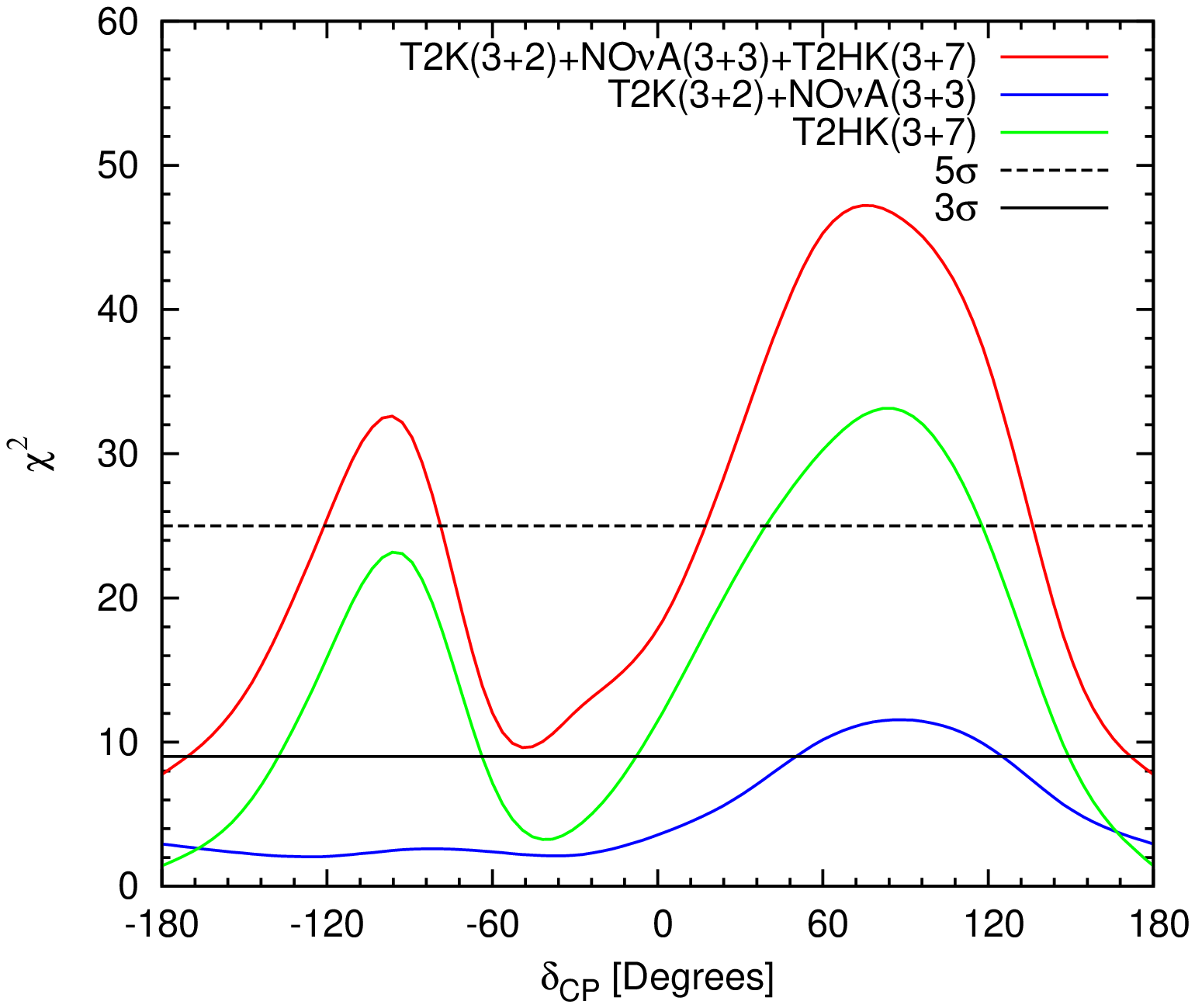}
\end{center}
\caption{Mass hierarchy significance as a function of true $\delta_{CP}$. In the left panel Normal hierarchy is considered as true hierarchy and inverted is taken as test hierarchy and in the right panel Inverted hierarchy is considered as true hierarchy and normal is taken as test hierarchy}
\end{figure}

\begin{figure}[htb]
\begin{center}
\includegraphics[width=7cm,height=5cm, clip]{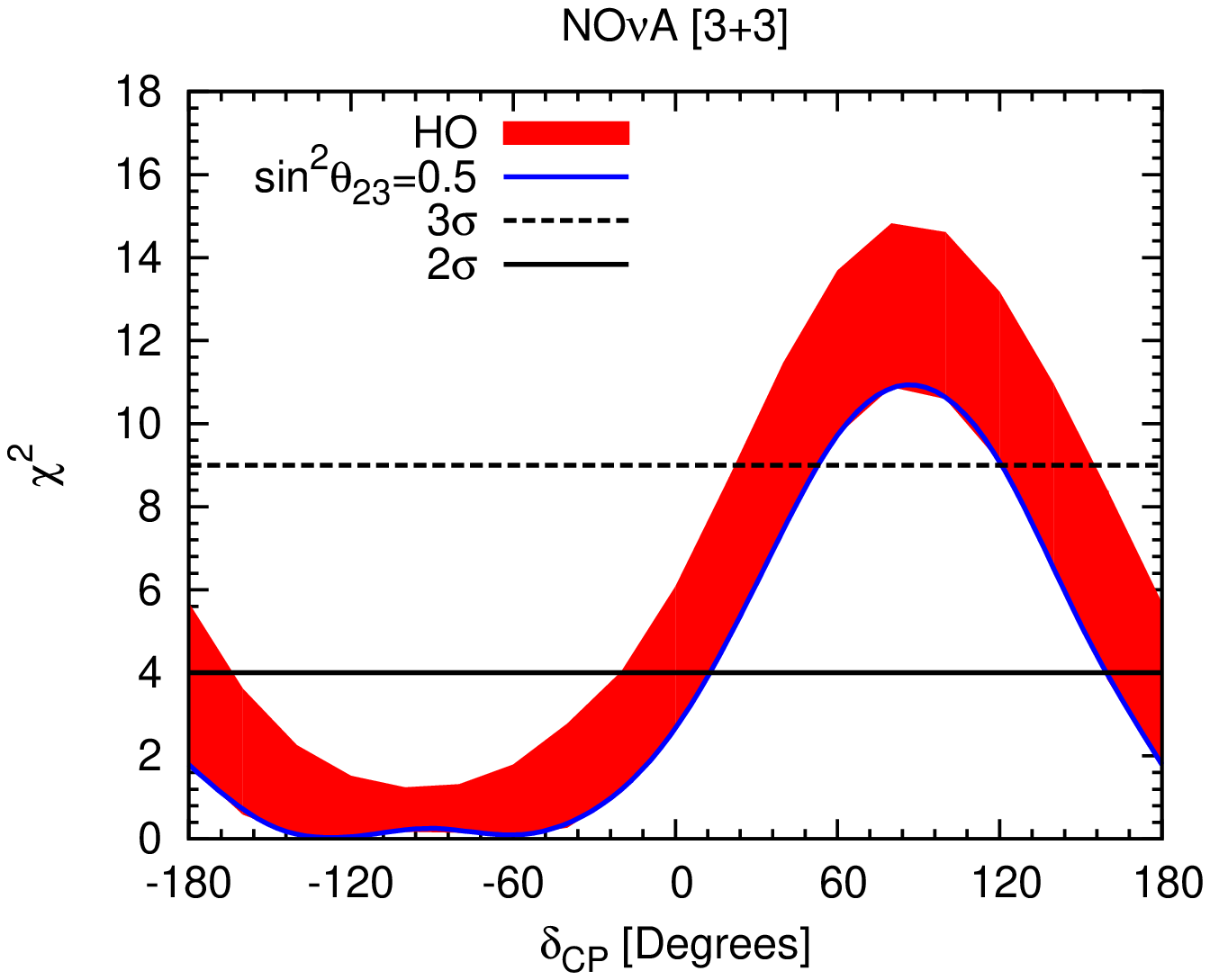}
\hspace{0.2 cm}
\includegraphics[width=7cm,height=5cm, clip]{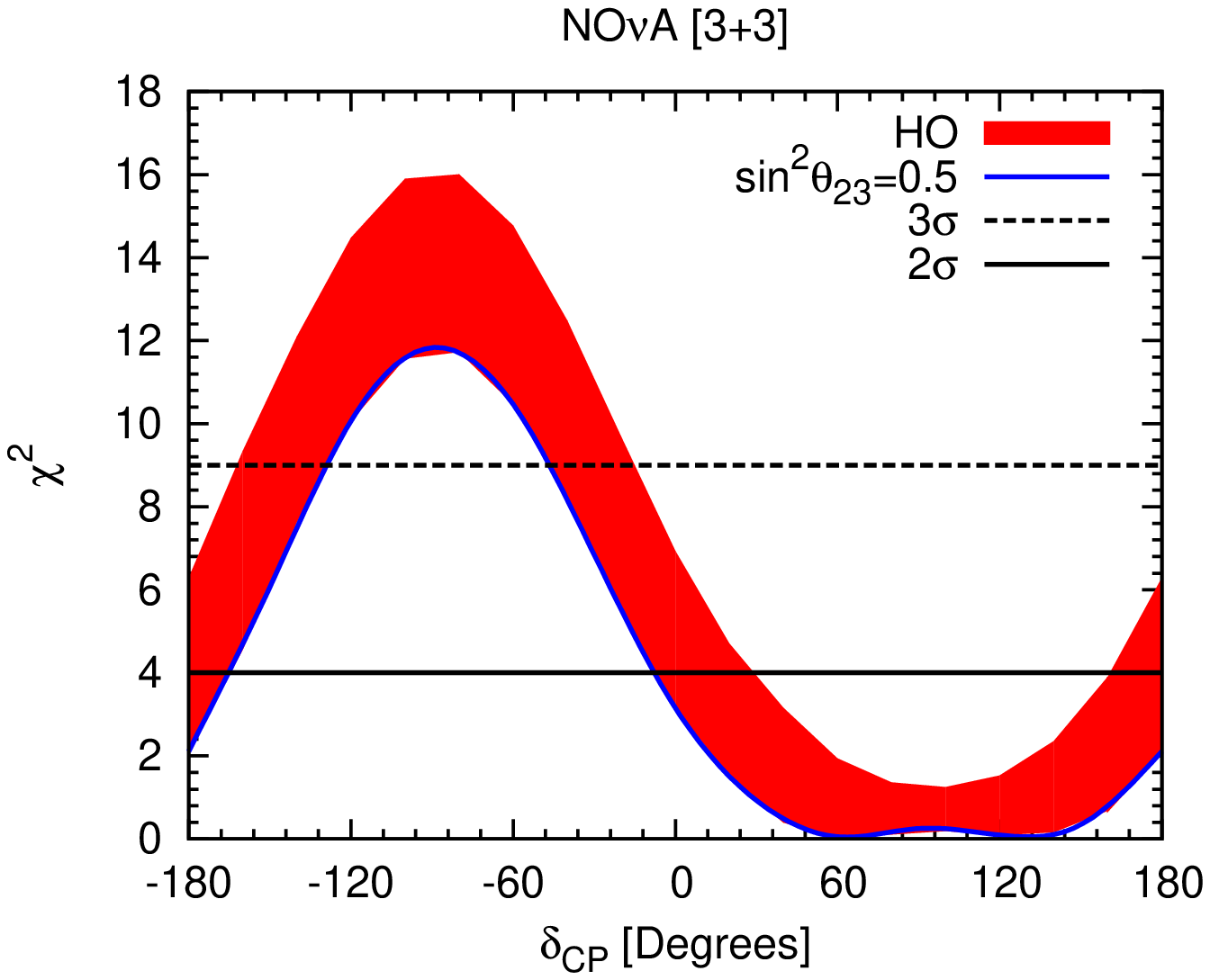}
\includegraphics[width=7cm,height=5cm, clip]{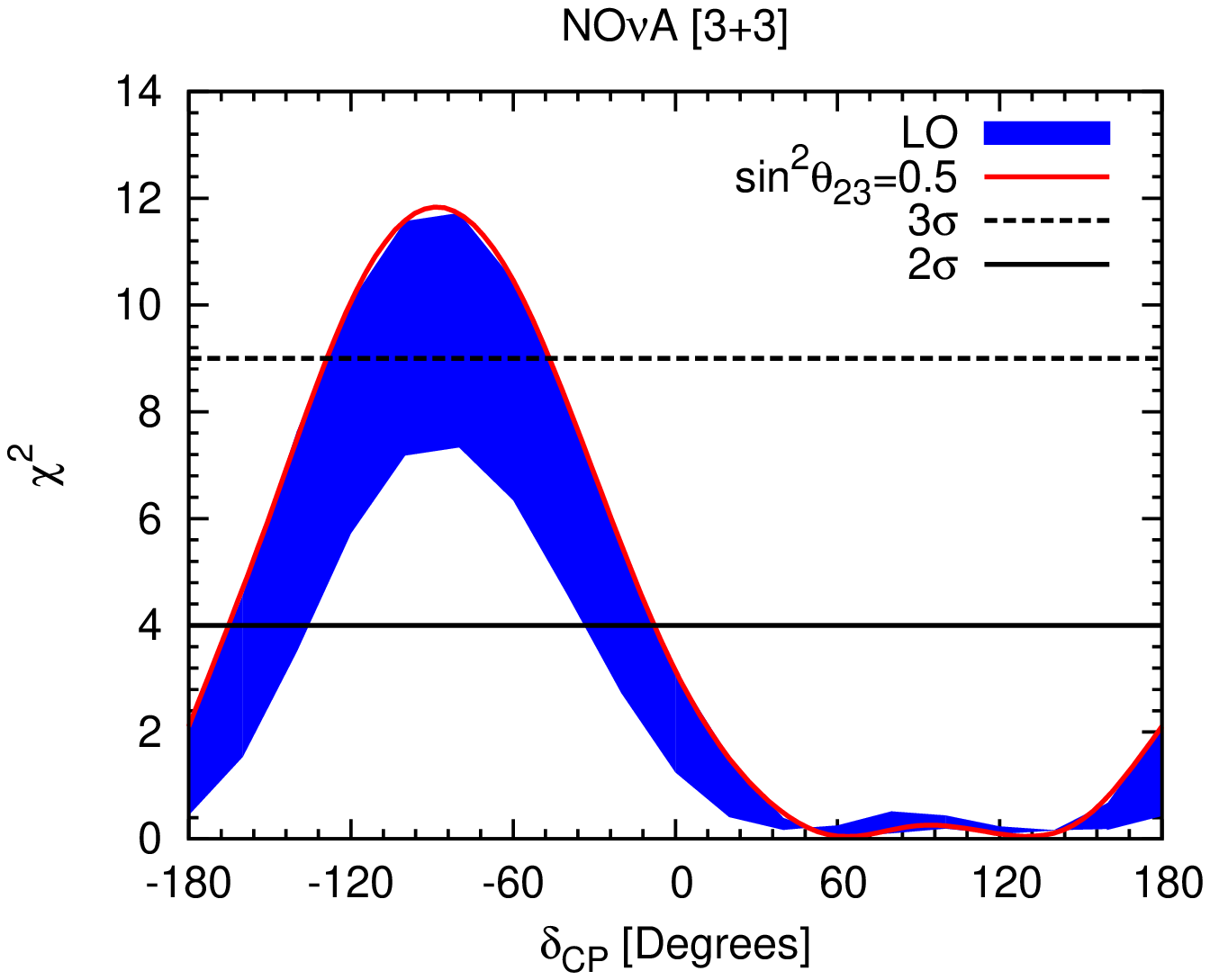}
\hspace{0.2 cm}
\includegraphics[width=7cm,height=5cm, clip]{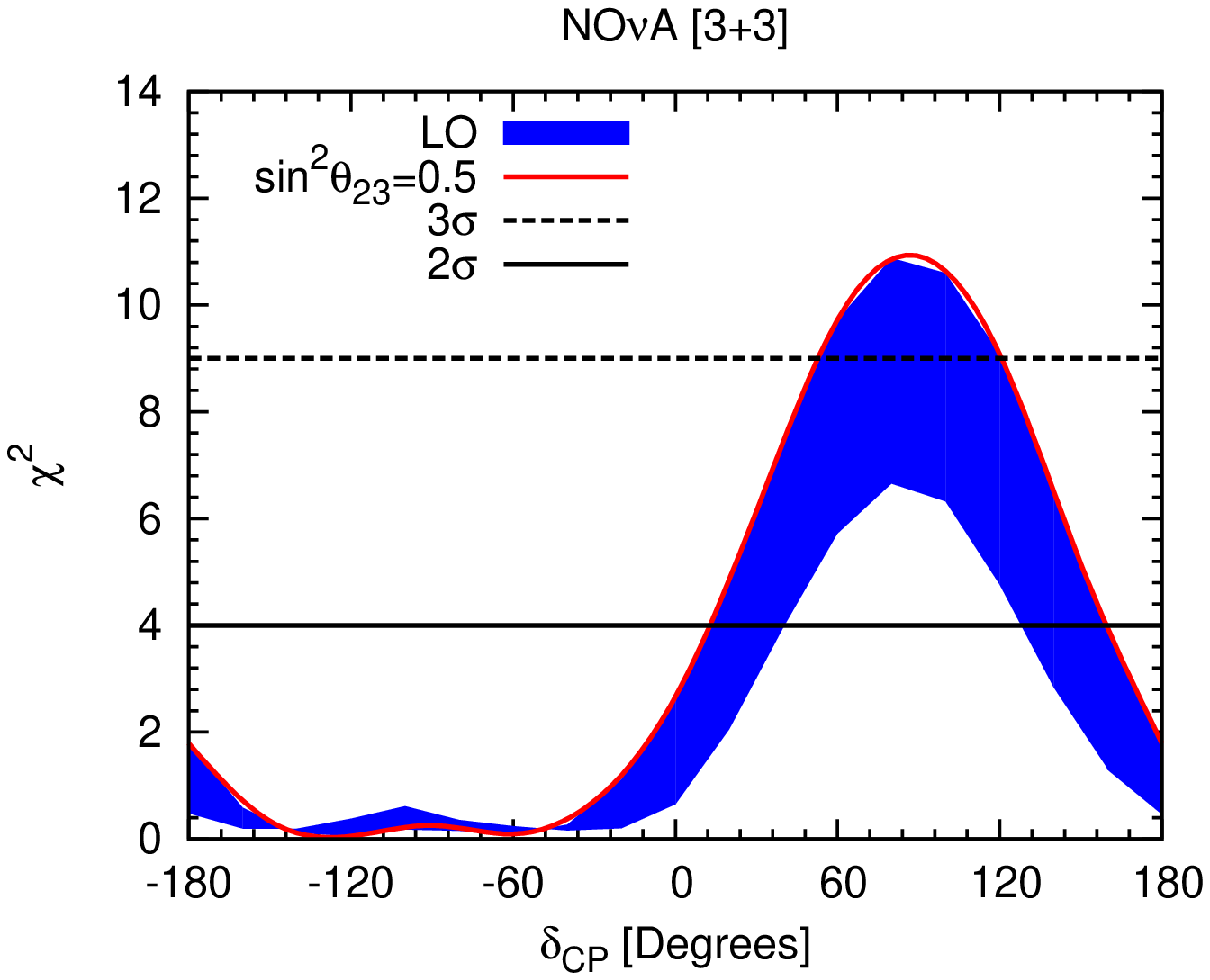}
\end{center}
\caption{Mass hierarchy significance with the octant of $\theta_{23}$ for the  scheduled run of NO$\nu$A experiment.}
\end{figure}

In Fig. 4, we present the hierarchy determination sensitivity of T2K, NO$\nu$A and T2HK as a function of true value of $\delta_{CP}$. 
We assume NH (IH) to be the true hierarchy in the left (right) panel. It can be seen that the wrong hierarchy can be ruled out quite effectively 
in the LHP (UHP) for NH (IH), which is basically the favorable half plane and in the other half plane the mass hierarchy can't  be determined  
effectively for T2K and NO$\nu$A experiments. However, the combined data from these two experiments (T2K (3+2) and NO$\nu$A (3+3)) improves 
the situation significantly and the sensitivity increases to more than 1$\sigma$ for all values of $\delta_{CP}$. The mass hierarchy significance 
above $3\sigma$, has a $\delta_{CP}$ coverage of 75\% for T2HK experiment  alone and 90\% for combined data of T2K, NO$\nu$A and T2HK experiments.

Next, we would like to study the effect of $\theta_{23}$ octant  on the MH sensitivity. We obtain the MH sensitivity by varying the true value of
$\sin^2 \theta_{23}$ in LO (HO) which has been shown in Fig. 5 where the red (blue) band in the top (bottom) panel corresponds to HO (LO). It is clear from
the figure that the MH sensitivity is significantly large if the value of $\sin^2 \theta_{23}$ is in higher octant.

\section{CP Violation Discovery Potential}
Accelerator based long-baseline neutrino oscillation experiments can address CP-violation problem through the appearance channels 
of $\nu_{\mu} \rightarrow \nu_{e}$ and $\bar \nu_{\mu} \rightarrow \bar \nu_{e}$. From Eq. (\ref{prob}) we can see that the CP violating 
effects due to $\delta_{CP}$ are modified by all the three mixing angles and their combinations, thus resulting in an eight fold parameter degeneracy. 
In order to obtain the significance of CP violation sensitivity, we simulate the true event spectrum by keeping the true values of oscillation parameters 
as in Table-III
except for $\sin^2 \theta_{23} =0.5$ and  vary the true value of $\delta_{CP}$ in the range $[-\pi,\pi]$. We then  compare those with test 
event spectrum for $\delta_{CP}$=0 or $\pi$ and thus, obtain the minimum $\chi^2$. We  consider the sign degeneracy of $\Delta m_{31}^2$ by 
marginalizing over it, in both NH and IH $3\sigma$ ranges, $\sin^2 2 \theta_{13}$ and $\sin^2 \theta_{23}$ in their $3\sigma$  and  added prior to $\sin^2 2 \theta_{13}$.

\begin{figure}[htb]
\begin{center}
\includegraphics[width=7cm,height=5cm, clip]{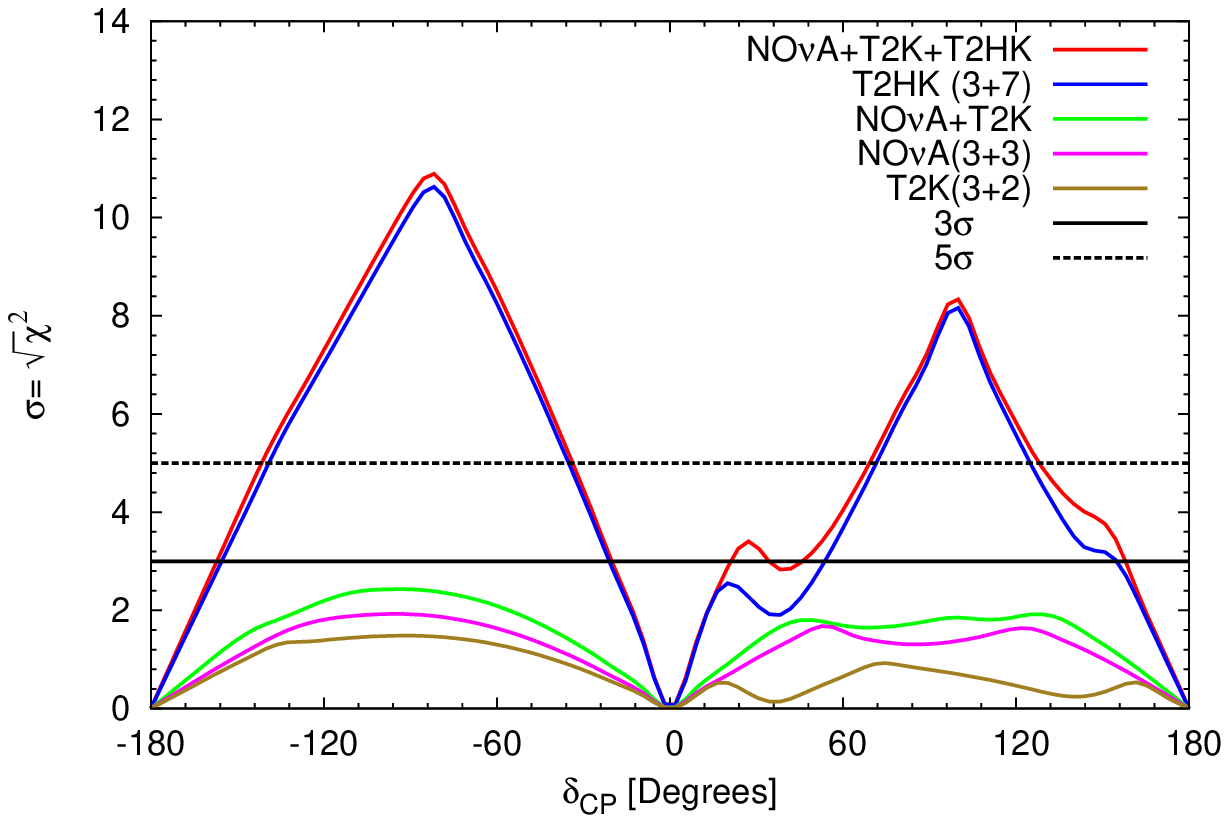}
\hspace{0.2 cm}
\includegraphics[width=7cm,height=5cm, clip]{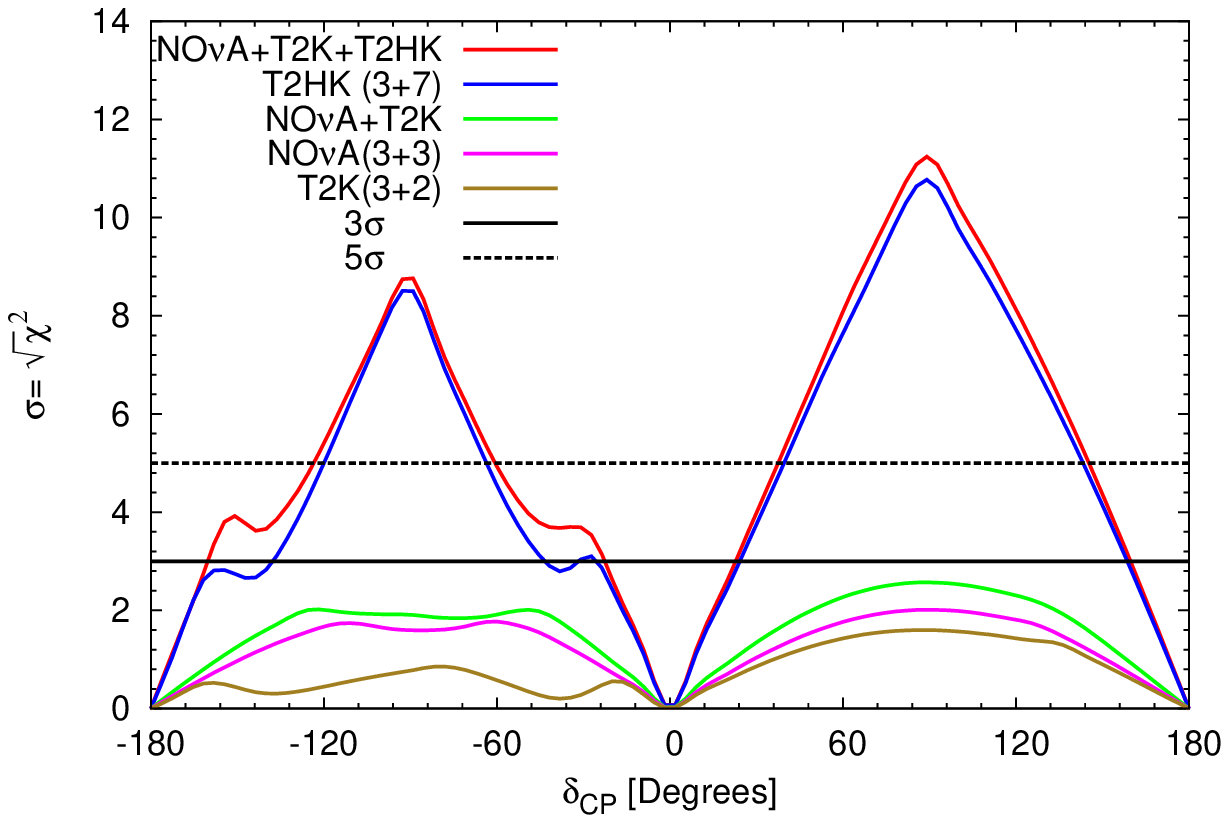}
\includegraphics[width=7cm,height=5cm, clip]{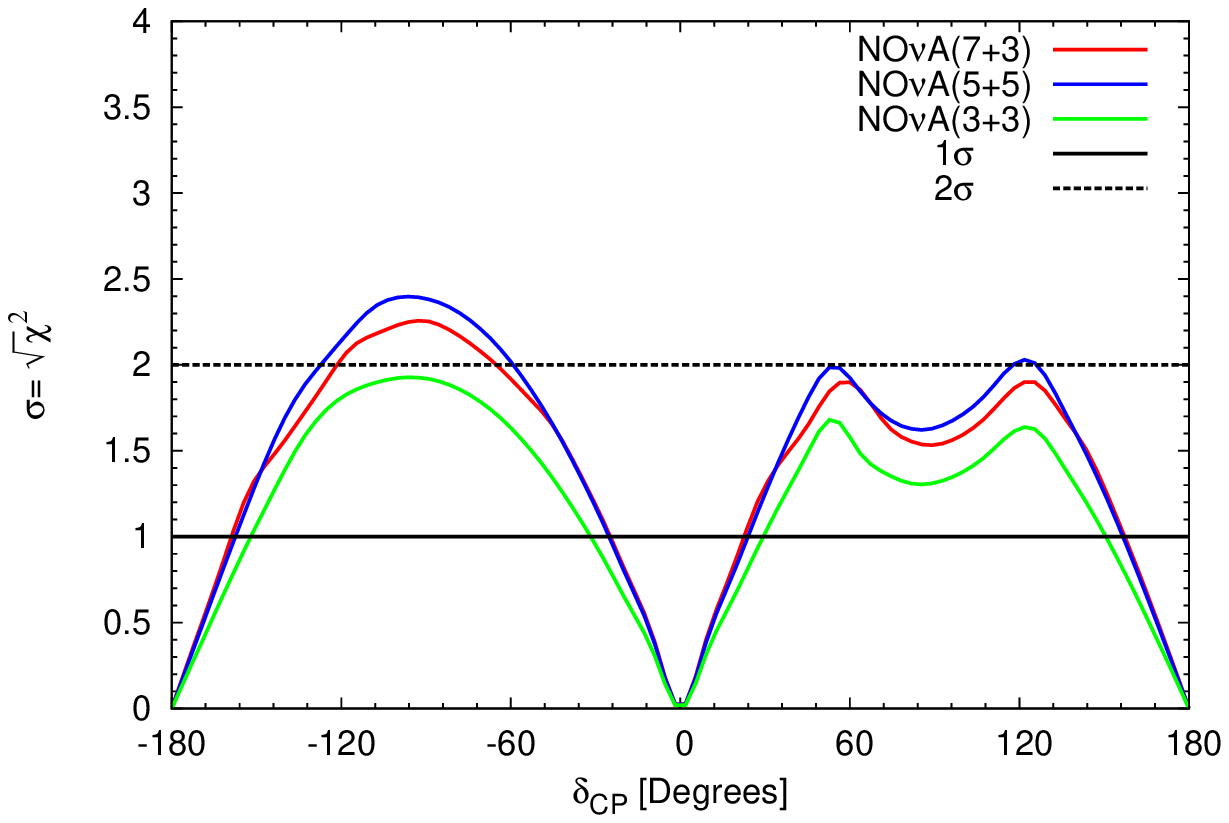}
\hspace{0.2 cm}
\includegraphics[width=7cm,height=5cm, clip]{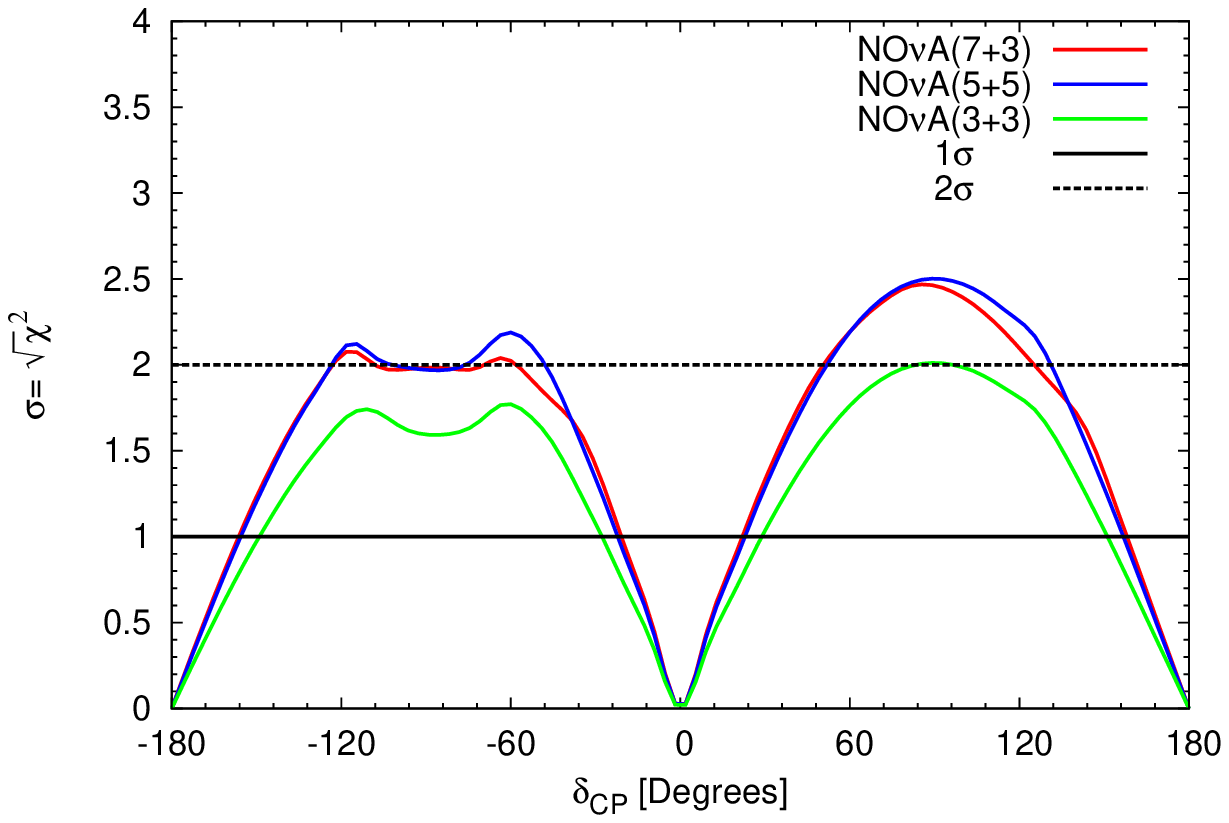}
\end{center}
\caption{CP violation sensitivity for  different combinations of run time of T2K, NO$\nu$A and T2HK experiment for NH (IH)  in the left (right) panel.}
\end{figure}

In Fig. 6, we plot the sensitivity to rule out the CP conserving scenarios, as a function of true $\delta_{CP}$ assuming  NH (IH) as the true hierarchy
in the left panel (right panel). 
From the figure one can notice that T2K by itself has no CP violation sensitivity at 2$\sigma$ C.L.. For NO$\nu$A  with (3+3) years of running, there will 
be CP violation sensitivity above 1.5$\sigma$ level for about one-third of the CP violating phase $\delta_{CP}$ space. Furthermore, 
the synergistic combination of NO$\nu$A and T2K leads to much better CP violation sensitivity compared to the individual capabilities. 
Even the combination of NO$\nu$A (3+3) and T2K (3+2) has comparable
sensitivity as for 10 years running of NO$\nu$A. Owing to the fact that main goal of T2HK experiment is to determine CP violation, one can see that 
it has a significance of above 5$\sigma$ C.L. for a fraction of two-fifth values of the CP violating phase $\delta_{CP}$ space. 
This in turn boosts up the sensitivity when its data is added to NO$\nu$A (3+3) yrs and T2K (3+2) yrs. From the plots in the lower panel, one can observe that the sensitivity of NO$\nu$A increases slightly for 10 years of run time, with (5$\nu$ + 5$\bar{\nu}$) combination has better sensitivity than that of  (7$\nu$ + 3$\bar{\nu}$) combination. The drop in the half planes of $\delta_{CP}$ i.e, in the region [0,180]$^{\circ}$ ([-180,0]$^{\circ}$) for NH (IH) is due to the fact that the hierarchy sensitivity  is highly sensitive to $\delta_{CP}$. As a result, of marginalization over hierarchy causes the CPV sensitivity to drop for unfavorable values of $\delta_{CP}$.

\begin{figure}[htb]
\includegraphics[width=7cm,height=5cm, clip]{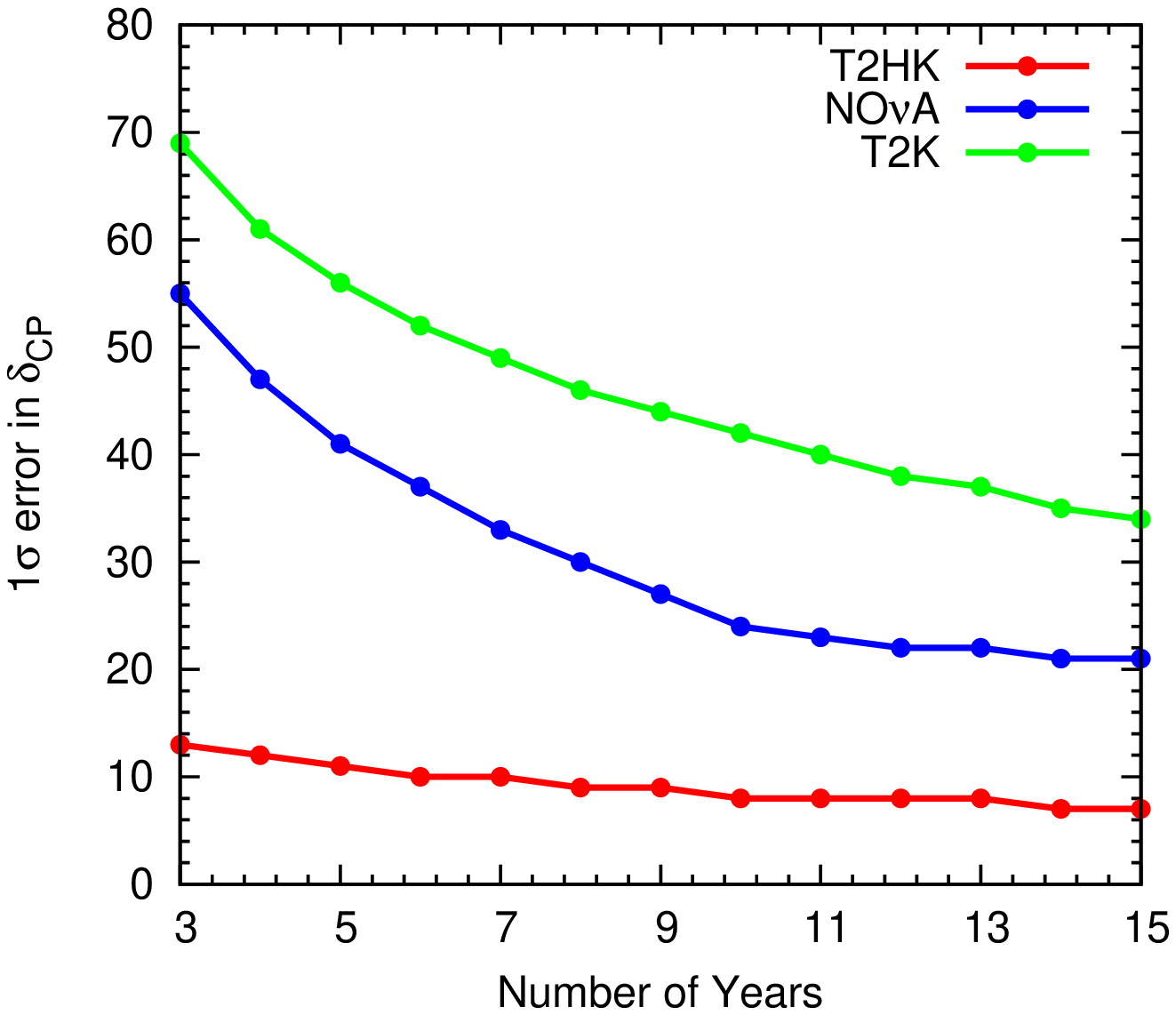}
\hspace{0.2 cm}
\includegraphics[width=7cm,height=5cm, clip]{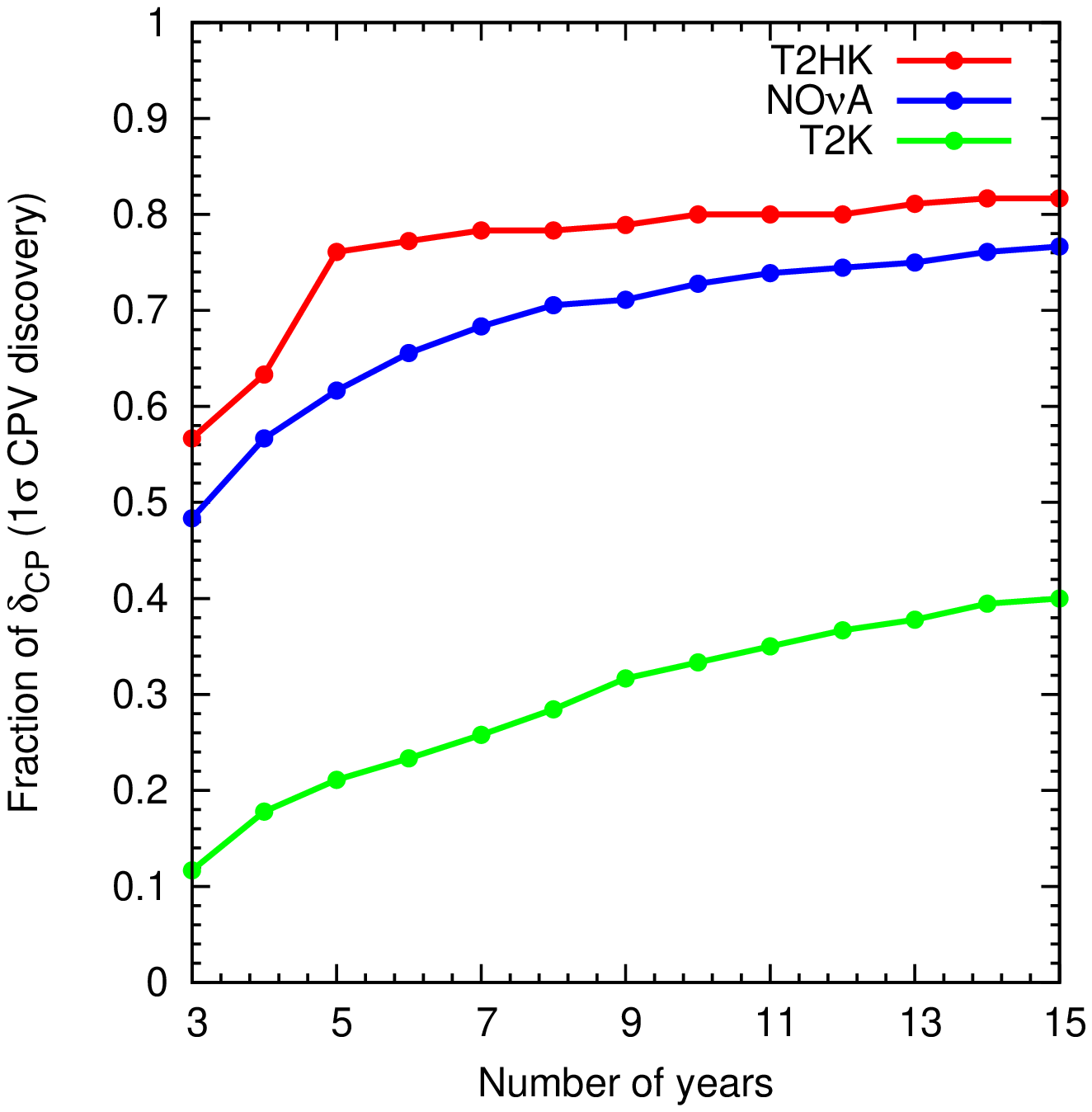}
\caption{Sensitivity vs running time: 1$\sigma$ error in $\delta_{CP}$ as a function of running time in years for true value of $\delta_{CP}$=0 (left panel). 
The fraction of $\delta_{CP}$ for which $\delta_{CP}$ =$0^\circ,180^\circ$ is excluded with 1$\sigma$ as a function of running time. }
\end{figure}

In Fig. 7, the left panel shows the 1$\sigma$ uncertainty of $\delta_{CP}$ as a function of running time (in years) for true value of $\delta_{CP}$=0
and  the right panel shows 
the CP violation sensitivity as a function of running time. In both cases,  the ratio of neutrino and antineutrino modes is fixed to 1:1 for T2K and NO$\nu$A and 
3:7 for T2HK. In this analysis, mass hierarchy is assumed to be unknown. Therefore, we have marginalized over both the  hierarchies, and the result 
shown in Fig. 7, are for the true  hierarchy as normal hierarchy.
From the left panel of the figure, we can see that the values of $\delta_{CP}$ can be determined to better than $35^\circ$ ($21^\circ$) for all values of 
$\delta_{CP}$ for T2K (NO$\nu$A). In the case of T2HK the values of $\delta_{CP}$ can be determined to better than $9^\circ$ for all values of $\delta_{CP}$. 
From the right panel, we can see that CP violation can be observed with more than 1$\sigma$ significance  for 40 (75)\% of the possible values of $
\delta_{CP}$ for T2K (NO$\nu$A). Whereas for  T2HK, CP violation can be observed with more than 1$\sigma$ significance  for 80\% of the possible 
values of $\delta_{CP}$.

\subsection{Correlation between $\delta_{CP}$ and $\theta_{13}$}
 
 The knowledge of reactor mixing angle $\theta_{13}$ plays a crucial role in the discovery potential of $\delta_{CP}$. The recent discovery of 
large value of $\theta_{13}$ has established the need to study and understand the dependency between $\delta_{CP}$ and $\theta_{13}$. In this subsection, we discuss the correlation between the oscillation parameters $\theta_{13}$ and $\delta_{CP}$.
In obtaining the confidence region, we have fixed the  true values as in Table-III and considered true $\sin^2\theta_{23}=0.59$ for Higher Octant 
and true $\sin^2 \theta_{23}=0.41$ for Lower Octant, since the octant of $\theta_{23}$ is not known. We have varied the  test value of $\sin^2 2\theta_{13}$ 
in its $3\sigma$ range. 
 In this analysis we have kept both true and test hierarchy as normal hierarchy.

Fig. 8, shows the confidence regions in the  $\sin^2 2\theta_{13}$ - $\delta_{CP}$ plane  for different combinations of T2K and NO$\nu$A experiments.
One can see from these figures that at the 2$\sigma$ confidence level, the uncertainty in the knowledge of $\theta_{23}$ octant has a
noticeable effect on the correlation between $\delta_{CP}$ and $\theta_{13}$.

\begin{figure}[htb]
\begin{center}
\includegraphics[width=5cm,height=5cm]{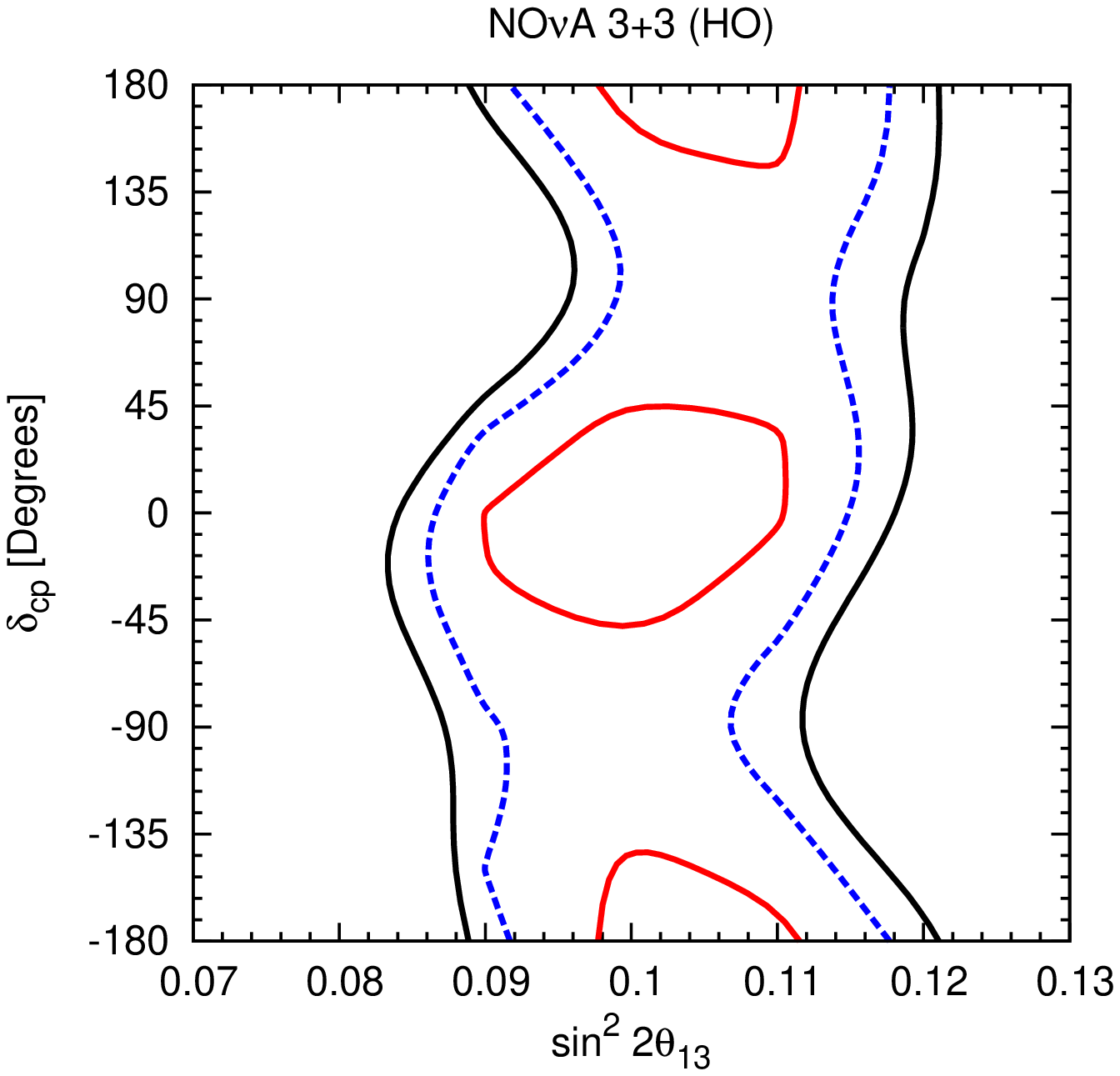}
\includegraphics[width=5cm,height=5cm]{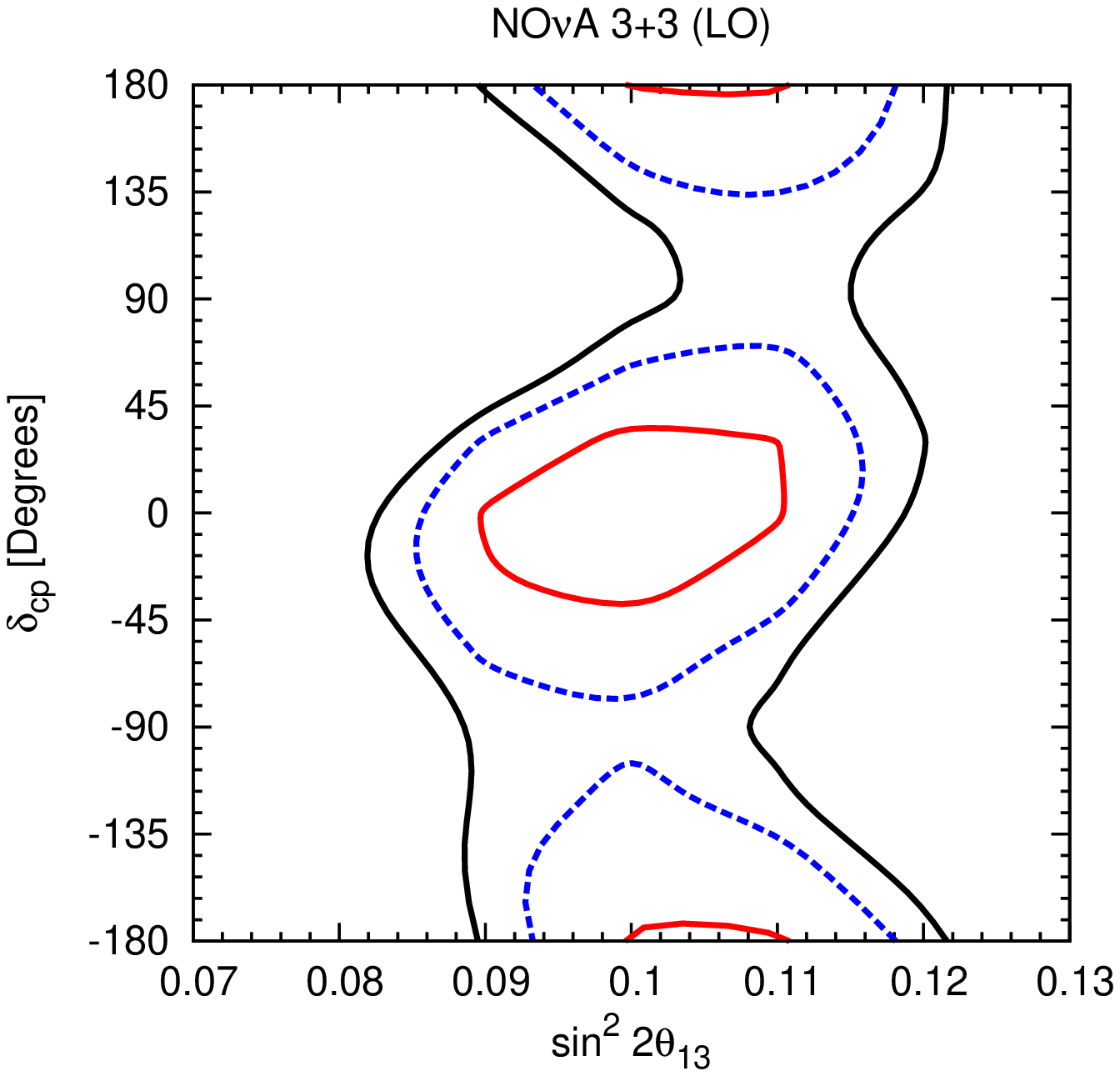}
\includegraphics[width=5cm,height=5cm]{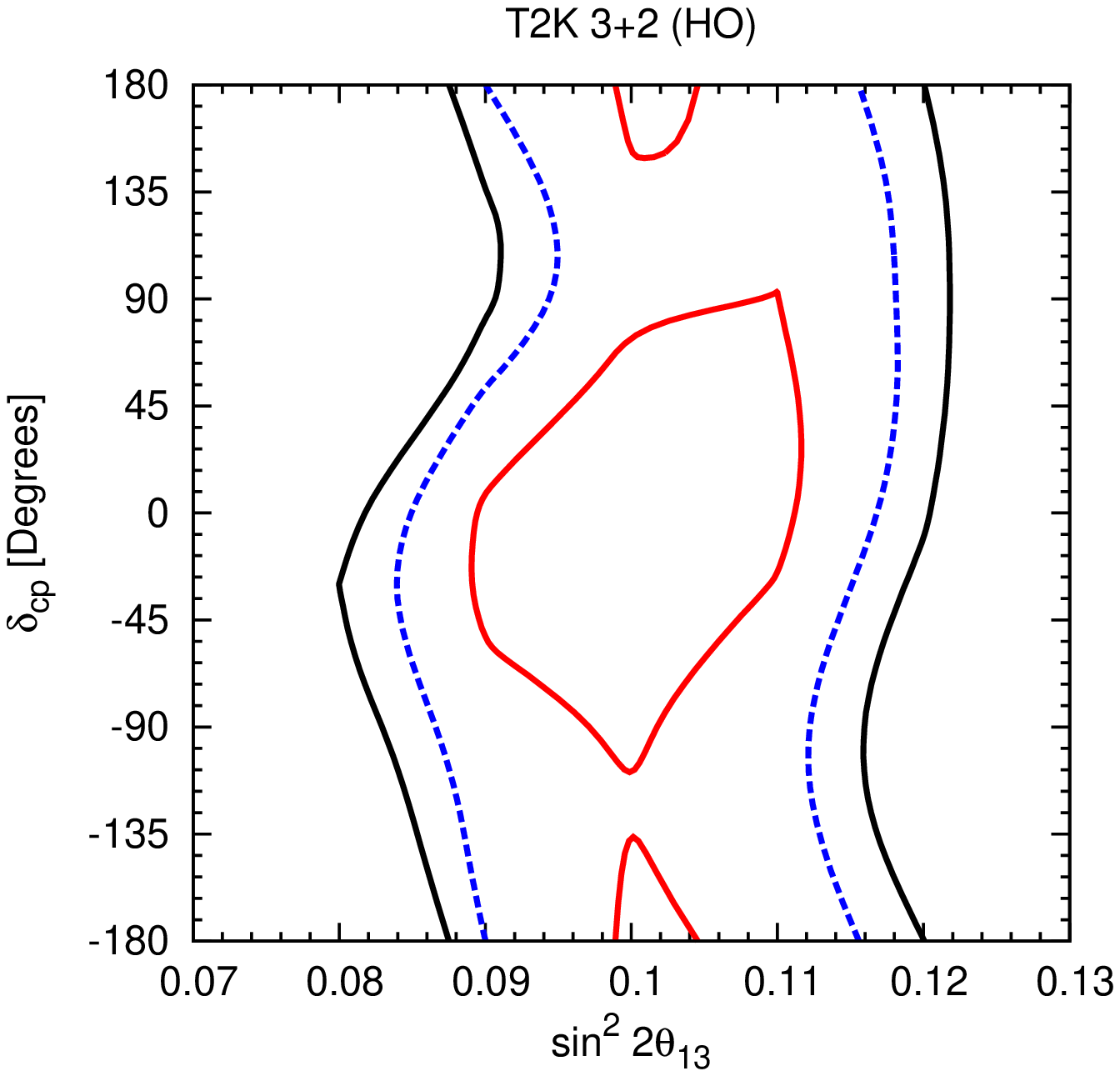}
\includegraphics[width=5cm,height=5cm]{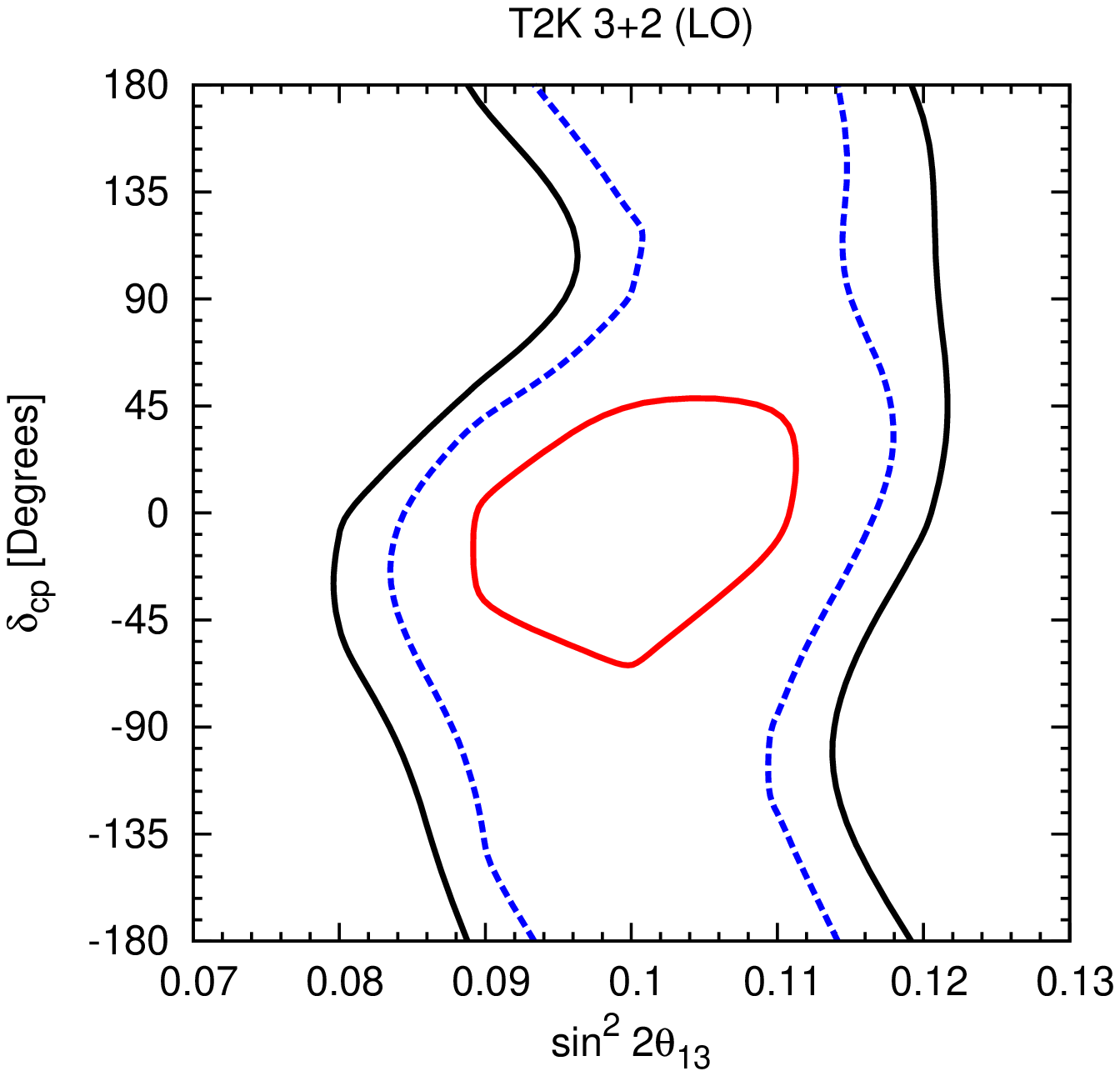}
\includegraphics[width=5cm,height=5cm]{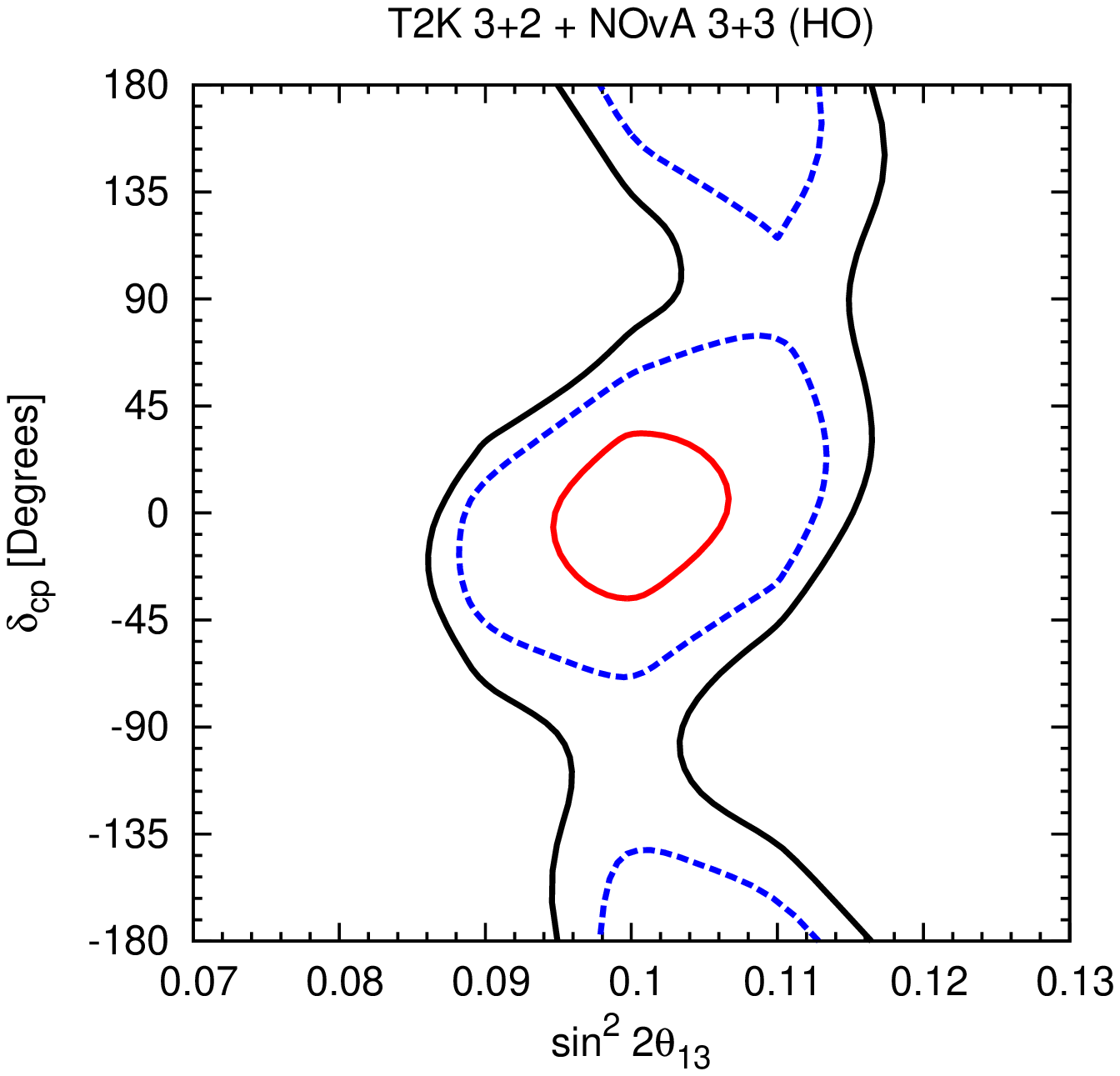}
\includegraphics[width=5cm,height=5cm]{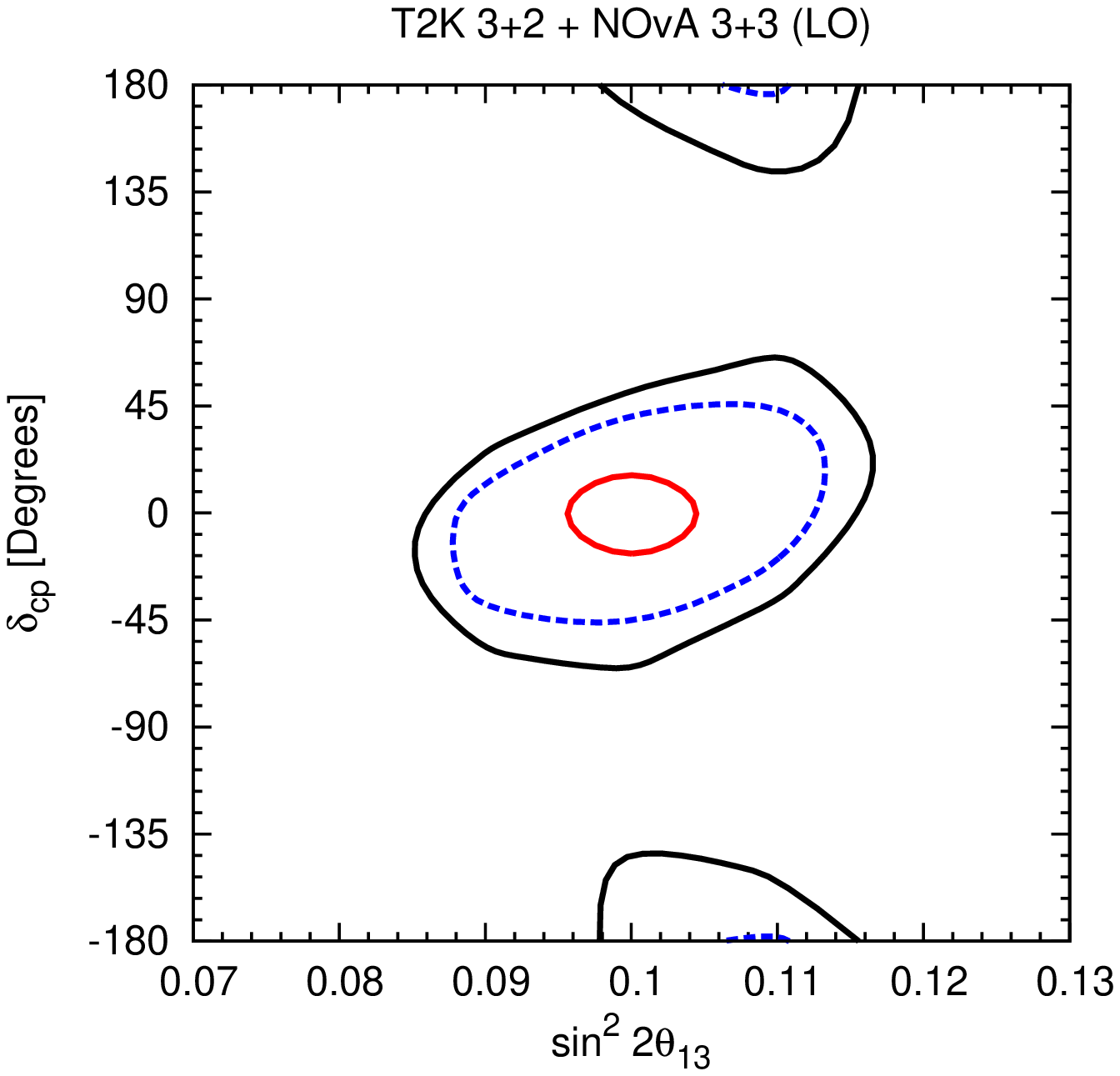}
\includegraphics[width=5cm,height=5cm]{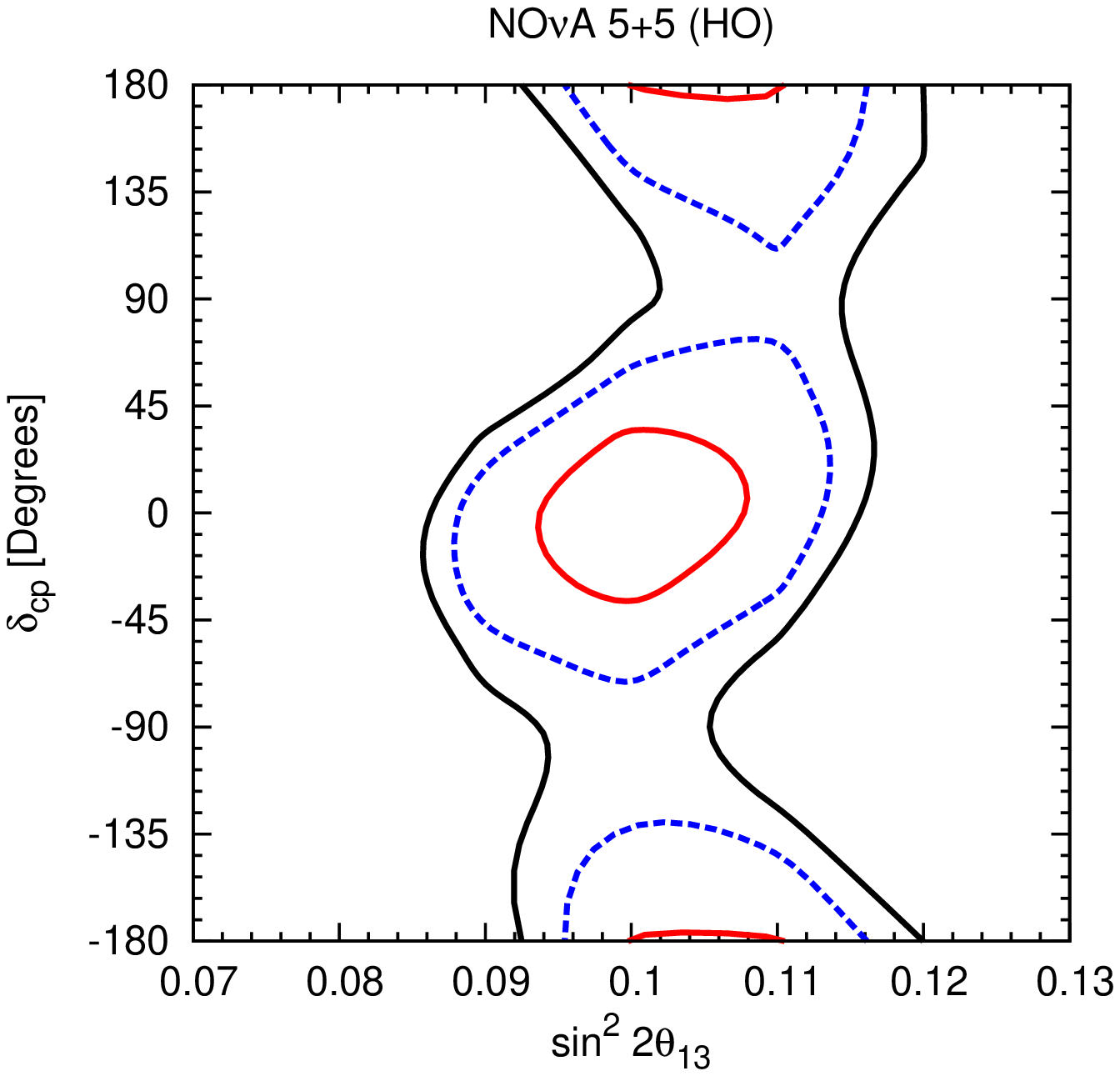}
\includegraphics[width=5cm,height=5cm]{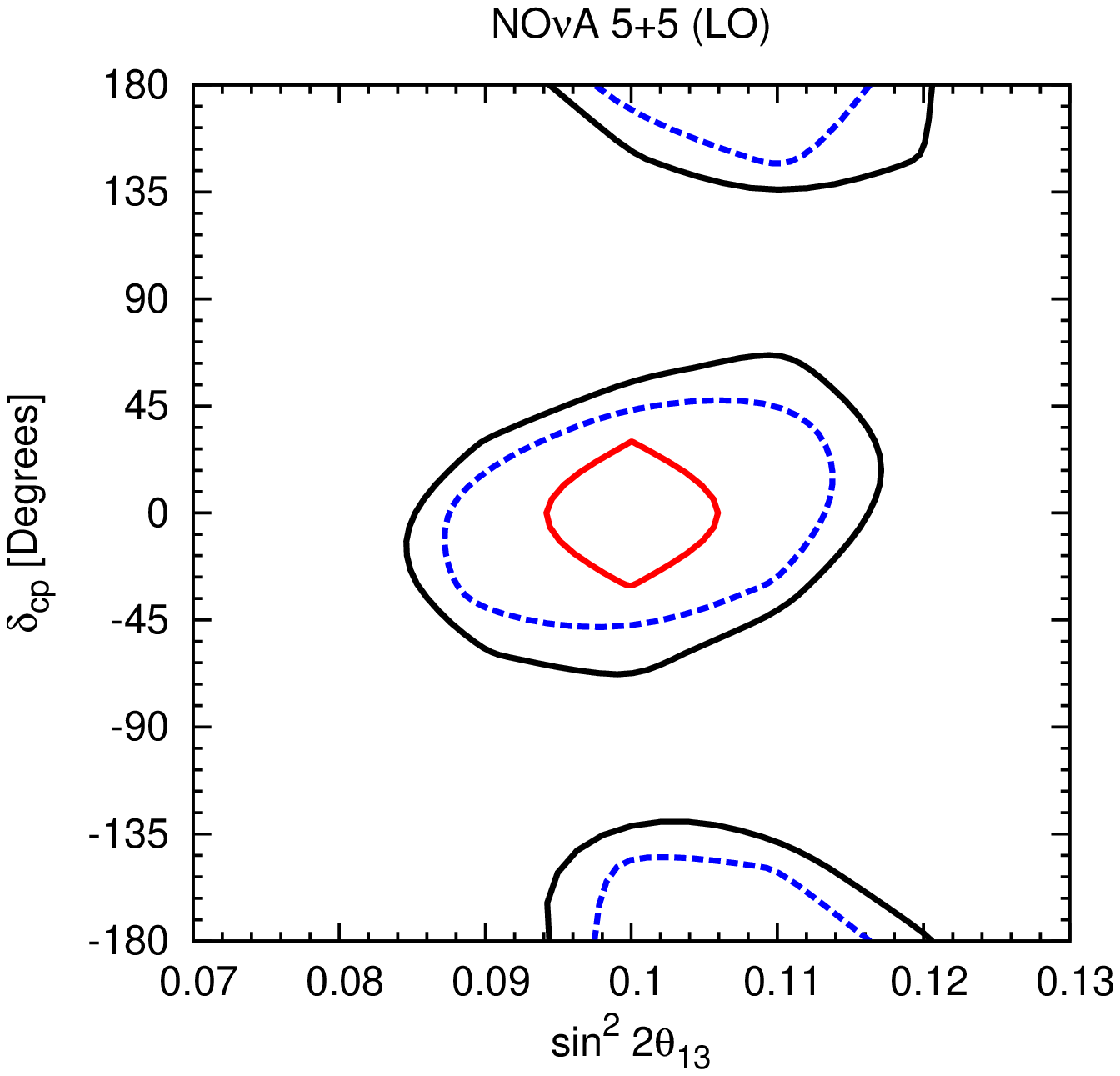}
\includegraphics[width=5cm,height=5cm]{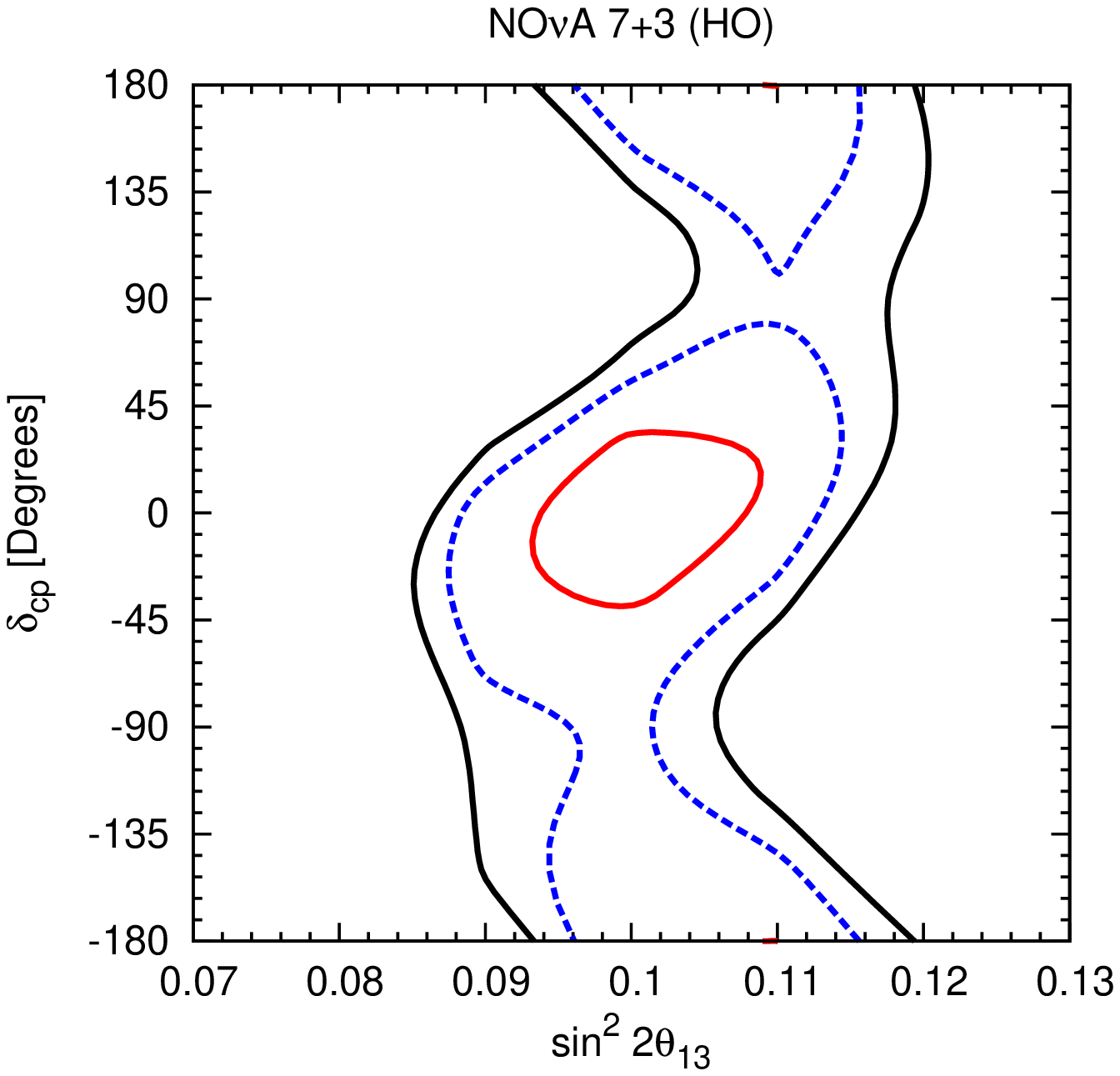}
\includegraphics[width=5cm,height=5cm]{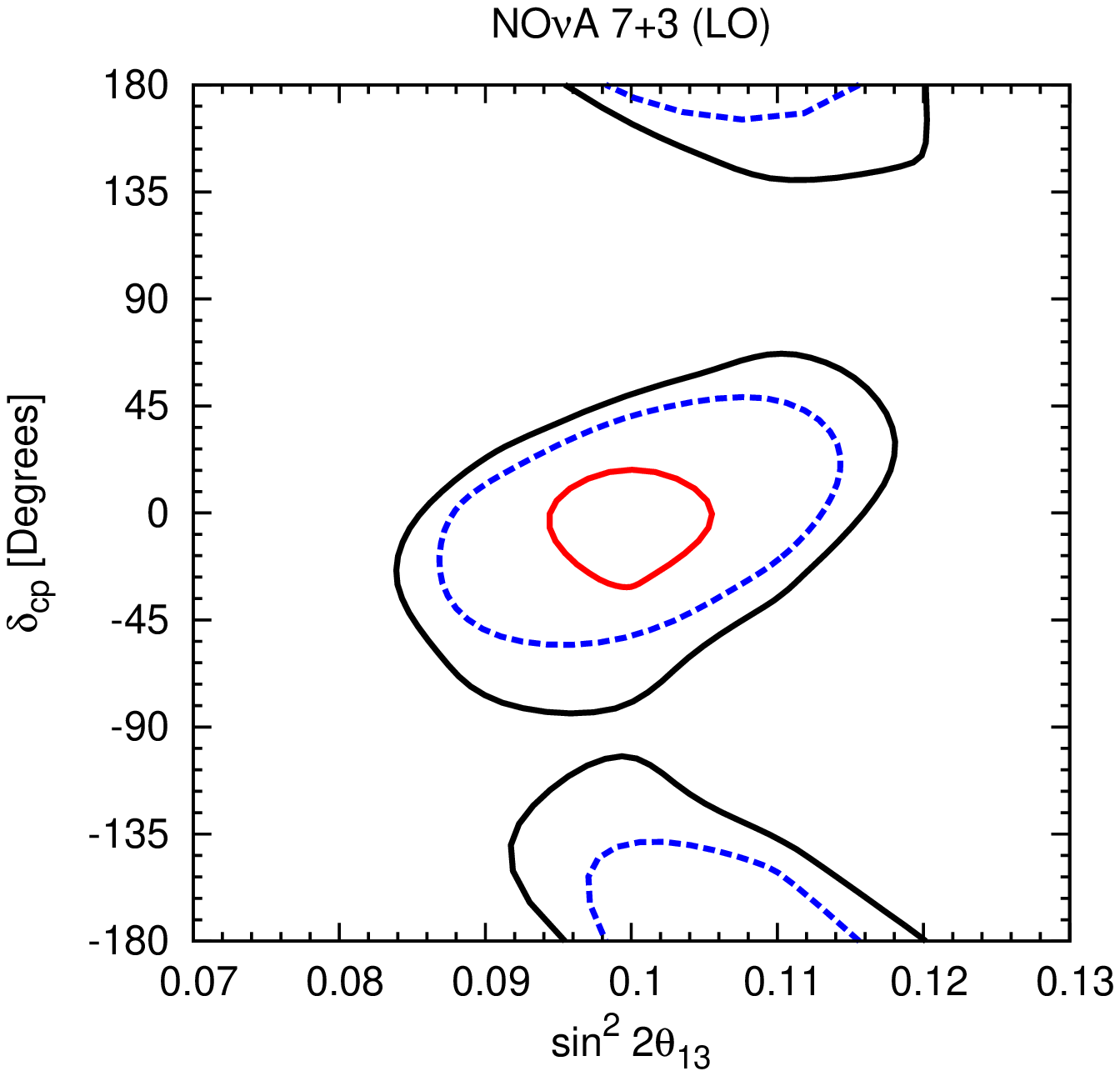}
\end{center}
\caption{Confidence region in  $\sin^2 2 \theta_{13}- \delta_{CP}$  plane for $\delta_{CP}=0$ and for different run combinations of T2K and NO$\nu$A experiments,
where the red, blue and black contours represent the 1$\sigma$ (68.3\% C.L.), 1.64$\sigma$ (90\% C.L.) and 2$\sigma$ (95.45\% C.L.) values respectively for 
two degrees of freedom.}
\end{figure}



\begin{figure}[htb]
\begin{center}
\includegraphics[width=5cm,height=5cm]{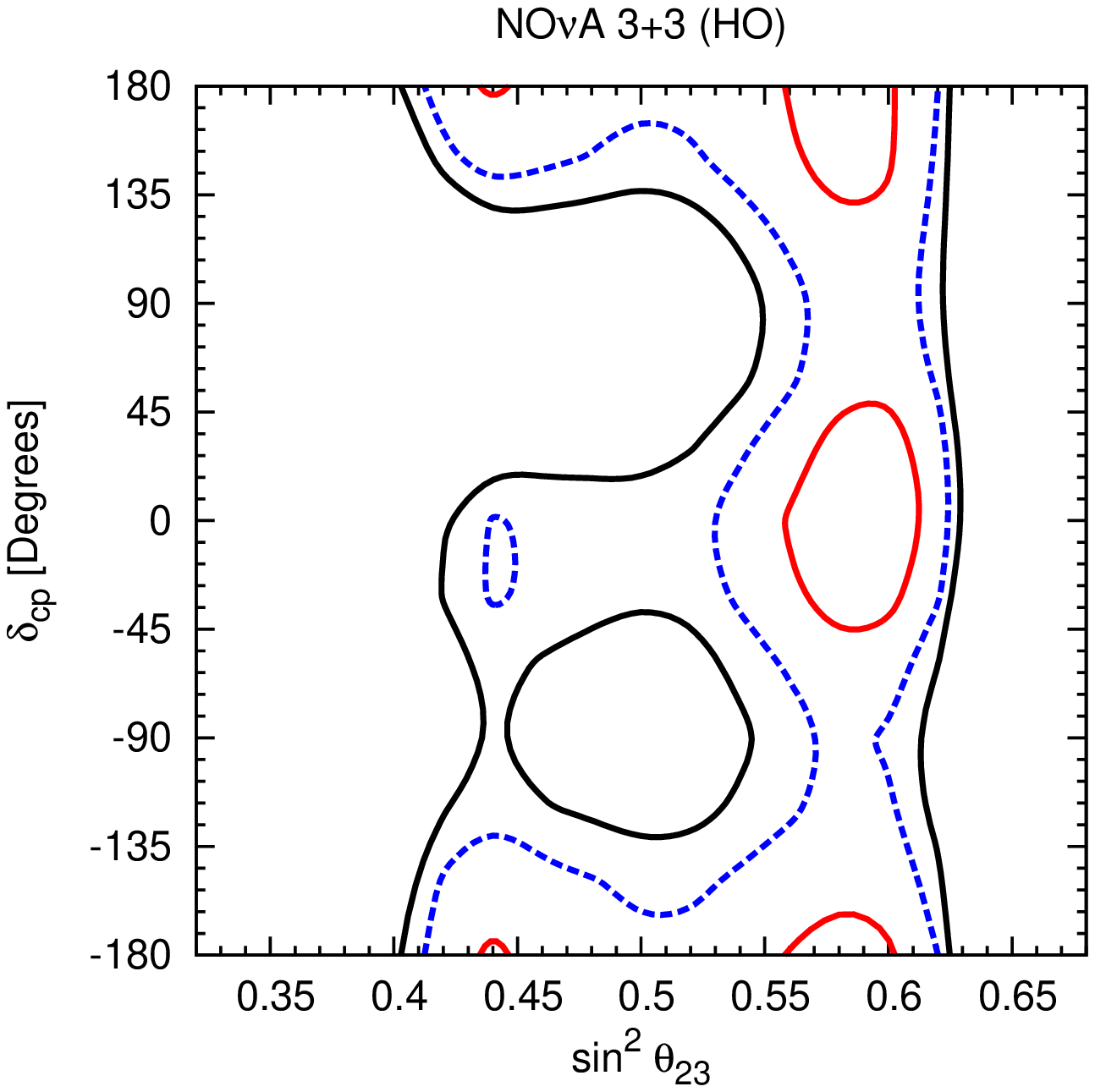}
\includegraphics[width=5cm,height=5cm]{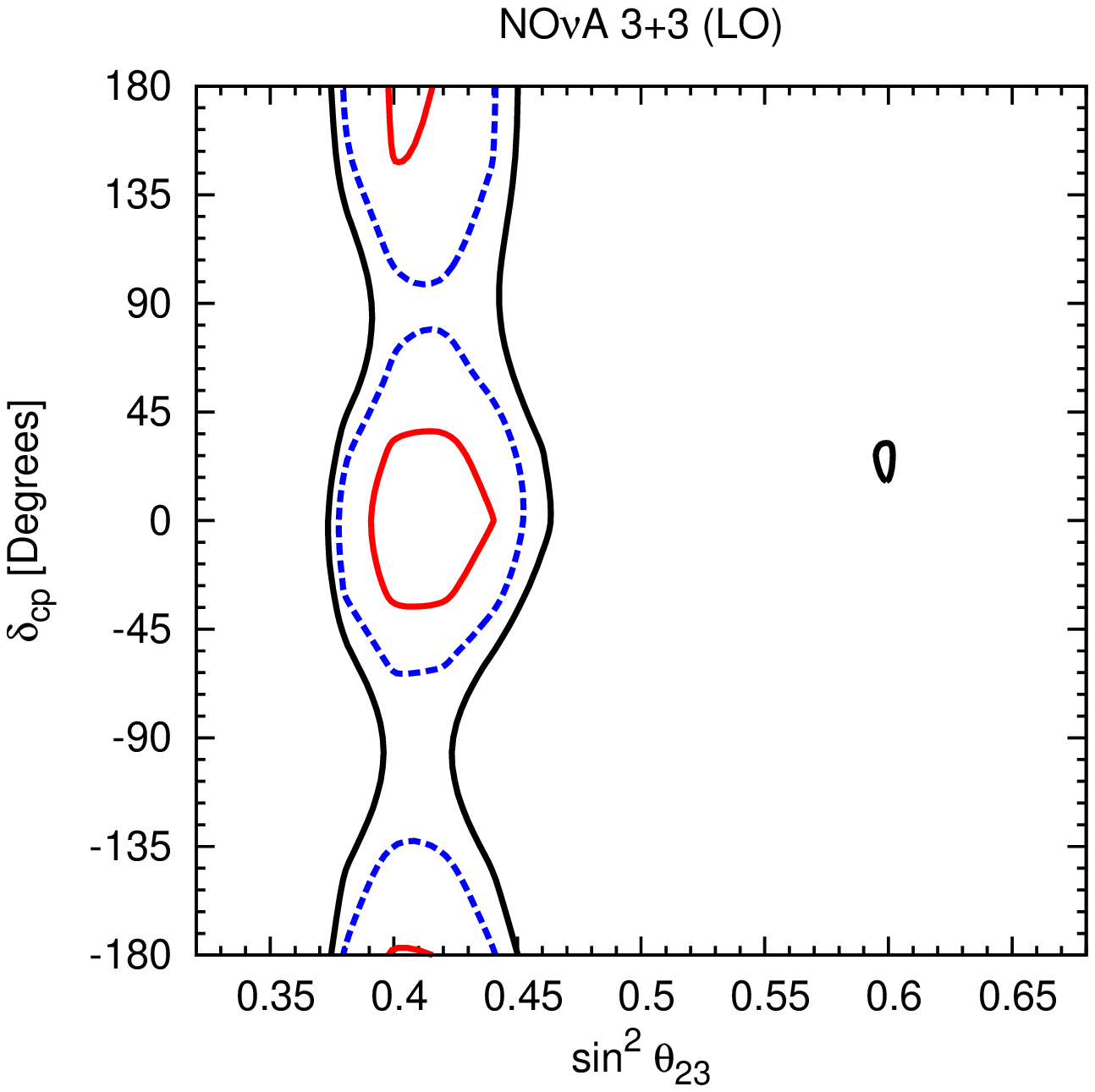}
\includegraphics[width=5cm,height=5cm]{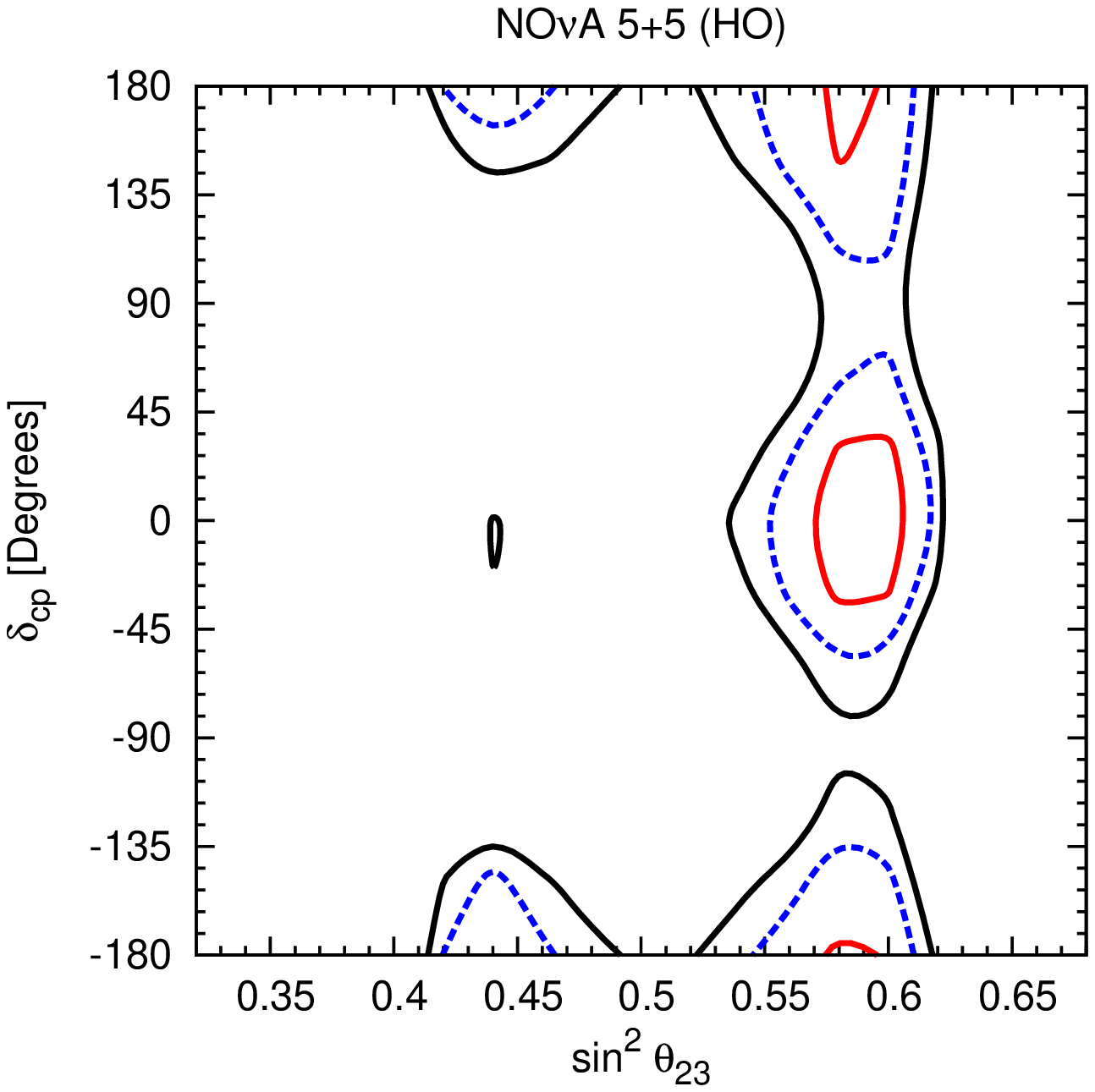}
\includegraphics[width=5cm,height=5cm]{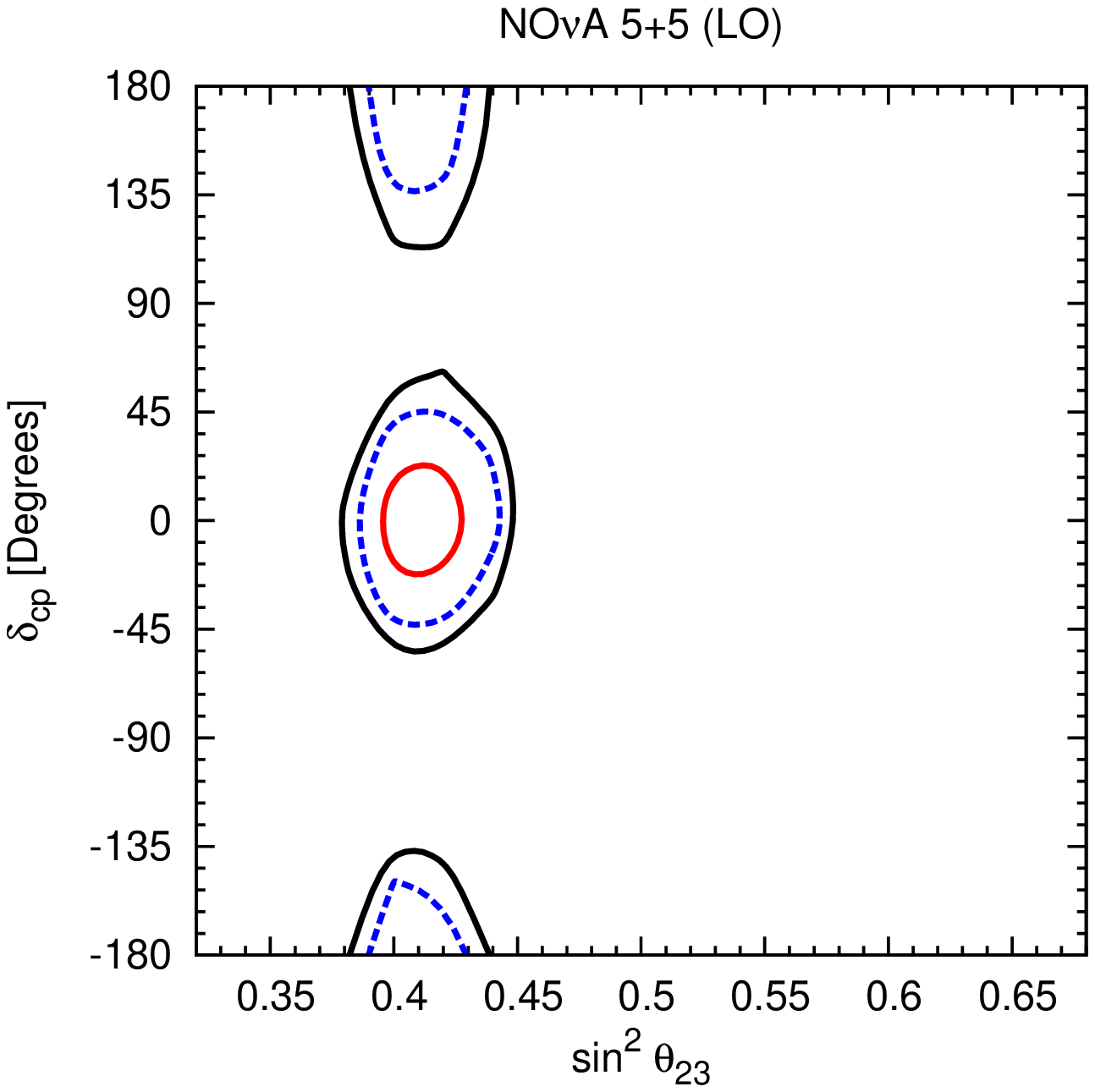}
\includegraphics[width=5cm,height=5cm]{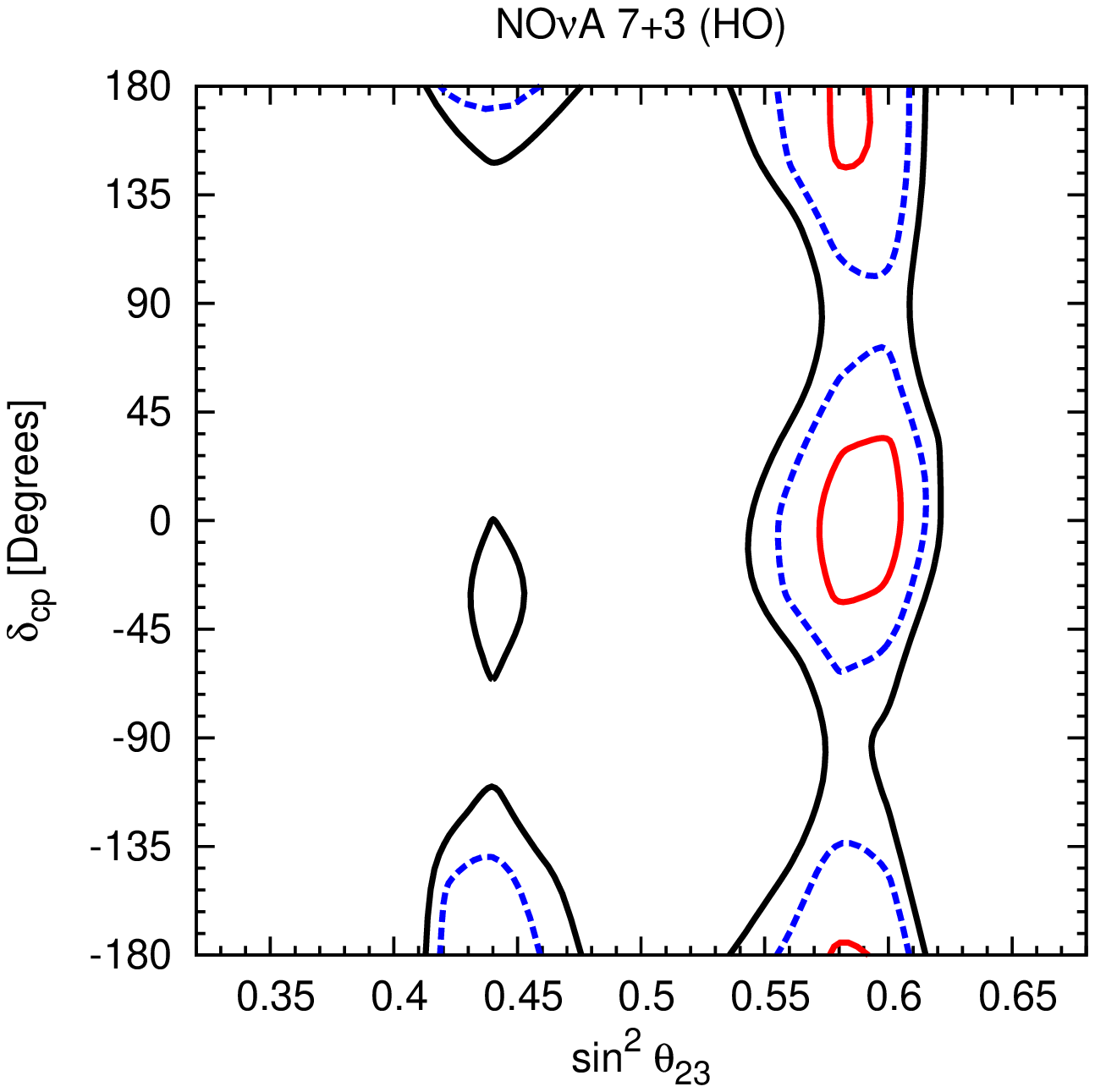}
\includegraphics[width=5cm,height=5cm]{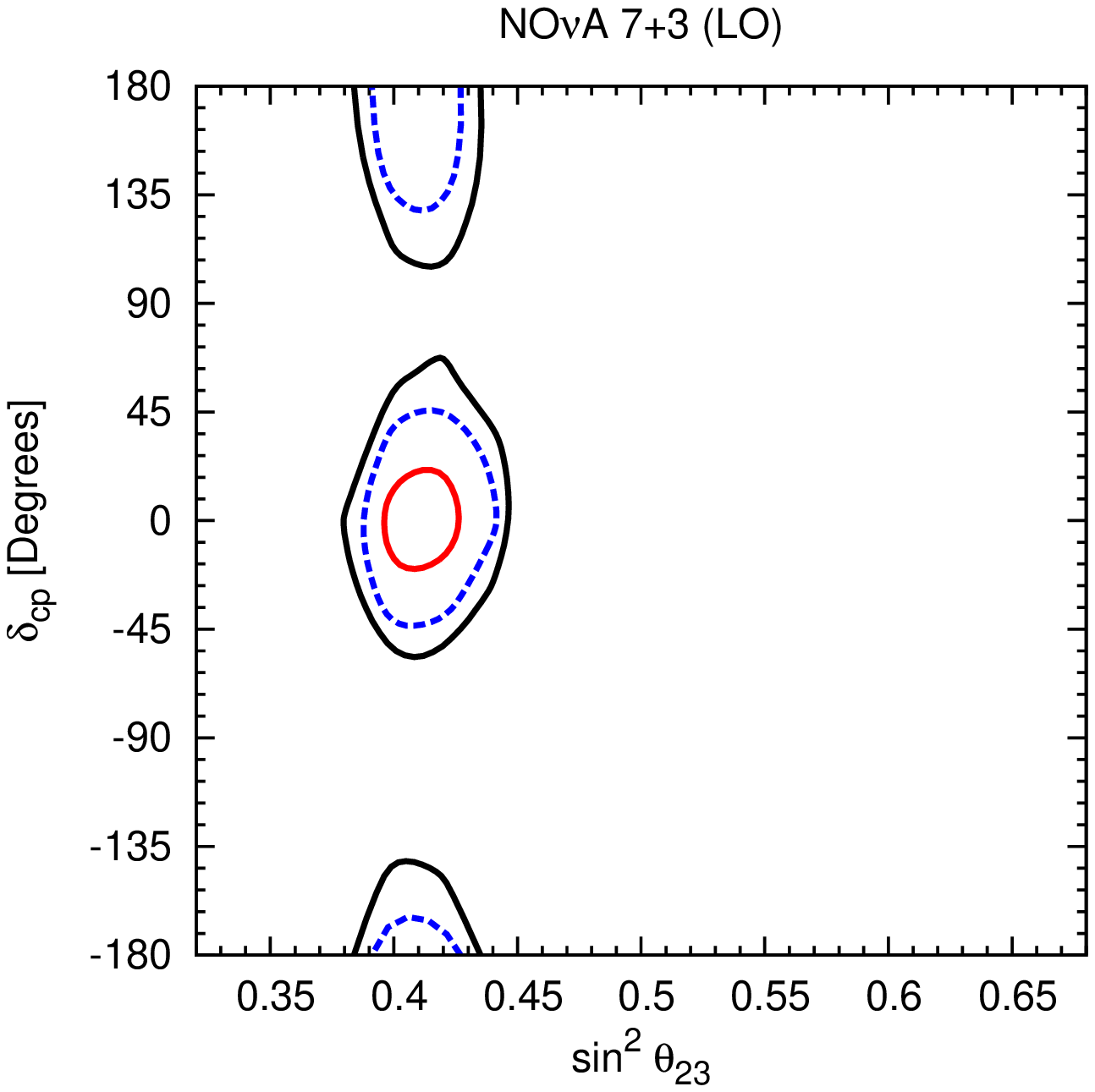}
\includegraphics[width=5cm,height=5cm]{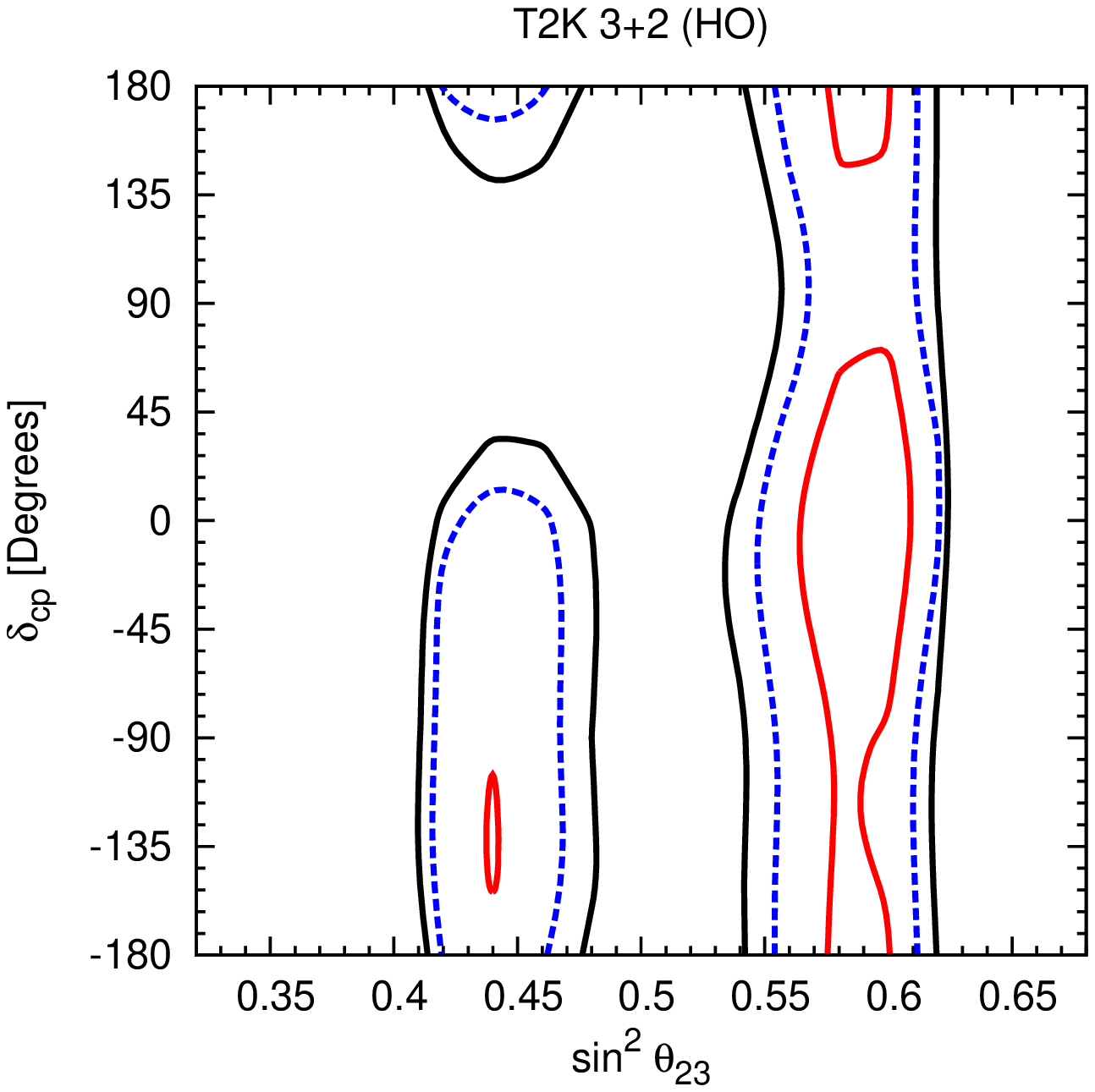}
\includegraphics[width=5cm,height=5cm]{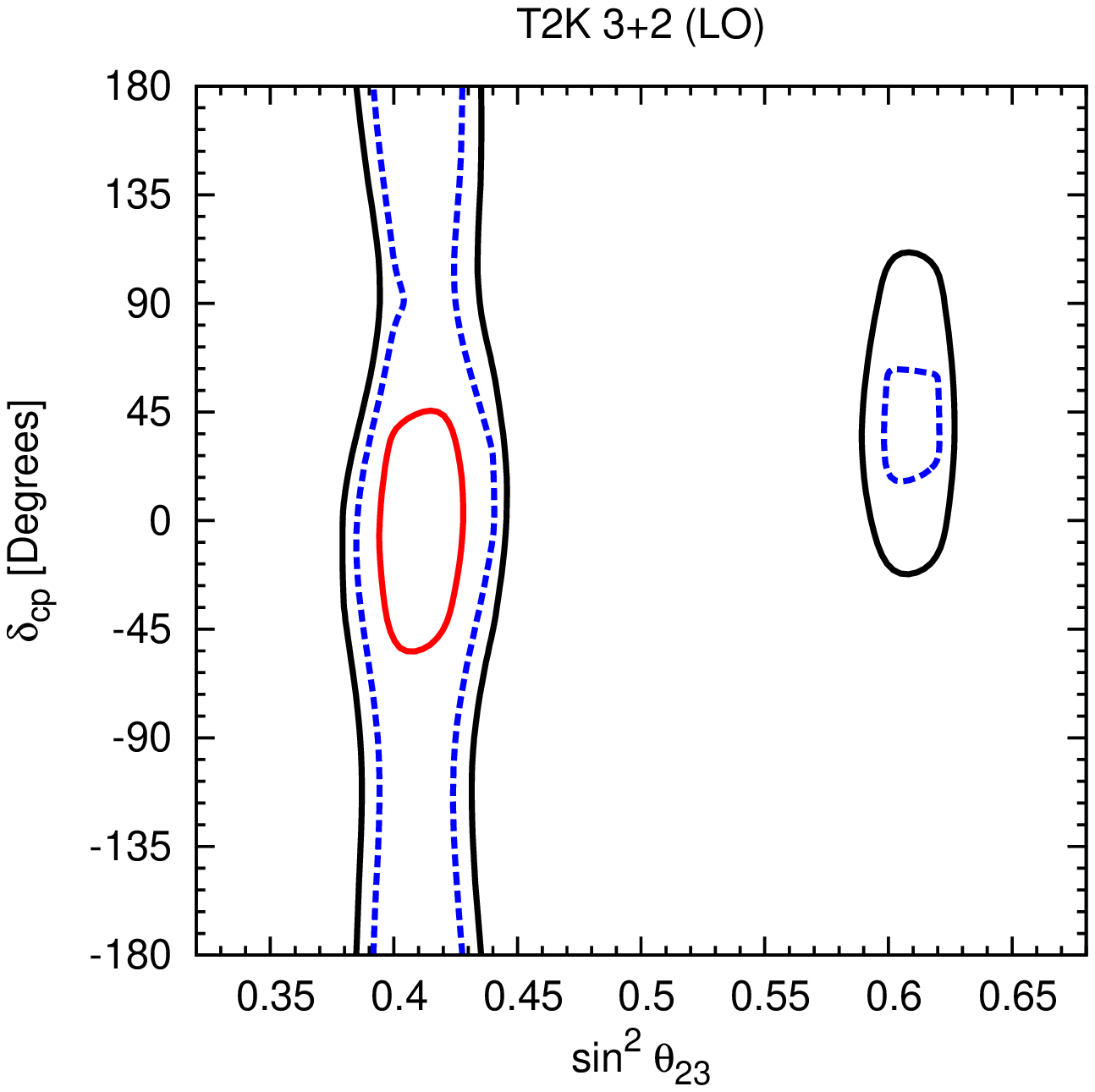}
\includegraphics[width=5cm,height=5cm]{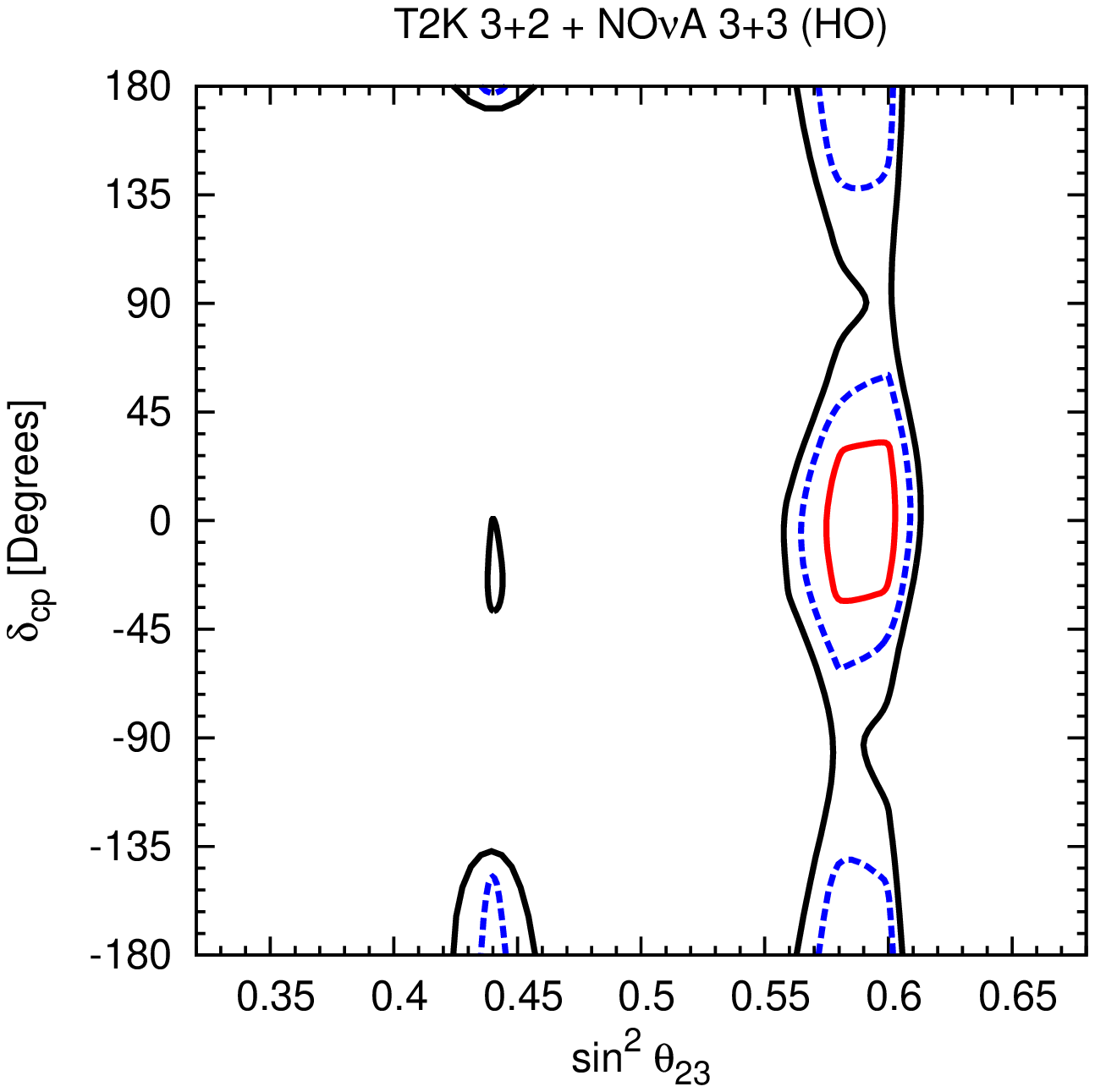}
\includegraphics[width=5cm,height=5cm]{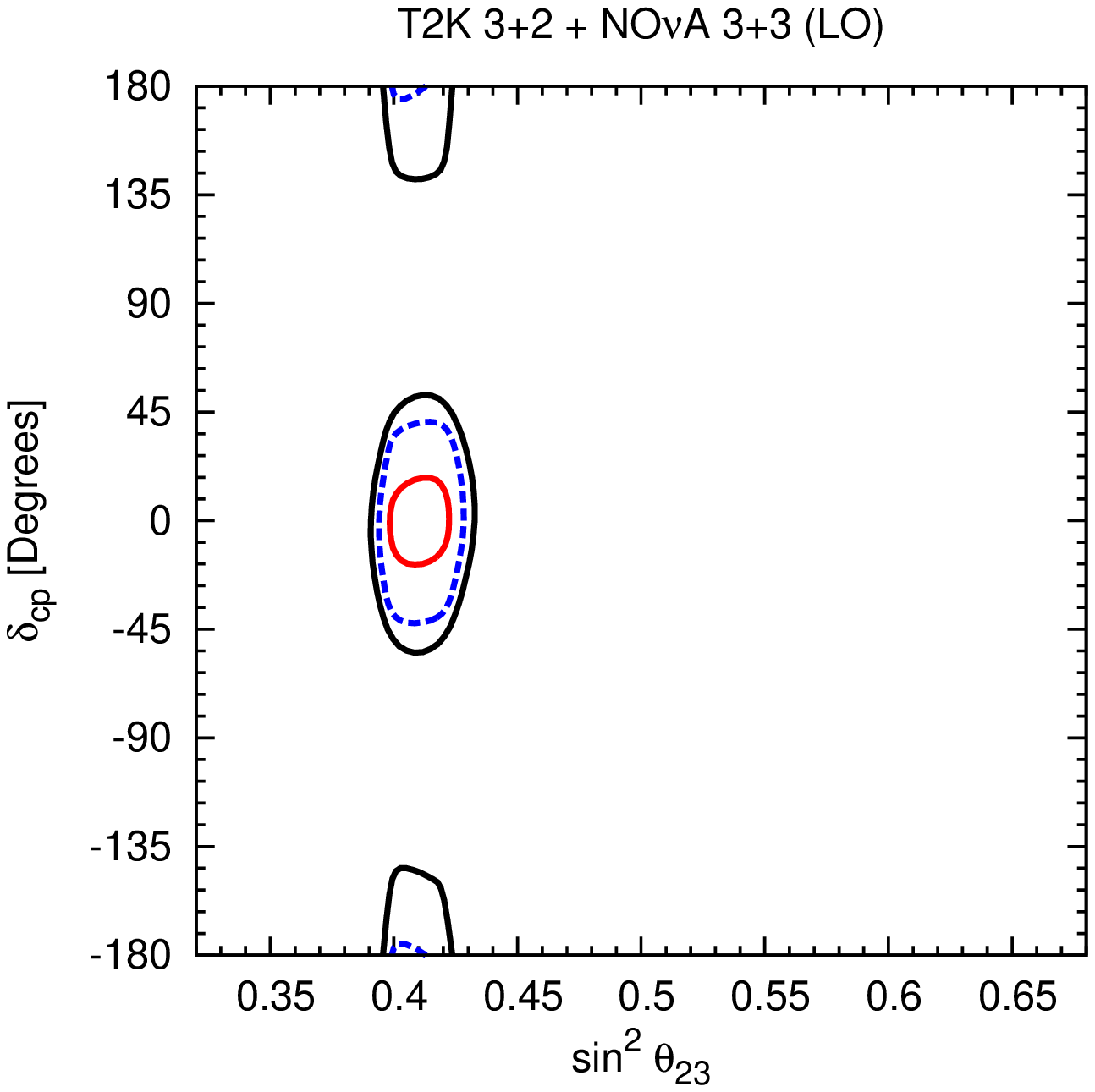}
\end{center}
\caption{Confidence region in   $\sin^2 \theta_{23}-\delta_{CP}$ plane for true  $\delta_{CP}=0$, where red, blue and black contours 
represent the 1$\sigma$ (68.3\% C.L.), 1.64$\sigma$ (90\% C.L.) and 2$\sigma$ (95.45\% C.L.) values respectively for 
two degrees of freedom. Here hierarchy is assumed to be IH.}
\end{figure}


\begin{figure}[htb]
\begin{center}
\includegraphics[width=5cm,height=5cm]{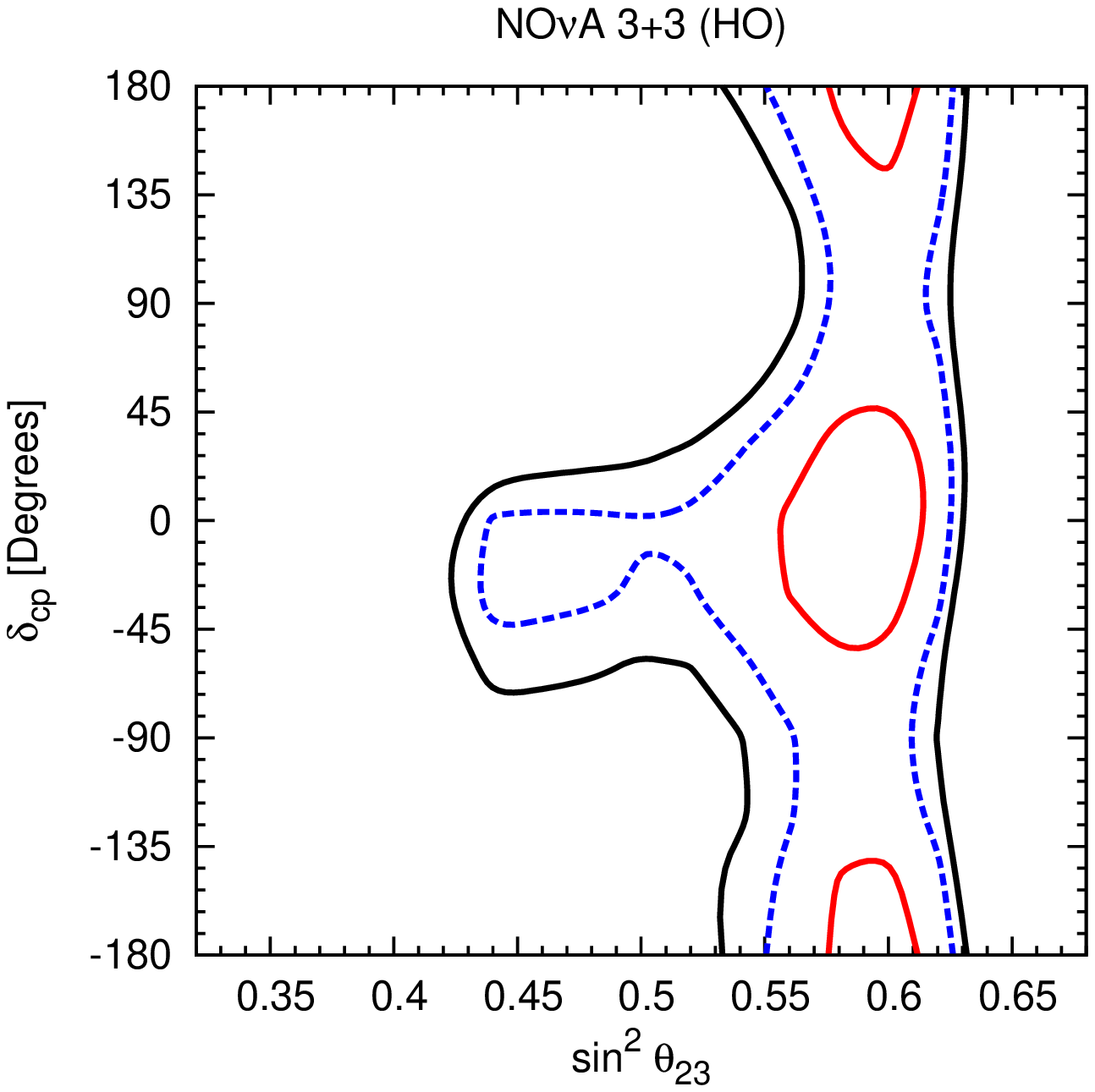}
\includegraphics[width=5cm,height=5cm]{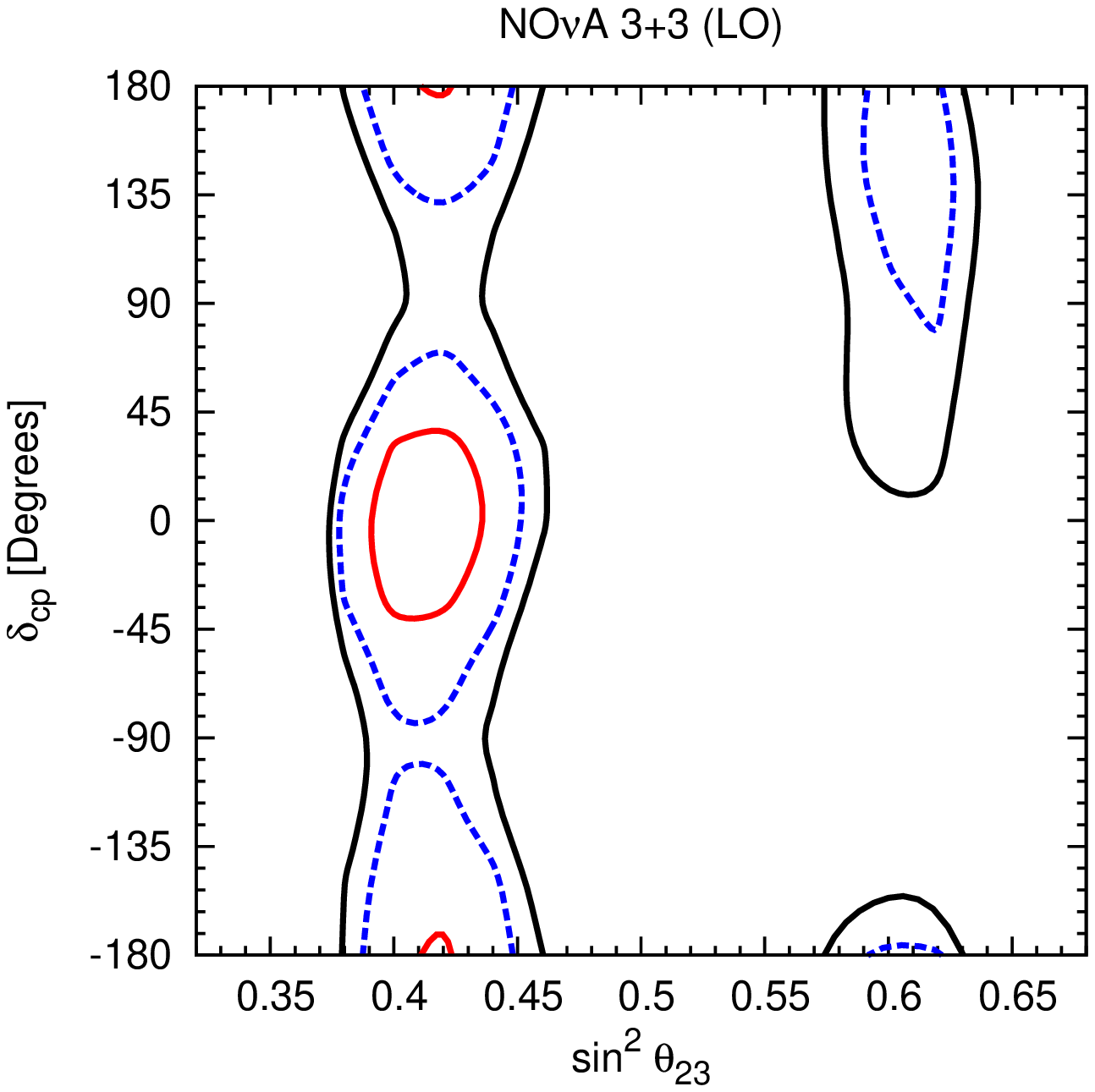}
\includegraphics[width=5cm,height=5cm]{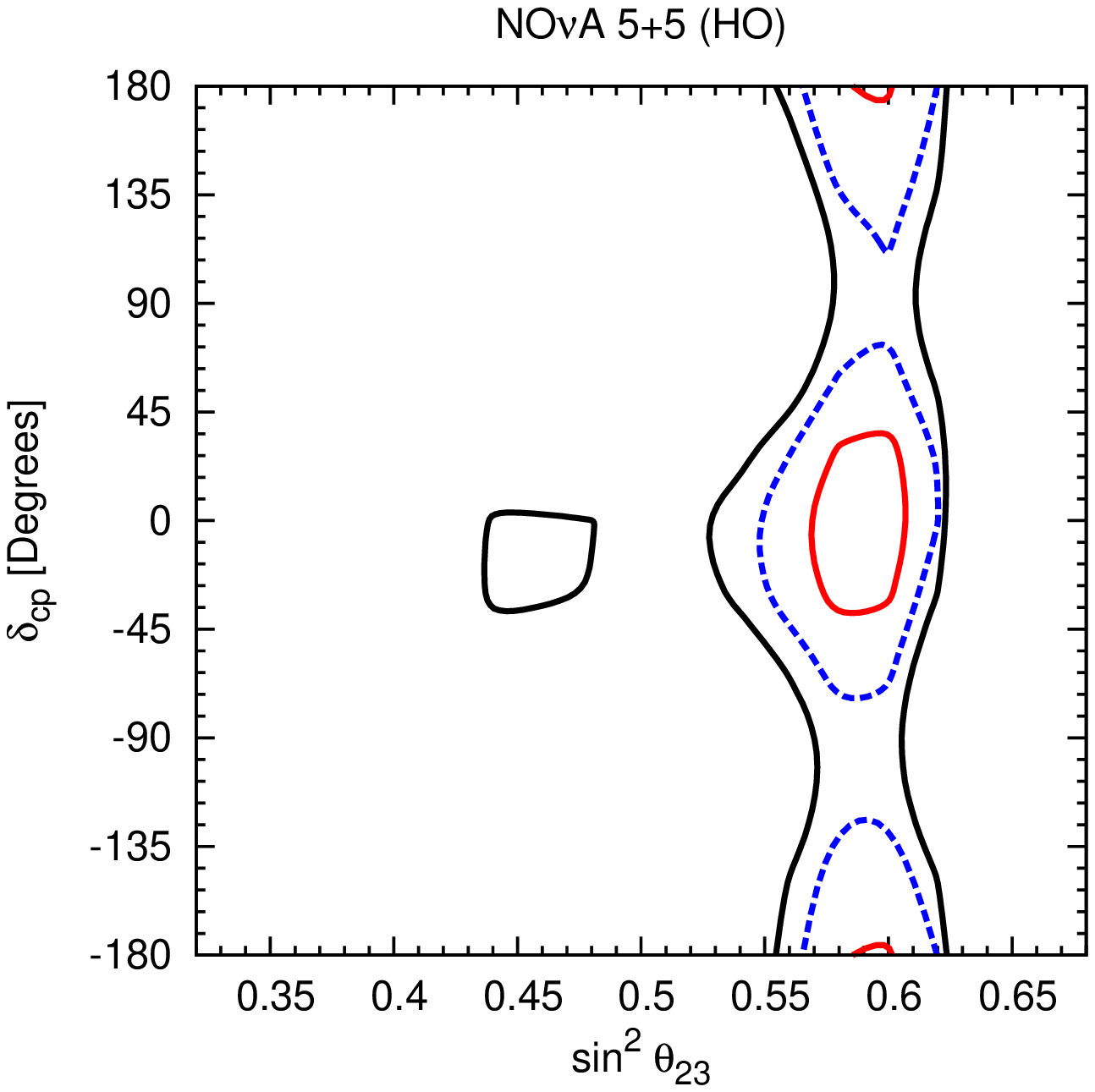}
\includegraphics[width=5cm,height=5cm]{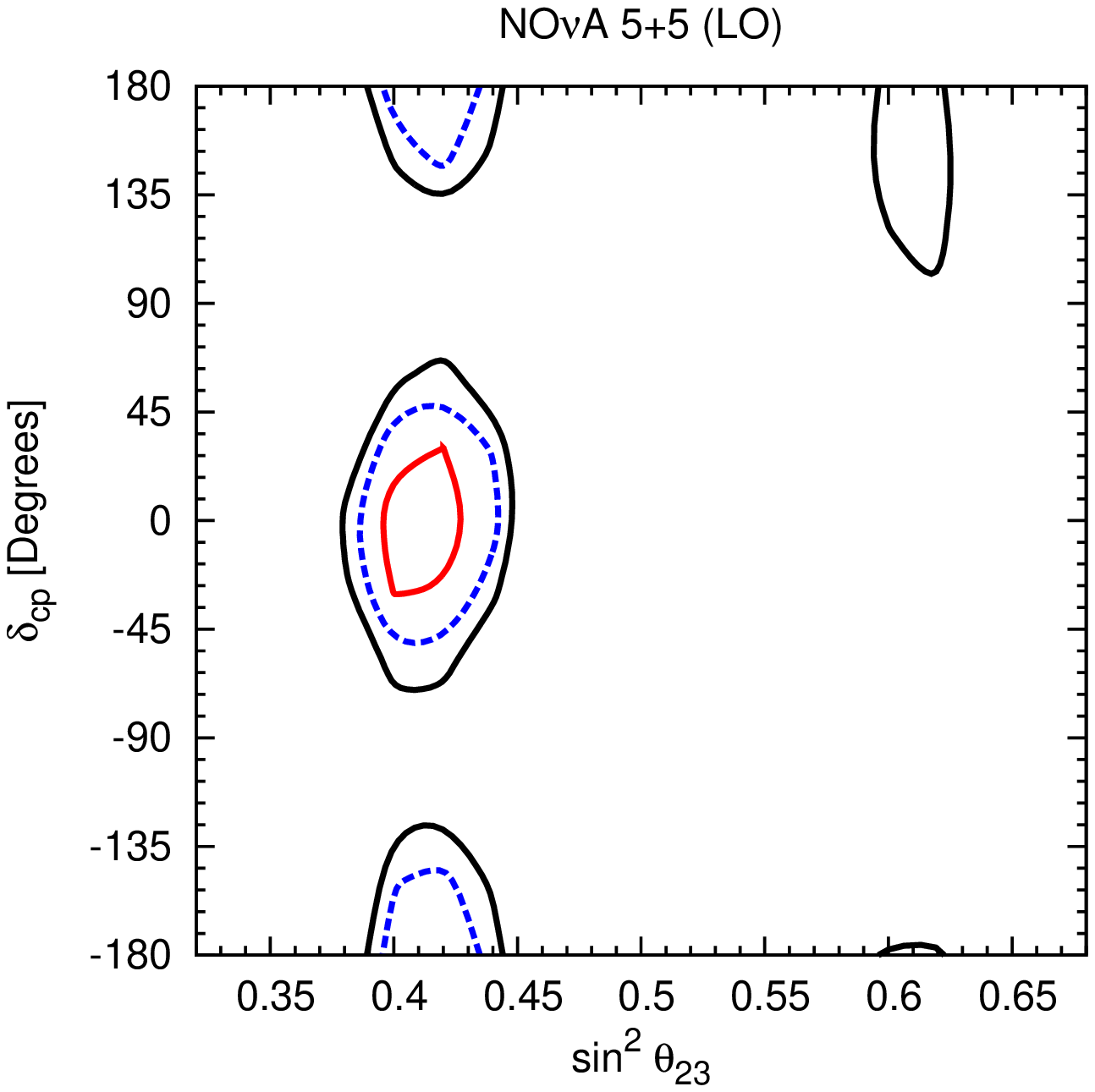}
\includegraphics[width=5cm,height=5cm]{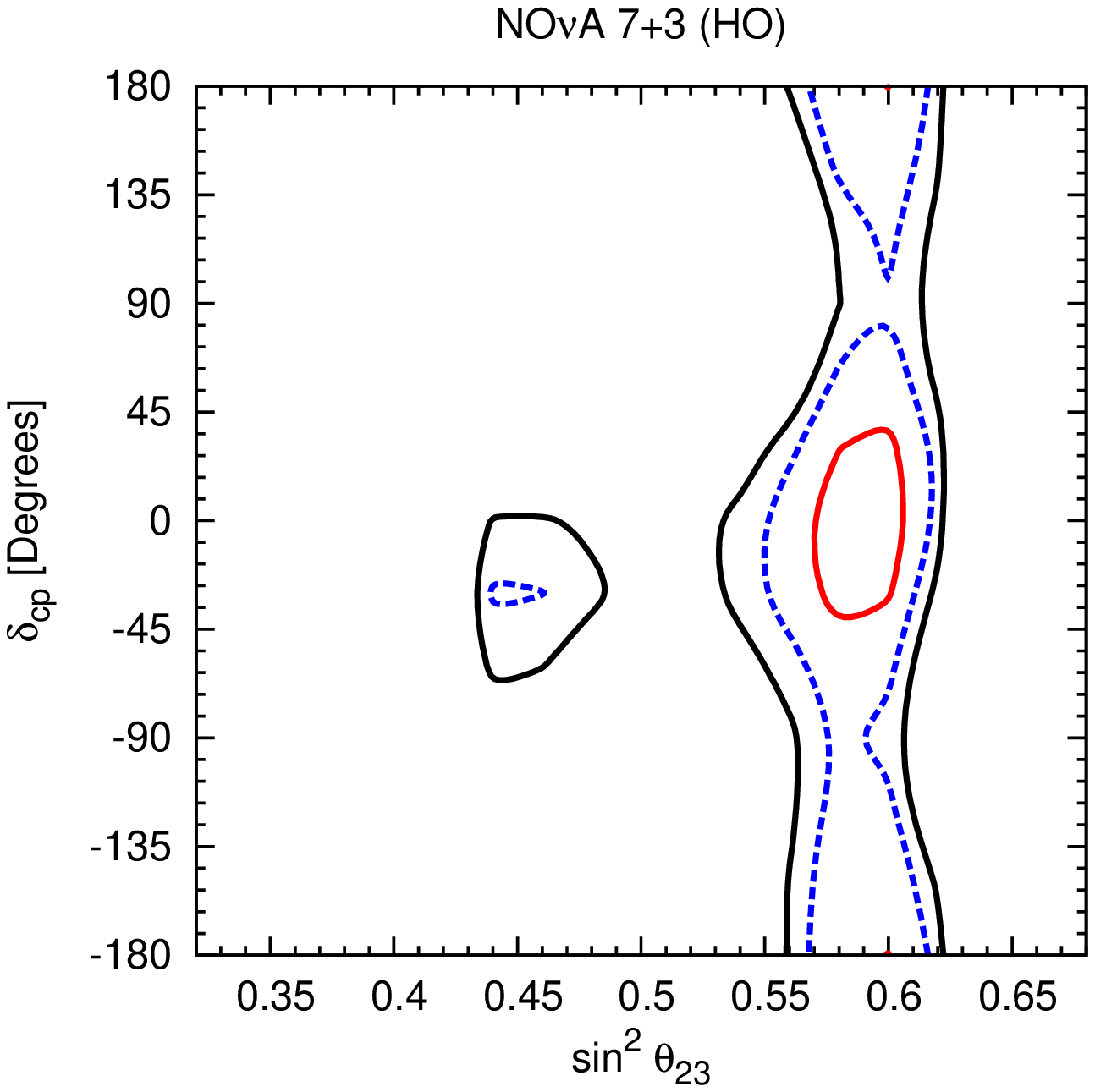}
\includegraphics[width=5cm,height=5cm]{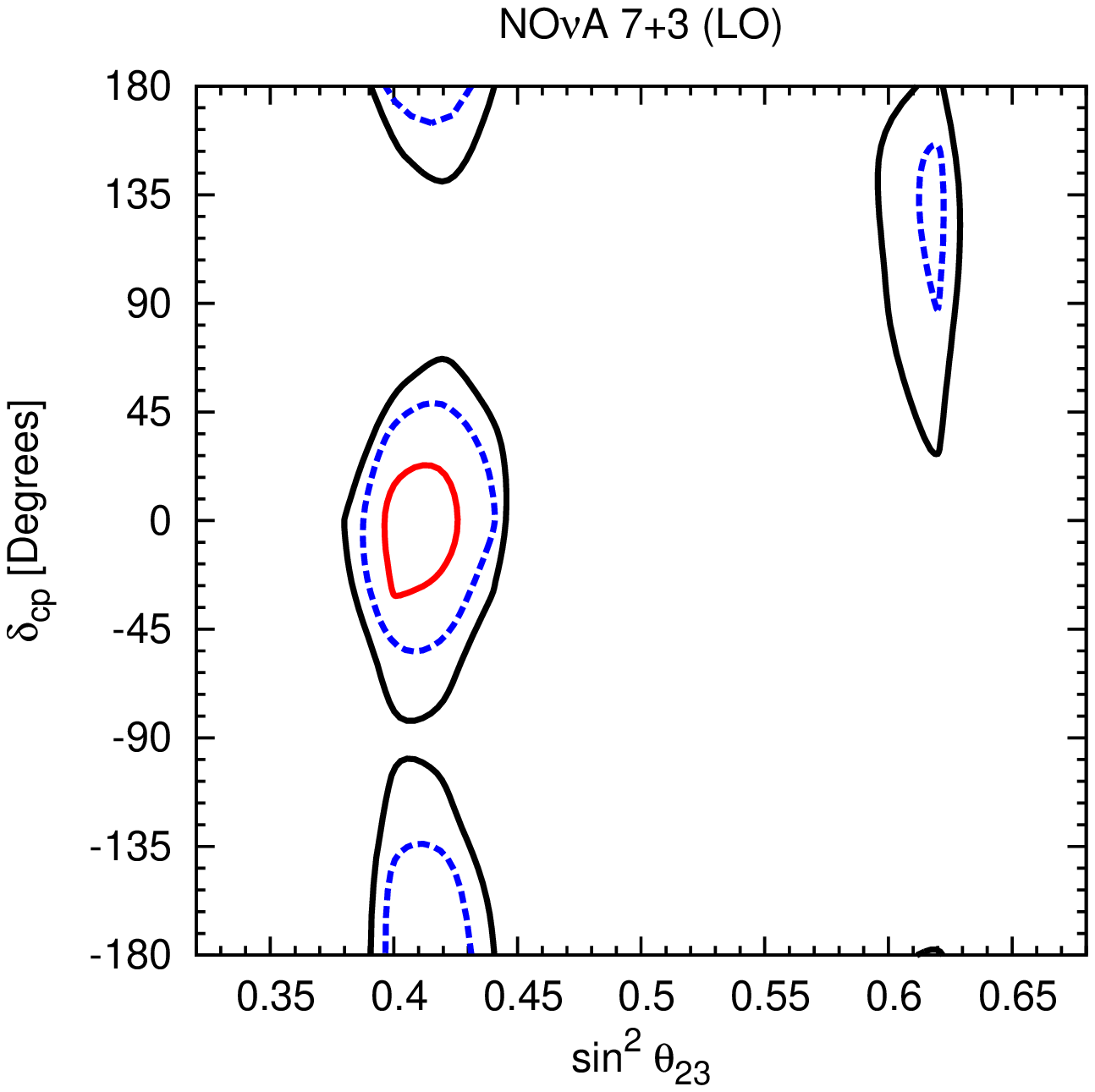}
\includegraphics[width=5cm,height=5cm]{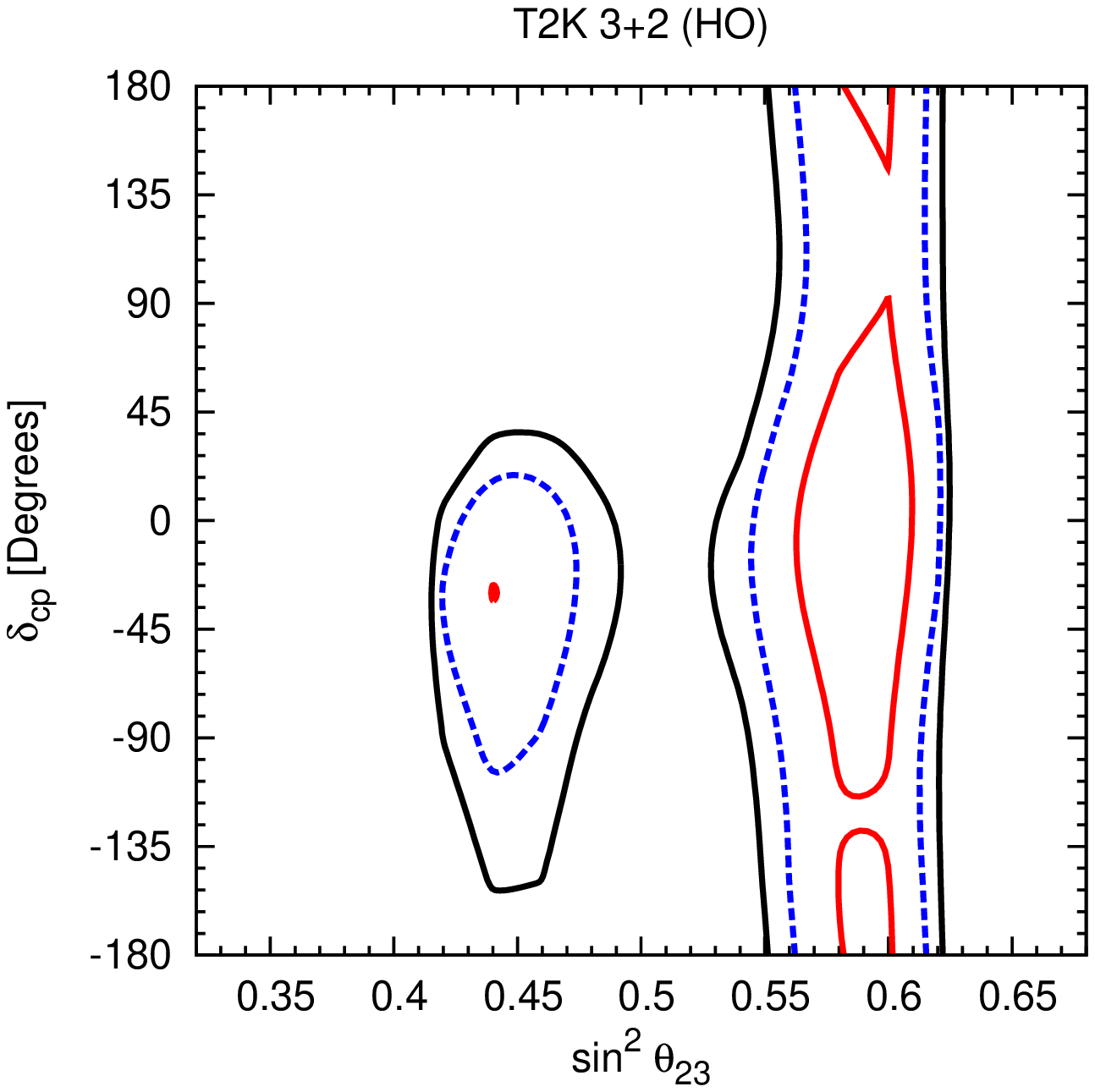}
\includegraphics[width=5cm,height=5cm]{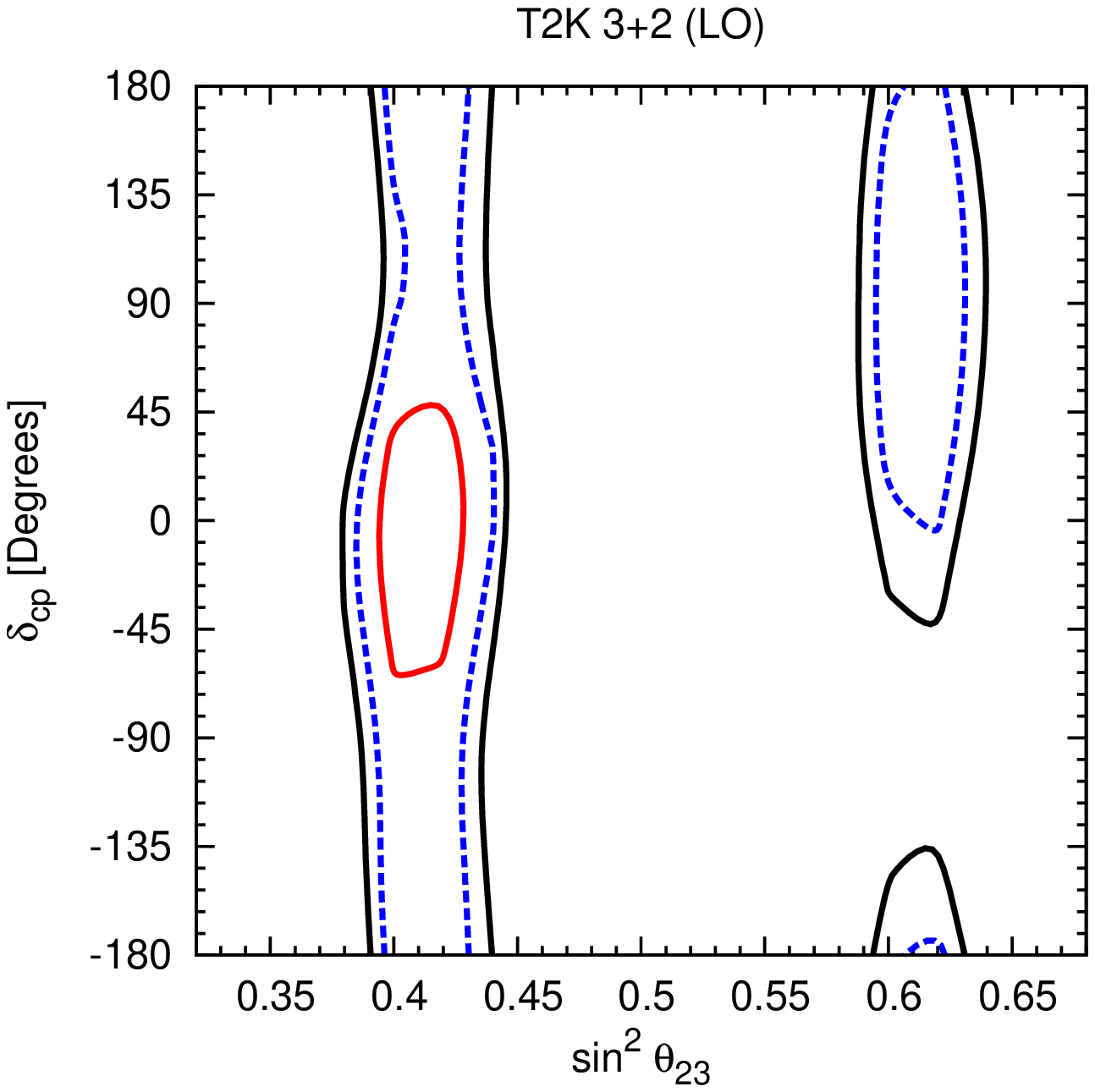}
\includegraphics[width=5cm,height=5cm]{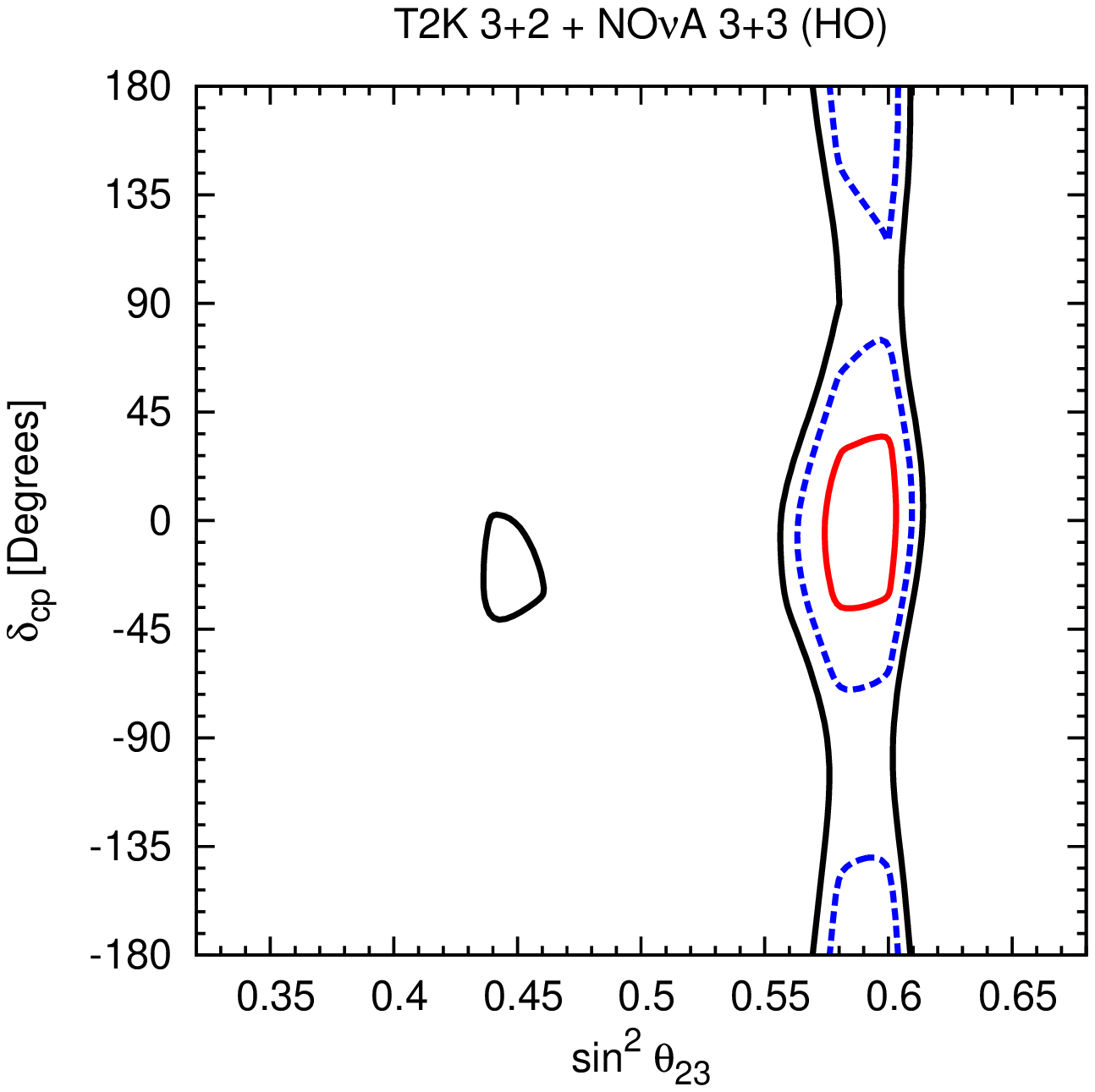}
\includegraphics[width=5cm,height=5cm]{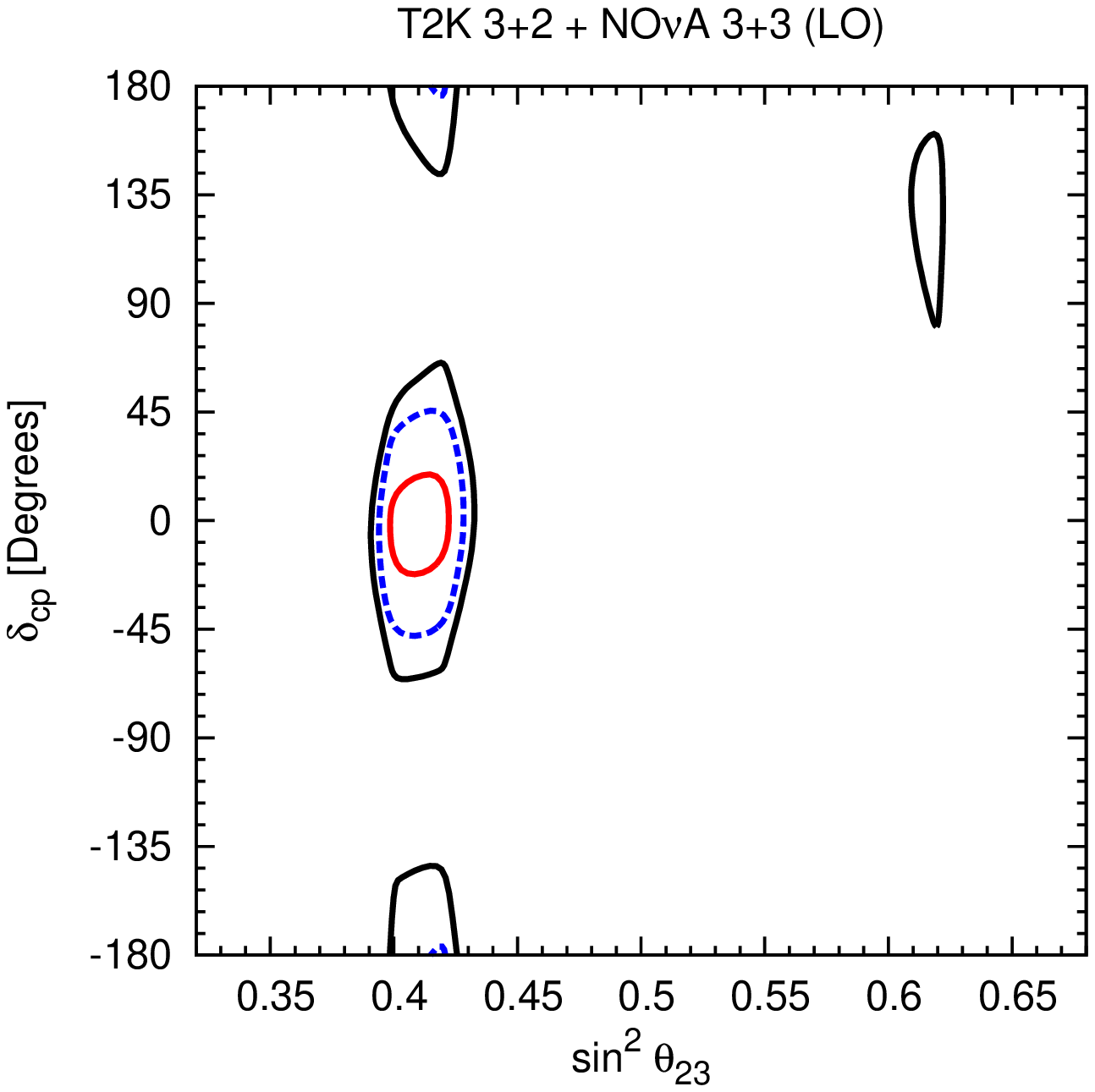}
\end{center}
\caption{Confidence region in   $\sin^2 \theta_{23}-\delta_{CP}$ plane for true  $\delta_{CP}=0$, where red, blue and black contours 
represent the 1$\sigma$, 1.64$\sigma$  and 2$\sigma$ values respectively for 
two degrees of freedom. Here hierarchy is assumed to be NH.}
\end{figure}

\subsection{Correlation between $\delta_{CP}$ and $\theta_{23}$}

From the previous subsection, we can see that the uncertainty in $\theta_{23}$ has a very large impact on determination of neutrino oscillation parameters.
Thus, it is important to understand the exclusive correlation between $\delta_{CP}$ and $\theta_{23}$ while keeping the true values of rest of 
the oscillation parameters to be fixed. In this subsection, we discuss the correlation between the oscillation parameters $\theta_{23}$ and $\delta_{CP}$. 
For our analysis, we have kept true values of oscillation parameters as in Table-III.
We vary the test values of $\sin^2 \theta_{23}$ and $\delta_{CP}$ in  their 3$\sigma$ ranges.
 The obtained results of the confidence regions in  $\sin^2\theta_{23}$ - $\delta_{CP}$ plane
are shown in Figs. 9 and 10 for all combinations of NO$\nu$A experiment.

\section{Summary and Conclusion}
With the recent discovery of the last unknown reactor mixing angle $\theta_{13}$, the mechanism  of three
flavor neutrino mixing pattern is now well established. But still
there are several issues related to neutrino oscillation parameters that  remain open, namely the
absolute mass scale of neutrinos, determination of the mass hierarchy,  octant of the atmospheric
mixing angle $\theta_{23}$, the magnitude of the CP violating phase $\delta_{CP}$ and the observation of CP violation in the neutrino sector. Therefore, the main focus of  the current and future oscillation experiments is to provide answers to some of these unsolved questions.

In this paper we have investigated the prospects of the determination of mass hierarchy, the octant of $\theta_{23}$ and the observation of CP violation in the neutrino sector due to $\delta_{CP}$ with the currently running accelerator based neutrino experiments NO$\nu$A and T2K and the forthcoming
T2HK experiment.
As the reactor mixing angle $\theta_{13}$ is now known  to be significantly large, the oscillation probability 
$P(\nu_\mu \to \nu_e)$ and its corresponding antineutrino counterpart are sensitive for the determination of mass hierarchy and $\theta_{23}$ octant. We found that T2K experiment with (3$\nu$ +2$\bar \nu$) years of running can resolve the octant degeneracy with nearly 2$\sigma$ C.L. if the true value of $\theta_{23}$
to be around $\sin^2 \theta_{23}=0.41$ (LO) or $\sin^2 \theta_{23}=0.59$ (HO). The sensitivity increases
to nearly $3\sigma$ with (3$\nu$ +3$\bar \nu$) years running of NO$\nu$A. However, if we combine the data from these 
two experiments the sensitivity increases significantly than the sensitivities of individual
experiments. Furthermore, if we assume that  NO$\nu$A continues data taking for 10 years then octant
degeneracy can be resolved with NO$\nu$A experiment alone with more than $3 \sigma$ significance.  For the determination of  mass hierarchy,
it is also possible to rule out nearly one-third of the $\delta_{CP}$ space at 3$\sigma$ C.L. if we use the synergy between NO$\nu$A and T2K experiments. 
In this case the sensitivity increases significantly for
ten years of running of NO$\nu$A with $(5+5)$ combination is found to be more suitable than
the combination of (7+3) years.

Measuring CP violation in the lepton sector is another important challenging problem today.
We have also performed a systematic study of the CP sensitivity of the current
long-baseline experiments T2K and NO$\nu$A. Although these experiments are not planned to study
leptonic CP violation, we analyze the synergies between these set-ups which may aid
in CP violation discovery by constraining the value of $\delta_{CP}$. Although dedicated long-baseline experiments like DUNE, 
T2HK are planned to study CP violation in neutrino sector, we may have the
the first hand information on $\delta_{CP}$ from these experiments much before those dedicated facilities are operational. 
We found that T2K by itself has marginal CP violation sensitivity at 1$\sigma$ CL. For NO$\nu$A  with (3+3) years of running
there will be CP violation sensitivity above 1.5$\sigma$ level for about one-third of the CP violating phase $\delta_{CP}$ space. The sensitivity
increases slightly for 10 years of run time, with (5$\nu$ + 5$\bar \nu$) combination having sensitivity than that of  (7$\nu$ + 3$\bar \nu$) combination. 
The data from T2HK experiment will improve the CPV sensitivity  significantly.
We have also found that the CP violating phase $\delta_{CP}$ can be determined to be better than 
 $35^\circ $, $21^\circ $ and $9^\circ $ for all values of $\delta_{CP}$ for T2K, NO$\nu$A and T2HK experiments.
We  also obtained the Confidence regions in the
$\delta_{CP}-\theta_{13} ~(\theta_{23})$ plane for both T2K and NO$\nu$A experiments.\\

{\bf Conflict of Interest}
The authors declare that there is no conflict
of interest regarding the publication of this article.\\

{\bf Acknowledgments}
SC and KND  would like to thank University Grants Commission for financial support.
The work of RM was partly supported by the Council of Scientific and Industrial Research,
Government of India through grant No. 03(1190)/11/EMR-II. We would like to thank Drs.
S.K. Agarwalla and S.K. Raut for many useful discussions regarding GLoBES.


\end{document}